\documentclass[article]{aa} 
\usepackage{epsfig,deluxe} 
\begin{document}

\newcommand{\gsim}{\hbox{\rlap{$^>$}$_\sim$}} 
  \thesaurus{06;  19.63.1} 
%
\authorrunning{S. Dado, A. Dar \& A. De R\'ujula} 
\titlerunning{Afterglows of GRBs} 
\title{On the Optical and X-ray Afterglows of Gamma Ray Bursts} 
 
\author{Shlomo Dado$^{^1}$, Arnon Dar$^{^1}$ and 
A. De R\'ujula$^{^2}$} 
\institute{1. Physics Department and Space Research Institute, Technion\\ 
               Haifa 32000, Israel\\ 
           2. Theory Division, CERN, CH-1211 Geneva 23, Switzerland}

\maketitle           
 
\begin{abstract} 
We severely criticize the consuetudinary analysis of the 
afterglows of gamma-ray bursts (GRBs) in the conical-ejection 
fireball scenarios. We argue that, instead, recent 
observations imply that the long-duration GRBs 
and their afterglows are produced by highly relativistic jets of 
cannonballs (CBs) emitted in supernova explosions. 
The CBs are heated by their collision with the supernova shell. 
The GRB is the boosted surface radiation the CBs emit as they reach 
the transparent outskirts of the shell. 
The exiting CBs further decelerate by sweeping up 
interstellar matter (ISM). The early X-ray afterglow is dominated by thermal 
bremsstrahlung from the cooling CBs, the optical afterglow by synchrotron 
radiation from the ISM electrons swept up by the CBs. 
We show that this model fits simply and remarkably 
well all the measured optical afterglows of the 15 GRBs with known 
redshift, including that of GRB 990123, for which unusually prompt 
data are available. We demonstrate that GRB 980425 was a normal 
GRB produced by SN1998bw, with standard X-ray and optical afterglows. 
We find that the very peculiar afterglow of GRB 970508
can be explained if its CBs  encountered a significant
jump in density as they moved through the ISM. 
The afterglows of the nearest 8 of the known-redshift GRBs 
show various degrees of evidence for an association 
with a supernova akin to SN1998bw. In all other cases such an 
association, even if present, would have been undetectable 
with the best current photometric sensitivities. 
This gives strong support to the proposition that most, maybe all, 
of the long-duration GRBs are associated with supernovae. 
Though our emphasis is on optical afterglows, we also 
provide an excellent description of X-ray afterglows. 
 
\end{abstract} 
 

\section{Introduction} 
 
Our information about the once totally mysterious gamma-ray bursts (GRBs) 
has increased spectacularly in the past few years. 
The rapid directional localization of gamma-ray bursts by the 
satellites BeppoSAX (e.g. Costa et al.~1997), Rossi (e.g. Levine et al.~1996) 
and by the Inter-Planetary Network (IPN) of the spacecrafts Ulysses, 
Konus-Wind and NEAR (e.g. Cline et al.~1999) led to 
a flurry of progress: the discovery of 
long-duration GRB afterglows (Costa et al.~1997; Van Paradijs et al.~1998); 
the discovery of the GRBs' host galaxies (Sahu et al.~1997a); 
the measurement of their redshifts (Metzger et al.~1997b) that verified their 
cosmological  origin (e.g. Paczynski 1986; Meegan et al.~1992); the 
identification of their birthplaces ---mainly star formation regions in 
normal galaxies (e.g. Paczynski 1998; Holland and Hjorth 1999)--- 
and the first evidence for a possible association between GRBs and supernova 
explosions (Galama et al.~1998a). 
 
The enormous isotropic energies inferred from the redshifts and fluences of GRBs 
and their short-time variability have indicated that 
the GRBs must be produced by gravitational stellar collapse (Goodman et 
al. 1987, Dar et al.~1992). The prevalent belief is that they are 
generated by synchrotron emission from relativistic fireballs produced by 
mergers of compact stars (Paczynski 1986; Goodman 1986; Goodman et al.~1987), 
by hypernova explosions (Paczynski 1998), or by relativistic 
``firecones'' (e.g. Rhoads 1997, 1999) from collapsars or failed supernovae 
(Woosley 1993;  Woosley \& MacFadyen 1999; MacFadyen 
\& Woosley 1999; MacFadyen et al.~2001). But various observations, 
including in particular the ones 
we shall extensively discuss here, strongly suggest that most of the 
long-duration GRBs are produced in supernova events (Dar \& De R\'ujula 
2000a and references therein) by highly collimated 
superluminal jets (e.g. Shaviv \& Dar 1995; Dar 1997; Dar \& Plaga 1999). 
 
Various authors (e.g. Rhoads 1997, 1999; MacFadyen and Woosley 1999; 
Sari et al.~1999) have merged the notion that GRBs are produced by 
highly relativistic jets (e.g. Brainerd 1992; Woosley 1993; 
Shaviv \& Dar 1995; Dar 1997; Dar 1998a; 
Dar \& Plaga 1999) with the popular fireball 
models of GRBs (see, e.g. Piran 1999  and references therein) 
to morphe the concept of ``firecones'' or similar denominations.  
Firecone considerations are used to analyze ``breaks'' in GRB 
afterglows (Rhoads 1997, 1999; Sari et al.~1999; Frail et al.~2001) and to 
extract properties of the GRB engine and ejecta (Frail et al.~2001, and 
references therein). 
The fireball idea that all radiation from GRBs originates from 
colliding shocks is certainly interesting and worth studying. 
But, concerning the evolution of AGs, 
the idea remains essentially untested, given the cavalier 
treatment it has received in much of the recent literature. In 
Section 2 we explain this harsh opinion. 
 
In recent papers the idea of jetted, supernova-associated 
 GRBs was made entirely explicit with the introduction of a 
relativistic cannonball (CB) model of GRB production in supernova 
explosions (Dar \& De R\'ujula 2000a, 2000b, 2001a, 2001b,  hereafter 
DD2000a, etc.). The CB model is completely different from the firecone scenarios, 
as we explain in Section 2. The CB model, we contend, explains the 
main observed features of long-duration GRBs and of their afterglows. In 
particular, in DD2000b, we have demonstrated that the CB model predicts 
the temporal and spectral properties of the bursts of 
$\gamma$-rays correctly. 
 
In this paper we derive the detailed predictions of the CB model for the GRB 
optical afterglows (AGs), which we only sketched in DD2000a. 
We compare the predictions, which are analytic in fair 
approximations, with the observed optical 
AG of all the GRBs with known redshift. We show that the CB model 
describes remarkably well these optical AGs, as well as the measured 
X-ray AGs of these GRBs. 
 
Our detailed analysis of the AGs allows us to show how the 
 nearest eight GRBs with measured redshift show varying degrees of 
evidence of an association 
with a supernova (SN) akin to SN1998bw: superimposed on the smooth 
AG of these GRBs one can discern the light curve of SN1998bw, 
adequately translated and red-shifted in luminosity distance, time-dependence 
 and spectrum (e.g. Dar 1998b; DD2000a). In all other 
cases, either there are no late-time measurements of the optical 
AG, or the SN contribution is too dim to be resolved from the late 
GRB afterglow or the host galaxy light, even by the HST or the most powerful 
ground-based telescopes.

In spite of the fact that we use similar vocabulary in what concerns 
the GRB engine (which is quite irrelevant to the AG properties 
discussed here) the CB model is completely different from the collapsar 
model of GRBs (Woosley 1993; Woosley \& MacFadyen 1999; MacFadyen and 
Woosley 1999; MacFadyen et al.~2001). The crucial differences are 
explained in DD2000a and DD2000b.  One of them is that 
we use empirical facts as our guiding line, rather than the results of 
simulations that fail to explode SNe and lack a proper treatment 
of features that are no doubt relevant (relativistic dynamics, 
angular momentum and its transport, magnetic fields). 
Concerning AGs, the CB model does have a predecessor, 
the ``plasmoid model'', 
of Chiang \& Dermer (1997). The CB model, however, differs in many 
crucial details; it is more complete and, unlike the plasmoid model, it 
is very successful in describing the observations.

We devote sections 3 to 6, to make the paper self-contained, 
to a brief review of the CB model. The novel theoretical core of 
the paper is in sections 7 to 10. We propose there, in particular, 
a simple mechanism governing the pace of radial expansion 
of a CB. The mechanism naturally explains the fact, observed in 
quasar and microquasar ejections, that {\it cannonballs}, faithful 
to their name, essentially stop expanding at some point of their 
voyage. We also derive predictions, which turn out to be in 
disagreement with observations, for the case of CBs that would 
continuously expand in an inertial manner. 
The remaining sections are devoted to the description of 
the AG data in the CB model, to a detailed comparison 
with observations, which turns out to be extremely successful, 
and to the extraction of conclusions.  
 
Three GRBs deserve special mention in their CB model interpretation. 
GRB 980425 turns out to be entirely ``normal'', it is uncommonly 
near ($\rm z=0.0085$) but its emitted CBs are observed at an unusually large 
angle, giving it a normal $\gamma$-ray fluence. This interpretation  (DD2000a) 
is strengthened by the fact that we successfully predict its 
optical AG (dominated up to 20 months by SN1998bw) from its 
X-ray AG, which is entirely due to the CBs, and normal. 
GRB 970508 has an extremely peculiar AG light curve, which 
can be easily explained, but only if its CBs encountered a significant jump
in density as they moved through the ISM.
There are uncommonly early 
data on the optical AG of GRB 990123, which 
fit the expectations for CBs that are moving through the wind 
of the parent star, in its Wolf-Rayet phase.

\section{Uses and abuses of fireballs} 
 
In this section we place the CB model of GRB afterglows in the perspective of 
the generally accepted views on the subject, which are based on the ``fireball'' 
model and on modifications thereof. This serves a double purpose: 
it clarifies how completely different the CB model is from the fireball 
ones, and it shows how unconvincing a fraction of the fireball literature is. 
 
In the fireball model, reviewed in Piran (1999, 2000) and Meszaros (2001), 
 both the $\gamma$-rays and the AG of a GRB are made by 
synchrotron radiation in inward- and outward-moving shocks, which 
are produced as relativistically expanding shells collide with each other 
and with interstellar material. The attitude is often espoused 
that the actual engine producing these colliding 
ejecta need not always be explicitly discussed. The possibility 
that the ejecta may not be spherically distributed has been repeatedly 
studied in the literature, but, in the fireball model, this was not done 
in detail prior to the influential papers of Rhoads (1997, 1999), who predicted 
abrupt breaks in the power law of the AG light curves. 
With the advent of GRB 990123, with its record equivalent spherical 
energy (see Table I) and an AG light curve through which it is 
possible to draw a broken power law (e.g. Figs. 1--4 of Holland et al.~2000a) 
the fireball advocates (see, e.g. Frail et al.~2001) adopted the 
arguments in favour of collimated GRBs (e.g. in the case of 
GRBs from quasars, Brainerd 1992; in the case of a funnel in 
an explosion, Meszaros \& 
Rees 1992; and in the case of jets in gravitational collapses, 
Shaviv \& Dar 1995, Dar 1997, Dar 1998a; Dar \& Plaga 1999, DD2000a and 
references therein). So have fireballs evolved into ``collimated fireballs'', 
``firecones'' or ``conical fireball jets'', while maintaining the ``fire'' 
lineage. 
 
\begin{figure} 
\begin{center} 
\vspace*{.003cm} 
\hspace*{-0cm} 
\epsfig{file=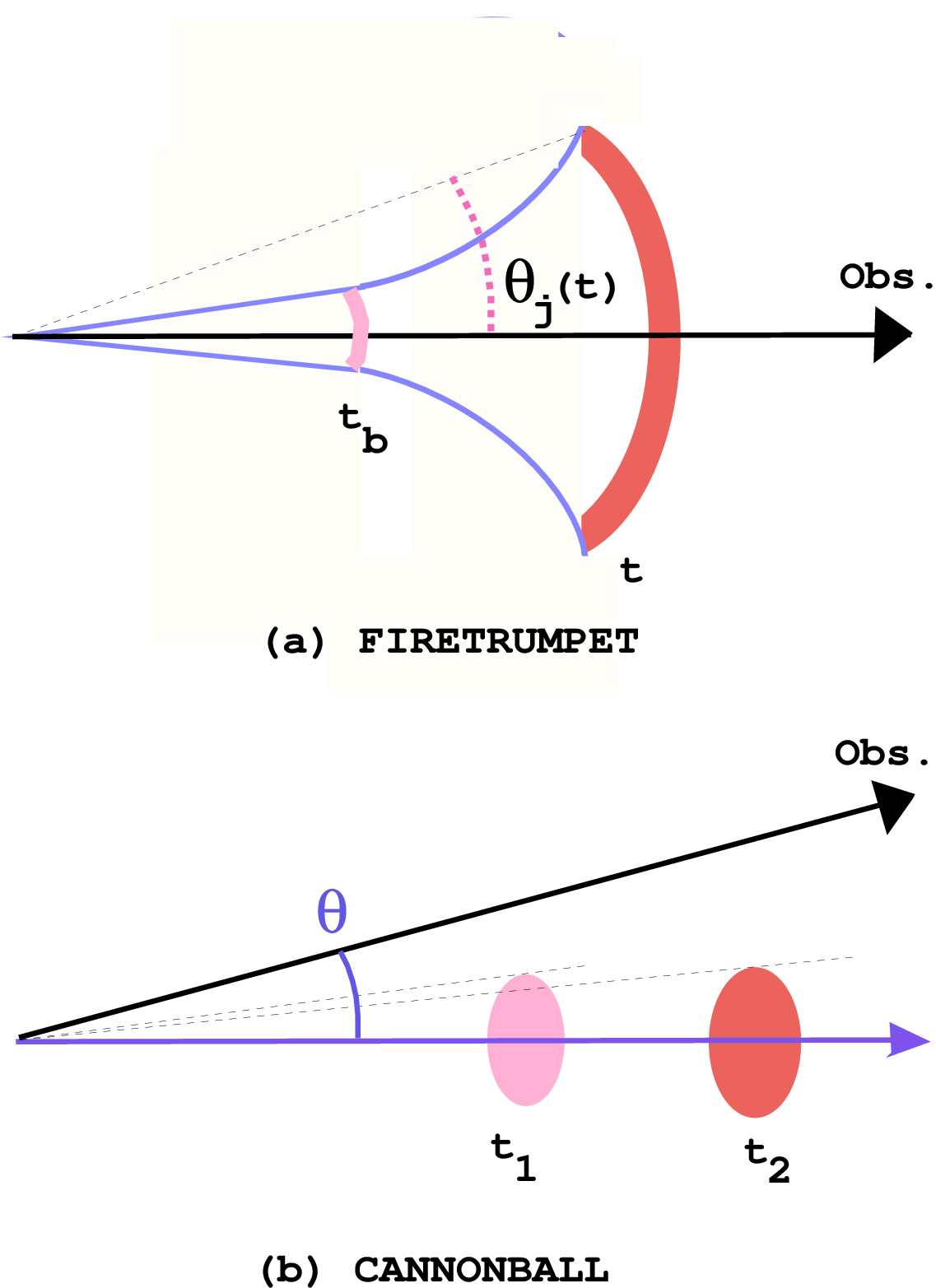,width=8.7cm} 
\caption{(a) A firecone or, more properly, a {\it firetrumpet}. 
In the scenario discussed in the text (initiated by Rhoads 1997), 
the cone expands 
conically for a distance, after which the jet angle 
$\rm\theta_j$ widens as its front travels. 
(b) Cannonballs (shown here, somewhat pedantically, 
a bit Lorentz-contracted) subtend decreasing angles as they travel. 
The only relevant angle in the CB model is the observer's 
viewing angle $\theta$.} 
\vspace*{-0.5cm} 
\label{figtrumpet} 
\end{center} 
\end{figure}

Consider first a proper (i.e. spherical) fireball expanding in a homogeneous 
(or spherically symmetric) medium. A conical section 
of this fireball would expand as a fixed-angle cone: a {\it firecone}. 
Consider next material that is ejected with a conical distribution. 
If the cone expands laterally 
at a transverse velocity $\rm v_T$, its opening angle, as viewed 
from the origin of the ejecta, increases with time. 
As illustrated in Fig.~(\ref{figtrumpet}a) the edges of the material 
describe a trumpet-shaped curve, not a fixed-aperture {\it cone}, 
as some of the names given to it may induce one to think. 
We call these ``firecones'' or ``conical fireball jets'' 
{\it firetrumpets}, since, for $\rm v_t\neq 0$, that is what they are. 
 
Let $\rm\gamma(t)$ 
be the Lorentz factor of the ejecta, that diminishes with time 
as they collide with ambient material. The light emitted by an 
element of a firetrumpet's surface is collimated by its motion 
into an angle of aperture $1/\gamma$. 
If $\rm v_T/\gamma$ were constant, the firetrumpet's opening angle 
would vary as: 
\begin{equation} 
\rm \theta_j(t)=\theta_j(0)+{v_T\over c\,\gamma}\, . 
\label{jetangle} 
\end{equation} 
At the time $\rm t=t_b$ at which $\rm \theta_j(t)=1/\gamma(t)$, the angle of 
emission of light becomes broader than the angle of the cone (Rhoads 1997, 1999). Thereafter the forward light-collimation 
is less efficient, an on-axis observer would see up to the 
edge of the cone, and no longer an increasing fraction of the ejecta 
(Meszaros et al. 1999). 
At early times $\rm t<t_{exp}$, the lateral expansion of the firetrumpet 
may not be important and the ejecta's deceleration as it plunges 
through constant-density material results 
---as it would for a fixed-angle cone--- in 
$\rm\gamma(r)\propto r^{-3/2}$, with $\rm r(t)$ the travelled distance, 
 while at later times 
$\rm\gamma(r)\propto exp[-r/r_{exp}]$. Rhoads assumes that these 
two transitions occur at the same time ($\rm t_b=t_{exp}$) and 
that they are abrupt, leading to a {\it break}: a sudden increase 
in the index $\alpha$ of an AGs' power-law evolution, $\rm F\propto t^{-\alpha}$. 
The break-time is estimated as the time at which a cone which 
is {\it not} laterally expanding decelerates to $\rm \gamma(t)=1/\theta_0$.

Sari et al.~(1999) change some of the parameters used by Rhoads 
(notably $\rm v_t=c/\sqrt{3}$ to $\rm v_T=c$) and invert 
$\rm t_b(\theta_j(0))$ to obtain $\rm \theta_j(t_b)$: 
\begin{equation} 
\rm \rm \theta_j(t_b) \simeq \theta_j(0) 
=0.1\,\left[{t_b\over 6.2\;h}\,{n_1\over E_{52}}\right]^{3\over 8}\, 
\label{Sari} 
\end{equation} 
where $\rm n_1$ is the local density in cm$^{-3}$ and $\rm E_{52}$ 
is the ejecta's energy in $10^{52}$ erg units. 
 
While these theoretical developments were taking place, more than a dozen 
(mainly R-band) optical AGs were being observed. They 
did not have abrupt breaks. The observers 
(GRB 990123: Fruchter et al.~1999a, Castro-Tirado et al.~1999b, 
Kulkarni et al.~1999;  GRB 990510:  Stanek et al.~1999, 
Harrison et al.~1999, Israel et al.~1999;  GRB 990705: Masetti et al.~2000;  
GRB 991208: Castro-Tirado et al.~2001;  
GRB 991216:  Halpern et al.~2000a;  GRB 000301c: Sagar et al.~2000b, 
Jensen et al.~2000;  GRB 000418:  Berger et al.~2001;  GRB 000926:  Fynbo 
et al.~2001, Sagar et al.~2001a;  GRB 010222:  Masetti et al.~2001, 
Stanek et al.~2001) fitted the 
slow steepening of AG fluences to phenomenological formulae such as: 
\begin{equation} 
\rm F_\nu = {2\,F_\nu^b \over 
\left[ (t/t_b)^{\alpha_1\,s}+(t/t_b)^{\alpha_2\,s}\right]^{1\over s}}\, , 
\label{pheno} 
\end{equation} 
which interpolate 
 between two power laws with a tunable ``abruptness'' $\rm s$, 
often set at $\rm s=1$. 
The values of $\rm t_b$ extracted from these fits, and their 
distributions, have no clear 
meaning, since different groups use different parametrizations, 
and none of them is theoretically justified. 
 
Moderski et al.~(2000), Huang at al. (2000a,b), 
Kumar \& Panaitescu ( 2000) and 
Panaitescu \& Kumar (2001) have modelled the 
light emitted by a firetrumpet without some of the approximations 
introduced by Rhoads. The evolution of the ejecta's deceleration 
is treated continuously. The emission is computed from isochronous 
points in the firetrumpet, so that light simultaneously received 
is light that had been simultaneously emitted 
(lifting the prior ``approximation'' that the speed of  light is infinite). 
Not having an abrupt break put in by hand, 
no abrupt break is predicted. The fair conclusion is that 
the light curves are too smooth to allow for a determination 
of a break time $\rm t_b$ (Moderski et al. 2000). 
 
{\it All the firetrumpet advocates place the 
observer precisely on the jet's axis}, as in Fig.~(\ref{figtrumpet}a), 
for no stated reason. It is obvious that the viewing 
angle is a relevant parameter that cannot be unceremoniously dismissed. 
In particular a non-vanishing viewing angle would contribute 
to erase even further any trace of a sharp break. 
Moreover, a distribution of viewing angles would completely erase 
a possible meaning to the distribution of specific $\rm t_b$ values extracted 
from expressions such as Eqs.~(\ref{Sari}) and (\ref{pheno}). 
 
The hypothetical firetrumpet ejecta behave in a 
 different way from most of the highly relativistic jets observed 
in quasars (e.g. radio jets: 
Bridle 2000; optical jets: Cranc et al.~1993; X-ray jets: Wilson et al.~2000) 
and microquasars (e.g. Mirabel \& Rodriguez 
1994, 1999). The ejecta of the real jets, as seen from their emission 
point up to the point where they eventually stop and expand, generally 
subtend angles that {\it decrease} with time, exactly the opposite 
of the assumed firetrumpet behaviour of Eq.~(\ref{jetangle}) 
and Fig.~(\ref{figtrumpet}a). 
In the analysis of these real objects (e.g. Pearsons \& 
Zensus 1987; Mirabel \& Rodriguez 1994, 1999; Ghisellini \& Celotti 
2001) it is the angle of observation ---and not the angle subtended by 
the ejecta--- that plays a key role. 
 
Frail et al.~(2001) 
use the published values of $\rm t_b$ ---fit to expressions such as 
Eq.~(\ref{pheno})--- to extract a set of values of $\rm \theta_j(0)$. 
In so doing they use a modified Eq.~(\ref{Sari}), in which the dependence 
on redshift and on the efficiency of light production are not overlooked. 
In this way they reach a series of conclusions that their analysis 
does not justify. 
 
The firetrumpet model may, to some extent, be correct. We have seen that, 
alas, the consequences of its basic assumptions have not been properly 
extracted. In particular, the ``anthropo-axial'' view that the 
ejecta of the observed AGs always point to the observer has not been shown 
to be a fair approximation. 
 
In Fig.~(\ref{figtrumpet}b) we illustrate the geometry of the CB model. 
In the AG phase, the CBs are expanding very slowly, or 
not at all (DD2000a), like the observed ejecta 
in quasars and microquasars. In contradistinction to 
the firetrumpet case of Eq.~(\ref{jetangle}), 
the angle with which the CBs are viewed from the origin 
{\it decreases} with time. But this angle is irrelevant, and 
negligible relative to the opening angle $1/\gamma$ of the 
emitted light. The angle at which the ejected CBs are viewed 
is obviously relevant and we do not 
set it to zero by fiat.

\section{The cannonball model of GRBs} 
 
In the CB model, long-duration GRBs and their AGs are produced in 
core collapse supernovae by jets of highly relativistic ``cannonballs'' 
that pierce through the supernova shell.  The detailed model is based 
essentially on the following analogies, hypotheses and explicit 
calculations: 
 
\subsection{Relativistic jets in astrophysics} 
 
Astrophysical systems, such as quasars and 
microquasars, in which periods of intense accretion onto a compact 
massive object occur, emit highly collimated relativistic jets of plasma. 
The Lorentz factor $\rm 
\gamma\equiv 1/\sqrt{1-v^2/c^2}$ of these jets ranges from mildly 
relativistic: $\gamma\sim 2.55$ for GRS 1915+105 (Mirabel \& Rodriguez 
1994, 1999), to quite 
relativistic: $\gamma={\cal{O}}\,(10)$ for typical quasars (e.g. Ghisellini 
et al.~1993), and even to highly relativistic: $\gamma\sim 10^3$ for PKS 
0405$-$385 (Kedziora-Chudczer et al.~1997). These jets are not continuous 
streams of matter, but consist of individual blobs, or ``cannonballs''.  
(e.g. Kraft et al. 2001). The mechanism producing these 
surprisingly energetic and collimated emissions is not understood, but it 
seems to operate pervasively in nature (In Section 7 we propose a 
mechanism capable of collimating CBs). We assume the CBs to be composed 
of ordinary ``baryonic'' matter (as opposed to $\rm e^+\, e^-$ pairs), as 
is the case in the microquasar SS 433, from which Ly$_\alpha$ and metal 
K$_\alpha$ lines have been detected (e.g. Margon 1984, Kotani et al.~1996). 
 
\subsection{The GRB/SN association} 
There is mounting evidence 
for an association of  supernova 
explosions of type Ib/Ic and GRBs (DD2000a). The first example was GRB 
980425 (Soffitta et al.~1998; Kippen et al. 1998), within whose error circle 
SN1998bw was soon detected optically  (Galama et al.~1998a) 
and at radio frequencies (Kulkarni et al.~1998a). The chance probability 
for a spatial and temporal coincidence is less than $10^{-4}$ 
(e.g. Galama et al.~1998a), or much smaller if the revised BeppoSAX position 
(e.g. Pian et al. 1999) is used in the estimate 
(we shall show that the observed X-rays originate in the CBs of this 
GRB, and not on the associated SN). The unusual radio 
(Kulkarni et al.~1998a; Wieringa et al.~1999) and 
optical (Galama et al.~1998a; Iwamoto et al.~1998) properties of SN1998bw 
support this association. The exceptionally small fluence 
and redshift of GRB 980425 make this event very peculiar, though 
not in the CB model (DD2000a). 
The energy supply in a SN event similar to SN 1998bw is too small to 
accommodate the fluence of cosmological GRBs, unless their $\gamma$-rays 
are highly beamed. SN 1998bw is a peculiar supernova, perhaps 
because it is observed close to the axis of its GRB emission. 
 
Evidence for a SN1998bw-like contribution to a GRB afterglow (Dar 1999a; 
Castro-Tirado \& Gorosabel 1999) was first found by Bloom et al.~(1999a) for 
GRB 980326, but the unknown redshift prevented a quantitative analysis. 
The AG of GRB 970228 (located at redshift $\rm z=0.695$) appears to 
be overtaken by a light curve akin to that of SN1998bw (located at $\rm 
z_{bw}=0.0085$), when properly scaled by their differing redshifts (Dar 
1999b; Reichart 1999; Galama et al.~2000).  Evidence of similar 
associations was found for GRB 990712 (Hjorth et al.~2000a; Sahu et 
al.~2000;  Bjornsson et al.~2001), GRB 980703 (Holland et al.~2000b), GRB 
000418 
(DD2000a), GRB 991208 (Castro-Tirado et al.~2001) and GRB 990510 (Sokolov et 
al. 2001a). For the remaining cases in Table I, corresponding to 
GRBs with larger redshifts, either no late 
observations of the AG are available, or the expected 
SN bump is an unobservably small effect. These conclusions 
will be strengthened by our detailed analysis of AGs 
in the CB model. All of the nearest GRBs with 
measured redshifts show various degrees of evidence for a 
supernova in their AG, suggesting the possibility of an association of 
{\it all} the long-duration GRBs with core-collapse SNe. 
 
\subsection{The SN/GRB association} 
 
By a SN/GRB association ---as opposed to the GRB/SN association 
we have discussed--- we mean the converse statement to that ending 
the last subsection: that most SNe of certain relatively 
frequent types may be associated with GRBs. This 
appears at first sight to be entirely untenable. The total rate of type 
II/Ib/Ic SNe has been estimated from their observed rate in the local 
Universe (e.g. Van den Bergh \& Tammann 1991) and the star formation rate 
as function of redshift, to be $\rm R_{SN}= 12\pm 5\, s^{-1}$, or $\rm 
R_{SN}\sim 3.8\times 10^8\, y^{-1}$, 
in the observable Universe (Madau et 
al. 1998), while the observed rate of GRBs is a mere 
$\rm R_{GRB}\simeq 10^3\, y^{-1}$ (see, for example, Lamb 2001).

The bolometric energy fluence from a CB 
moving with a Lorentz factor $\gamma\gg 1$ and 
seen by a stationary observer at an angle $\theta\ll 1$ 
relative to the CB's direction of motion  (e.g. DD 2000a) is: 
\begin{equation} 
\rm  {dF\over d\Omega} \propto 
\left[{2\,\gamma \over 1+\gamma^2\,\theta^2}\right]^3\, . 
\label{fluence} 
\end{equation} 
Barring the case of GRB 980425 (whose exceptionality and 
interpretation we shall discuss) the equivalent spherical 
energies of the GRBs with measured redshifts, as listed in Table I, range 
between, approximately, $2 \times 10^{54}$ erg (GRB 990123) and $6.6 \times 
10^{51}$ erg (GRB 970508). The $\theta$ dependence is the steepest 
parameter dependence of the CB model (DD2000b). It is therefore 
reasonable to attribute the range of observed equivalent spherical energies 
mainly to the $\theta$ dependence (as if GRBs were otherwise approximately 
standard candles). The observed  spread in 
equivalent energy then corresponds, according to Eq.~(\ref{fluence}), 
to a spread of 
viewing angles between $\theta \approx 0$ and $\theta\approx 2.4/\gamma$. 
Thus the geometrical fraction of GRBs which are observable (with the 
current or past sensitivity) is approximately $\rm f(\gamma)=2\pi\, \theta^2 / (4 
\pi) \approx 2.84 / \gamma^2 $,  where we have taken two jets of 
CBs per event. Compare 
$\rm R_{SN}$ and $\rm R_{GRB}$ 
to conclude that an approximately one-to-one GRB/SN 
association would require beaming into a solid angle 
that is a fraction $\rm \sim\! 2.8\times 10^{-6}$ of $4\pi$. 
For CBs moving with $\gamma\!\sim\! 10^3$, $\rm f(\gamma)=2.84\times 10^{-6}$: 
precisely the required beaming factor. That is, 
for a one-to-one SN/GRB association: 
\begin{equation} 
\rm R_{GRB} = f(\gamma)\, R_{SN}= (1082 \pm 450)\, 
\left[{10^3\over \gamma}\right]^2 \; y^{-1}\,, 
\label{GRB} 
\end{equation} 
in agreement with observation. 
Moreover, if the recent claims that the $\rm\sim (1+z)^3$ dependence of the 
star formation rate continues to $\rm z\!>\! 1$ (Fenimore \& Ramirez-Ruiz 2000; 
Ramirez-Ruiz and Fenimore 2000; Reichart et al.~2001) 
were correct,  SN of types Ib/Ic 
would by themselves suffice to explain the observed GRB rate. Thus, 
relativistic beaming solves the energy crisis of GRBs and may allow 
an approximately one-to-one 
SNIc/GRB association (Dar 1999b;  Dar \& Plaga 1999; DD2000a). 
The above considerations leading to a GRB/SN association that may 
be as biunivocal as indicated by Eq.~(\ref{GRB}) are weakened by 
the fact that we have not taken into account effects such as the efficiency of 
GRB identification as a function of fluence. It is clear, however, 
that for the high beaming factors we have advocated, the GRB and SN 
rates are quite comparable. In the CB model, the ``special'' character of  
SN 1998bw is due to the fact that it is observed very close to the 
GRB axis. 
 
Previous analyses of GRB/SN associations, except that in DD2000a, 
were based on  a power-law extrapolation to late times 
of the early-time GRB afterglows.  Here we shall use the CB model, 
instead, to calculate the 
GRB-afterglow light curves at all times. This procedure leads, as we shall 
see, to a much better exposition of the GRB/SN association.

\subsection{The GRB engine}

\begin{figure} 
\begin{center} 
\vspace*{.003cm} 
\hspace*{-0cm} 
\epsfig{file=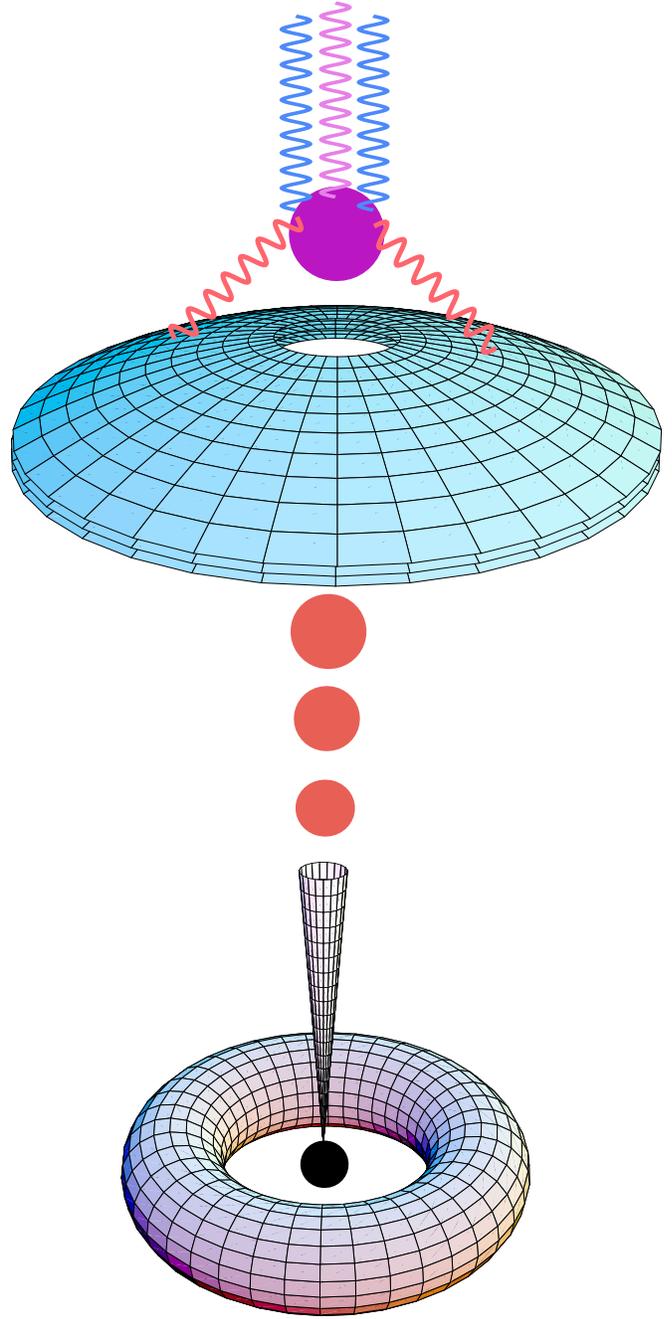,width=8.7cm} 
\caption{An ``artist's view'' (not to scale) of the CB model 
of GRBs and their AGs. A core-collapse SN results in 
a compact object and a fast-rotating torus of non-ejected 
fallen-back material. Matter (not shown) catastrophically accreting 
into the central object produces 
a narrowly collimated beam of CBs, of which only some of 
the ``northern'' ones are depicted. As these CBs pierce the SN shell, 
not precisely on the same spot, 
they heat and re-emit photons, that are 
Lorentz-boosted and collimated by the CBs' relativistic motion.} 
\vspace*{-0.5cm} 
\label{model} 
\end{center}  
\end{figure}

We assume that in  core-collapse SN events 
a fraction of the parent star's 
material, external to the newly-born compact object, 
falls back in a time very roughly of the 
order of one day (De R\'ujula 1987, DD2000a). Given the considerable 
specific angular momentum of stars, 
it settles into an accretion disk and/or torus 
around the compact object. 
The subsequent sudden episodes 
of accretion ---occurring with a time sequence that we cannot predict--- 
result in the emission of CBs. These emissions last till 
the reservoir of accreting matter is exhausted. 
The emitted CBs initially expand in the SN rest system at a speed 
$\rm\beta\,c/\gamma$, with $\rm\beta\,c$ presumably of the same order as 
the speed of sound in a relativistic plasma 
($\beta=1/\sqrt{3}$), or smaller. 
The solid angle a CB subtends is so extremely small that presumably 
successive CBs do not hit the same point of the outgoing SN shell, 
as they catch up with it. These considerations 
are illustrated in Fig.~(\ref{model}). 
 
\subsection{The GRB}From this point onwards, the CB model is not 
based on analogies or assumptions, but on processes whose outcome can be 
approximately worked out in an explicit manner. The violent collision of 
the CB with the SN shell heats the CB (which is not transparent at this point 
to $\gamma$'s from $\pi^0$ decays) to a surface temperature that, by the 
time the CB reaches the transparent outskirts of the SN shell, is 
$\sim 150$ eV, further 
decreasing as the CB travels (DD2000b). The resulting CB surface 
radiation, Doppler-shifted in energy and forward-collimated by the CB's 
fast motion, gives rise to an individual pulse in a GRB (DD2000b). The GRB 
light curve is an ensemble of such pulses, often overlapping one another. 
The energies of the individual GRB $\gamma$-rays, as well as their typical 
total fluences, indicate CB Lorentz factors of ${\cal{O}}$(10$^3$), as the 
SN/GRB association does (DD2000a). In the CB model, unlike in 
the shocked-fireball models, the photons of the GRB proper 
are not made by synchrotron radiation which, as we shall see, 
 is subdominant at this stage of the evolution of a CB. 
 
\section{Afterglow components} 
 
In the CB model, the persistent radiation in the direction of an observed GRB 
has three origins: the ejected CBs, the concomitant SN explosion, and the 
host galaxy. These components are usually unresolved in the 
measured ``GRB afterglows'', so that the corresponding light curves and 
spectra refer to a cumulative energy flux density: 
\begin{equation} 
\rm    F_{AG}=F_{CBs}+F_{SN}+F_{HG}\, , 
\label{sum} 
\end{equation} 
with $\rm F\equiv\nu\,dN_\gamma/(dt\,d\nu\,dA)$. 
 
The emission of the GRB's host galaxy is usually determined from 
measurements at times late enough for the CB's afterglow and 
the SN light to have become comparatively weak (e.g. 
Sokolov et al.~2001b and references therein). This assumes that the host 
galaxy's emission is steady  on periods of a few months. 
There is no indication of GRB host-galaxy variability 
on such time scales. 
 
Core-collapse supernovae (SNII/Ib/Ic) 
are far from being standard candles. But if their explosions 
are fairly asymmetric ---as they would be if a fair fraction of 
them emitted jets of CBs---  much of the variability could be a reflection 
of the varying angles from which we see their 
non-spherically expanding shells. 
Exploiting this possibility to its extreme, we shall use 
SN1998bw as an ansatz standard candle, associated with every 
GRB of known $\rm z$ (Dar 1999b; DD2000a). The adequacy 
of this bold hypothesis can be judged from its rather surprising success. 
 
Let the energy flux density of SN1998bw be $\rm F_{bw}[\nu,t]$. 
For a similar SN placed at a redshift $\rm z$: 
\begin{eqnarray} 
{\rm F_{SN}[\nu,t] = } && 
{\rm{1+z \over 1+z_{bw}}\; 
{D_L^2(z_{bw})\over D_L^2(z)}}\, \times\nonumber \\ 
&&{\rm F_{bw}\left[\nu\,{1+z \over 1+z_{bw}},t\, 
{1+z_{bw} \over 1+z}\right]\; A(\nu,z)}\, , 
\label{bw} 
\end{eqnarray} 
where $\rm D_L(z)$ is the luminosity distance\footnote{The cosmological 
parameters we use in our calculations are: 
$\rm H_0=65$ km/(s Mpc), ${\rm \Omega_M}=0.3$ and 
${\rm \Omega_\Lambda}=0.7$.} 
and $\rm A(\nu,z)$ is the extinction along 
the line of sight. The extinction in our Galaxy 
is reasonably well measured, but  for the GRBs' environments it must be 
estimated from the spectra of each particular AG and host galaxy.  
 
The contribution of CBs to the GRB afterglows requires 
 a much more detailed discussion.

\section{Times and frequencies} 
 
Four ``clocks'' ticking at different paces and three different scales 
of frequency need be considered in the cannonball model of 
GRBs and their afterglows. 
 
Let $\rm \gamma=1/\sqrt{1-\beta^2}={E_{CB}/(M_{CB}c^2)}$ be 
the Lorentz factor 
of a CB, which diminishes with time as the CB hits the SN shell 
and as it subsequently ploughs through the interstellar medium. 
Let $\rm t_{SN}$ be the 
local time in the SN rest system, $\rm t_{CB}$ the time in the CB's 
rest system, $\rm t_{Ob}$ the time   measured by 
a nearby observer viewing the CB at an angle $\theta$ 
away from its direction of motion, and $\rm t$ the time 
measured by an earthly observer viewing the CB at 
the same angle, but from a ``cosmological'' distance 
($\rm z\neq 0$). 
Let x be the distance travelled by the CB in the SN rest system. 
The relations between the above quantities are: 
\begin{eqnarray} 
&&\rm 
dt_{SN}=\gamma\,dt_{CB}=\rm{dx\over\beta\, c}\, ; 
\nonumber \\ 
&&\rm 
dt_{CB}=\delta\,dt_{Ob}\, ;\nonumber\\ 
&&\rm 
dt=(1+z)\,dt_{Ob}={1+z\over \gamma\,\delta}\;dt_{SN}\;, 
\label{times} 
\end{eqnarray} 
where the Doppler factor $\delta$ is: 
\begin{equation} 
\rm 
\delta\equiv\rm{1\over\gamma\,(1-\beta\cos\theta)} 
\simeq\rm {2\,\gamma\over (1+\theta^2\gamma^2)}\; , 
\label{doppler} 
\end{equation} 
and its approximate expression is valid for $\theta\ll 1$ and $\gamma\gg 1$, 
the domain of interest here. 
Notice that for large $\gamma$ and not large $\theta\gamma$, 
there is an enormous ``relativistic aberration'': 
$\rm dt\sim dt_{SN}/\gamma^2$, and the observer sees 
a long CB story as a film in extremely fast motion. 
 
The frequency of the photons radiated by a CB 
in its rest system, $\rm \nu_{CB}$, their frequency 
in the direction $\theta$ 
in the local SN system, $\rm \nu_{SN}$,  and the photon 
frequency $\nu$ measured by a cosmologically distant observer, 
are related by: 
\begin{equation} 
\rm \nu_{CB}=   {\nu_{SN}\over \delta} 
\, ;\;\;\;\;\;\nu_{SN}=(1+z)\,\nu\; , 
\label{energies} 
\end{equation} 
with $\delta$ as in Eq.(\ref{doppler}).

\section{The cooling of CBs} 
 
As a CB pierces through the SN shell, its surface 
is heated by the collisions with the shell's constituents, and 
cools down from an early maximum temperature 
because of the decreasing density of the shell's material 
it collides with (a detailed description of the CB--shell 
collision can be found in DD2001b). At this early point of a CB's avatars, 
the internal radiation pressure is very large. Thus, 
in studying the properties of the $\gamma$ rays (DD2000b), we assumed 
the CBs to expand (in their rest system and at early times) at a speed 
comparable to that of sound in a relativistic plasma ($\rm c/\sqrt{3}$). 
This fast expansion implies that it is a good approximation 
to treat CBs, in their rest system, as spherical objects. 
 
Let $\rm N_{jet}$ be the baryon or electron number of 
the ensemble of CBs in a jet, which we have estimated 
to be $\rm N_{jet}\sim 6\times 10^{51}$ (e.g Eq.~(5) of 
DD2001b, for $\rm \gamma_{in}=10^4$), which is close to that of the 
Earth ($\rm N_\otimes\simeq 3.6\times 10^{51}$). 
On average, GRBs consist of 5 to 10 significant pulses, so 
that a single CB may have one order of magnitude fewer constituents. 
As they exit the shell and enter the interstellar medium (ISM), CBs 
become transparent to their enclosed radiation 
when they reach a radius: 
\begin{equation} 
\rm R_{trans}\sim \left[{3\over 4\pi}\,N_{CB}\,\sigma_T\right]^{1/2}\! 
\simeq (10^{13}\;cm)\;\left[{N_{CB}\over 6\times 10^{50}}\right]^{1/2}\!\!\! , 
\label{Rtrans} 
\end{equation} 
where $\rm\sigma_T=6.65\times 10^{-25}$ cm$^2$ is the Thomson 
cross section. We can use Eqs.~(\ref{times}) to conclude that, if 
the CBs are expanding at a fraction $\rm\beta_{trans}$  of the 
speed of light\footnote{The quantity $\rm\beta_{trans}$ is nearly 
identical to $\rm\beta_{out}$, the transverse speed as the 
CB exits the SN shell, introduced in our previous work on the CB model 
(e.g. DD200a,b).}, 
they reach a size $\rm R_{trans}$ in an observer's time: 
\begin{equation} 
\rm t_{trans}={1+z\over \delta}\, t_{trans}^{CB}= 
{(1+z)\,R_{trans}\over\delta\,\beta_{trans}\,c}. 
\label{ttrans} 
\end{equation} 
For the reference value of $\rm N_{CB}$ in Eq.~(\ref{Rtrans}), 
$\gamma=1/\theta=10^3$ and $\rm\beta_{trans}=1/(3\sqrt{3})$, 
CBs become transparent in a mere $\rm t_{trans}\sim 3.5$ s. 
 
The GRB is emitted by the CBs from a distance 
of ${\cal{O}}(1)$ radiation length from the 
exterior of the SN shell, 
when their temperature is $\rm T_\gamma\sim 150$ eV 
and their radius, for our typical parameters, is 
$\rm R_\gamma\sim 2.5\times 10^{11}$ cm (DD2000b, DD2001b). 
Soon thereafter, travelling in a thin environment and expanding fast, the CBs 
should cool in an approximately adiabatic way. Their temperature 
at $\rm t_{trans}$ is then: 
\begin{equation} 
\rm T_{trans} \sim {R_\gamma\over R_{trans}}\, T\simeq 4.0\, eV. 
\label{TTh} 
\end{equation}
>From about one third of $\rm t_{trans}$ onwards, the CBs would appear 
to be ``collisionless'' to the ISM hadrons piercing through  
them\footnote{ISM particles that get entangled in the CB's 
magnetic field would not be collisionless after such a very short time.}, 
since the high-energy nucleon--nucleon cross section ($\rm\sigma_N\sim 
4\times 10^{-26}$ cm$^2$) is about one order 
of magnitude smaller than $\rm \sigma_T$ and the 
condition for ``transparency'' to the ISM particles is, up to 
a numerical factor of ${\cal{O}}(1)$, analogous to Eq.~(\ref{Rtrans}). 
 
\section{The expansion of CBs} 
 
When a CB, in a matter of (observer's) seconds, becomes transparent to 
radiation, it loses its internal radiation pressure. If it has been 
expanding up to that moment at a speed comparable to that of relativistic 
sound, should it not inertially continue to do so? The fact that it is 
collisionless makes the conclusion seem unavoidable. But the CBs 
emitted by many quasars appear, within the resolution of the 
observations, not to expand laterally for most of their trajectory, 
before their forward motion nearly stops. What may the reason be? 
 
\begin{figure} 
\begin{center} 
\vspace*{.003cm} 
\hspace*{-0cm} 
\epsfig{file=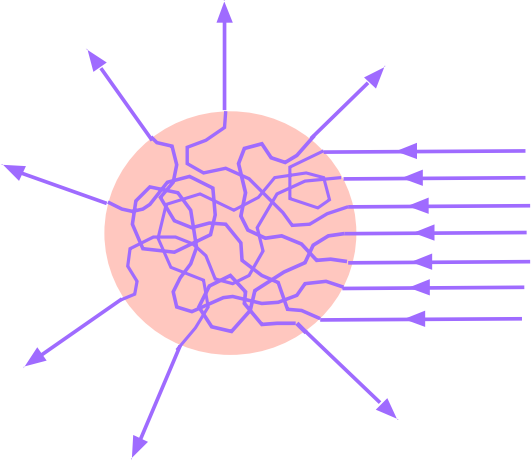,width=8.7cm} 
\caption{A CB, in its rest system, sees the constituents of the ionized 
ISM impinge in one direction. The CB's chaotic magnetic field 
disperses these particles so that they come out isotropically. 
Electrons, but not protons, lose their energy by synchrotron radiation.} 
\vspace*{-0.5cm} 
\label{CBrest} 
\end{center}  
\end{figure}

The ISM the CBs traverse has been previously partially 
ionized by the forward-beamed GRB radiation. The neutral ISM fraction 
is efficiently ionized by Coulomb interactions as it enters the CB. 
In analogy to processes occurring in quasar and microquasar 
ejections, the bulk of the swept-up ionized ISM particles are multiply 
scattered, in a ``collisionless'' way, by the CBs' turbulent magnetic fields. 
As illustrated in Fig.~(\ref{CBrest}), in 
the rest system of the CB these particles are isotropically 
re-emitted into the ISM. In the rest system of the parent SN 
they are forward collimated and boosted to an energy 
$\rm\sim\! m\, c^2\gamma^2$ (Dar 1998b). The isotropic re-emission implies 
an inwards force on the CB's surface. Assume that the bulk of the 
ISM particles are {\it not} re-energized by the CB's turbulent fields. Let $\rm R$  
be the CB's radius and let $\rm n_p$ be the proton ISM number density. The 
rate at which the ISM protons impinge on the CB is $\rm 
r=\pi\,R^2\,c\,\gamma\,n_p$, with $\rm \gamma\,n_p$ the ISM proton density  
seen from the CB's rest system. The momentum (or, for large $\gamma$, the  
energy) 
of these protons isotropically leaving the CB is, per unit surface, 
\begin{equation} 
\rm P=r\,{E_p\,c\over 4\pi\,R^2}={1\over 4}\,m_p\gamma^2\,n_p\,c^2\, . 
\label{pressure} 
\end{equation} 
 
During the first hours after the GRB time, the CBs are still fully ionized 
and cooling rapidly by expansion and 
bremsstrahlung (DD2001a). Their full constituency of 
relatively cold ions, electrons, cosmic rays and entangled magnetic fields is 
electromagnetically coupled, and subject to the very large inwards 
pressure of Eq.~(\ref{pressure}). This stabilizes the CB's radius to an 
asymptotic value $\rm R_{max}$. To estimate it, since the initial expansion 
velocity is not fully relativistic ($\rm \beta_{trans}^2\ll 1$), we may 
use Newton's equation: 
\begin{equation} 
\rm P=-{M_{CB}\over 4\pi\,R^2}\;{d^2 R\over dt^2_{CB}} 
\label{Newton} 
\end{equation} 
with $\rm P$ as in Eq.~(\ref{pressure}), and 
integrate, to obtain, for a constant\footnote{A density distribution 
falling with distance to the progenitor 
star as $\rm x^{-2}$ (in a certain distance-domain) gives 
similar results, but in terms of more parameters. It suffices for the moment 
to use a constant value representing an average density close 
to the progenitor, particularly at the considerable distances from 
the progenitor at which the value of $\rm R_{max}$ is reached.} $\rm n_p$: 
\begin{equation} 
\rm R_{max}^3\sim 
R_{trans}^3+{3\,N_{CB}\,\beta_{trans}^2\over 2\,\pi\,n_p\,\gamma_0^2}\, , 
\label{Rinfinity} 
\end{equation} 
where we have approximated $\gamma$ by its initial value 
because the asymptotic radius, as we shall see, is reached much 
before the CB has had the time to decelerate significantly. 
For a value $\rm n_p=1\, cm^{-3}$ 
of the ISM density close to the progenitor, $\rm N_{CB}=6\times 10^{50}$, 
$\gamma_0=10^3$, Eq.~(\ref{Rinfinity}) gives 
$\rm R_{max}=2.2\times 10^{14}$ cm [$10^{14}$ cm] for 
$\rm\beta_{trans}=1/(3\sqrt{3})$ [$1/(10\sqrt{3})$]. 
 
\begin{figure} 
\begin{center} 
\vspace*{.003cm} 
\hspace*{-0cm} 
\epsfig{file=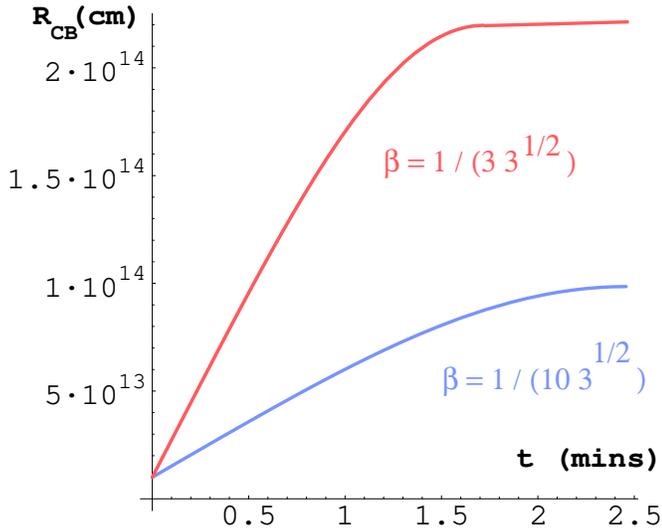,width=8.7cm} 
\caption{The radius of a CB as a function of observer's time, 
after the CB becomes transparent to radiation, for two choices 
of the initial transverse expansion velocity $\rm\beta_{trans}\,c$.} 
\vspace*{-0.5cm} 
\label{RCB} 
\end{center}  
\end{figure}

The interval (in the CB's rest system) 
between the times when the CB has 
radius $\rm R_{trans}$ and radius $\rm R$ can be deduced from 
Eqs.~(\ref{pressure}) and (\ref{Newton}) to be: 
 
\begin{equation} 
\rm 
t(R)=\int^R_{R_{trans}}\;{dx\over c\,\beta_{trans}}\; 
\sqrt{{R_{max}^3-R_{trans}^3\over R_{max}^3-x^3}}\; . 
\label{tofR} 
\end{equation} 
The observer's time is shorter by a factor $\rm (1+z)/\delta$. 
In Fig.~(\ref{RCB}) we invert Eq.~(\ref{tofR}) to 
show the CB's radius as a function of 
observer's time (in minutes), for $\rm z=1$, $\delta=10^3$, and the other 
typical parameters quoted in the previous 
paragraph, for two choices of $\rm\beta_{trans}$. 
The CB is seen to expand linearly 
at a speed close to the initial $\rm \beta_{trans}\,c$, and then 
to settle fast into an approximately constant radius. 
To a good approximation the steady radius is reached 
in an observer's time: 
 
\begin{equation} 
\rm 
t_\infty\sim {1+z\over \delta}\;{R_{max}\over \beta_{trans}\,c}\, , 
\label{tapprox} 
\end{equation} 
which yields $\sim 1.2\;[1.9]$ minutes for 
$\rm \beta_{trans}=1/(3\sqrt{3})$ [$1/(10\sqrt{3})$], 
the examples in Fig.~(\ref{RCB}). 
Thus, typically, a few observer's minutes after the GRB, the CBs 
are expanding very very slowly. 
 
To estimate the internal magnetic field of the CB after it stopped 
expanding, $\rm B_\infty\, , $ we conjecture that the bulk of the kinetic  
energy of the CB's expansion after it becomes transparent is converted 
to internal magnetic energy. Since $\rm R_{max}^3 \gg R_{trans}^3$ and 
(by hypothesis) $\rm\beta_{trans}^2 \ll 1$, this means: 
\begin{equation} 
\rm {4\, \pi\over 3}\;  R_{max}^3\;  {B_{\infty}^2 \over 8\, \pi} 
        \sim {N_{CB}\, m_p\, \beta_{trans}^2\, c^2\over 2}  \, , 
\label{Bequiv} 
\end{equation} 
which, upon substitution of Eq.~(\ref{Rinfinity}), yields: 
\begin{equation} 
\rm B_{\infty } 
\sim 100\, 
\left[{n_p\over 1\, cm^{-3}}\right]^{1/2}\, 
\left[{\gamma\over 10^3}\right]\; \, Gauss\; . 
\label{Binfty} 
\end{equation} 
The very large field of Eq.~(\ref{Binfty}) is consistent with the fact 
that, in order to be able to sustain the inwards pressure of the isotropically 
re-emitted protons that it ejects, 
the magnetic field within the CB must have a pressure (or energy 
density) comparable to the pressure $\rm P$ of Eq.~(\ref{pressure}). 
The condition $\rm P=B^2/(8\pi)$ exactly reproduces Eq.~(\ref{Binfty}). 
We could also have added the building-up magnetic pressure 
to Eq.~(\ref{pressure}), to obtain a result for $\rm R_{max}$ differing from that 
of Eq.~(\ref{Binfty}) by $2^{1/3}$. This affects our conclusions insignificantly 
and, in any case, we cannot pretend to have a detailed understanding 
of the magnetohydrodynamics of turbulent plasmas. 
 
We have argued that the re-emission of the ISM 
protons, isotropic in the CB's rest frame, is what makes 
CBs stop expanding  at a speed that {\it ab-initio} was semirelativistic. 
This may also explain the surprising quasar and microquasar observations. 
There may be other reasons ---such as Coulomb-interaction ram pressure 
from the ambient material--- for CBs to stop expanding significantly at 
some point of their voyage. The strength of our conclusion that CBs 
expand slowly ---or not at all--- during the AG phase should be judged 
from the ability of the CB model to describe the AG observations. 
 
\section{The dominant afterglow mechanisms}

In the CB model the GRB emission in $\gamma$ rays is mainly of 
thermal origin (although it does not have a thermal spectrum) 
and, in a fixed energy interval, it decreases exponentially 
with time (DD2000b). A few seconds after the last GRB pulse 
(the last CB), this pseudothermal emission becomes a 
subdominant effect. For the next few hours, the evolution 
of a CB is interestingly complicated. In particular, its originally 
ionized material should recombine into  hydrogen and 
emit Lyman-$\alpha$ lines that are seen Doppler-boosted 
to keV energies (DD2001a). Later, the 
CBs settle down to a much simpler phase, which typically lasts 
for months, till the CBs finally stop moving relativistically. 
 
Because of the CBs' large Doppler factors, radio emission in their rest 
frame is boosted to optical light in the observer's frame while their 
emitted optical light is boosted to the soft X-ray band. Radio emission 
from astrophysical plasmas at eV temperatures 
is mainly due to synchrotron radiation from 
relativistic electrons, whereas their optical glow is 
usually due to bremsstrahlung and line emission. For parameters in the 
general vicinity of the ones we have argued to be ``typical'' of the CB 
model, the X-ray AG is initially dominated by thermal 
bremsstrahlung (and line emission) and by synchrotron radiation 
thereafter, while the optical AG, generally observed later, 
is dominated by synchrotron 
radiation.  In this section we analyse these two dominant mechanisms, 
relegating to Appendix 1 the discussion of various subdominant ones. 
 
\section{Thermal bremsstrahlung: the early X-ray AG} 
 
When it becomes transparent, a CB cools down mainly by thermal-electron 
bremsstrahlung (TB) in $\rm e\,p$ collisions and by expansion. The comoving 
TB emission  rate (e.g. Peebles 1993) is: 
\begin{equation} 
\rm L_{brem}\simeq \eta\, {\overline n_e}^2\, T^{1/2} 
\, erg\, cm^3\, s^{-1}\, , 
\label{Lbrem} 
\end{equation} 
where $\rm \eta = 1.435\times 10^{-27}$, 
$\rm \overline n_e$ is the electron density 
of the CB and, here and in the rest of this section, 
$\rm T$ is its temperature in Kelvin and the remaining quantities 
are in c.g.s. units. Thus, as long as the CB is fully ionized, 
its total comoving TB energy-loss rate is: 
\begin{equation} 
\rm {dE_{CB}\over dt_{CB}} \simeq - \eta \, {3\, N_{CB}^2 
\, T^{1/2}\over 4\, \pi\, R^3} \, erg\,s^{-1}\, , 
\label{CBbrem} 
\end{equation} 
while its thermal energy is: 
\begin{equation} 
\rm E_{CB} \simeq 3\, N_{CB}\, k\, T\, , 
\label{Etotal} 
\end{equation} 
with $\rm k=1.38\times 10^{-16}\, erg/deg$ Boltzmann's constant. 
As long as TB dominates the cooling and for 
a constant expansion rate 
$\rm R(t_{CB})\sim  \beta_{trans}\, c\, t_{CB}$, 
Eqs.~(\ref{Rtrans}, \ref{ttrans}, \ref{CBbrem}, \ref{Etotal}) yield 
 $\rm T$ for $\rm t_{CB}\geq t_{trans}^{CB}$: 
\begin{eqnarray} 
\rm T^{1/2}(t_{CB})&\simeq&\rm 
 K+{\eta\, N_{CB} \over 16\, \pi\, k\, \beta_{trans}\, c\, R^2}\, 
\nonumber\\ 
\rm K &=& \rm T_{trans}^{1/2} - {\eta \over 12\, k\, \beta_{trans}\, 
                 c\, \sigma_T}\, . 
\label{Tbrem} 
\end{eqnarray} 
 
A distant observer at an angle 
$\theta$ relative to the CB's direction of motion receives this radiation 
in a Doppler-boosted, collimated and time-aberrant form 
(e.g. DD2000a). At a luminosity distance 
$\rm D_L(z)$, the total power (integrated over frequencies) 
per unit area is: 
\begin{equation} 
\rm {dF\over dt\, d\Omega}  \simeq {3\, \eta\,  N_{CB}^2\, T^{1/2}\, 
\delta_0^4 \over 16\, \pi^2\, R^3\,  D_L^2}\,  erg\, cm^{-2}\, s^{-1}\, , 
\label{fbrem} 
\end{equation} 
where $\delta_0=\delta[\gamma_0,\theta]$ is given by Eq.~(\ref{doppler}) 
with the initial $\gamma$ value, which does not change in the 
very short time during which TB dominates the X-ray AG. At 
the transparency radius and reference baryon number of Eq.~(\ref{Rtrans}), 
the temperature is $\sim 4.0$ eV $\simeq 4.6\times 10^4$ K, 
as in Eq.~(\ref{TTh}), and for 
$\delta=10^3$ and $\rm z=1$, the predicted energy flux, per CB, is 
$\rm 1.7\times 10^{-8}\; erg\, cm^{-2}\, s^{-1}$. 
This radiation's spectrum  is that of bremsstrahlung 
(a flat $\rm E\,dn_\gamma/dE$), typically extending 
to  $\rm E\sim 3\,T \,\delta/(1+z)\sim 12$ keV, in the X-ray domain. 
 
For our typical CB parameters, synchrotron emission takes over, as we 
shall see in the next section, 
 before the CB reaches its asymptotic radius and before the $\rm K$ term 
in Eq.~(\ref{Tbrem}) becomes important. During this early phase 
(for $\rm t\!\sim\! t_{trans}$), the TB emission by the CB's electrons 
declines with time, if cooling is dominated by bremsstrahlung losses, as 
$\rm R^{-3}\,T^{1/2}\!\sim\!R^{-5}\!\sim t^{-5}$.  
Should the temperature decrease be dominated by adiabatic cooling 
($\rm 3\,N_{CB}\,k\,dT=-P\, dV$), the substitution 
$\rm P=2\,\overline n_e\,k\,T$ results in $\rm T\propto R^{-2}$ and in 
a TB emission declining as $\rm R^{-3}\,T^{1/2}\!\sim\!R^{-4}\!\sim t^{-4}$. 
A decline as fast as $\rm t^{-4}$ or $\rm t^{-5}$ 
has been observed in early X-ray afterglow observations, 
e.g. GRB 920723: Burenin et al.~1999; GRB 970508: Piro et al. 1998;  GRB 
970828: Smith et al.~2001; GRB 990510:  Pian et al.~2001; GRB 010222: In 't 
Zand et al.~2001. In less than a minute of observer's time, this emission 
mechanism is overtaken by synchrotron radiation, whose decline is much 
slower, as we proceed to discuss. 
 
\section{Synchrotron radiation: the optical afterglow} 
 
In this section we study the various effects resulting from 
the interaction of a CB's entangled magnetic field with the 
ISM particles that it sweeps as it travels. 
The CBs lose momentum by sweeping the nuclei of the ionized ISM, and 
re-emitting them isotropically (in the CB's rest system) at an energy 
comparable to their incoming one. This allows us to predict the 
law of CB deceleration: the behaviour of its decreasing 
Lorentz factor $\rm\gamma(t)$. The incoming electrons, suffering 
collisions with the magnetic domains, and losing energy effectively 
by synchrotron radiation, acquire a predictable power-law energy 
spectrum, which implies a given distribution of the emitted photons. 
The emitted energy rate is equal to the rate at which the ISM electrons 
bring energy into the CB in its rest system; this provides the 
absolute normalization of the AG light curve. The dynamical time 
for the energy supply by the swept-up ISM electrons is much longer than  
the time it takes the electrons to acquire a power-law energy 
distribution and to emit synchrotron radiation, justifying a 
quasi-steady-state analysis. 
 
The rate at which the energy of the ISM electrons enters the CB 
(in its rest frame) is: 
\begin{equation} 
\rm {dE_{CB}\over dt_{CB}}\simeq \pi \, R^2\, n_e\, m_e\, c^3\,\gamma^2\, , 
\label{depo} 
\end{equation} 
where the incident 
ISM electron energy is $\rm \gamma \, m_e\, c^2$, 
and the extra power of $\gamma$ originates in the Lorentz 
contraction of the ISM electron density,  $\rm n_e$. 
An observer at a luminosity distance $\rm D_L$ and at an angle 
$\theta$ relative to the CB's direction of motion receives a total 
power per unit area (integrated over frequencies): 
\begin{equation} 
\rm {dF\over dt\, d\Omega} \simeq  {\pi \, R^2\, n_e\, m_e\,c^3\, 
\gamma^2\,\delta^4  \over 4\, \pi\, D_L^2}\, . 
\label{flux} 
\end{equation} 
At the very early time of CB transparency, for $\rm R=R_{trans}=10^{13}$ 
cm, $\rm n_e=1$ cm$^{-3}$, 
and for $\rm z=1$, $\gamma=\delta=10^3$, the above expression yields 
$1.27\, \times 10^{-9}$ erg/(cm$^2$ s), which 
is comparable to the bremsstrahlung emission of 
Eq.~(\ref{CBbrem}). But, as we shall see anon, the synchrotron 
radiation has a much softer spectrum than that of bremsstrahlung, and 
the latter mechanism dominates at early times in the X-ray domain. 
We have seen that in a matter of minutes the 
asymptotic radius $\rm R_{max}$ of Eq.~(\ref{Rinfinity}) is reached, so 
that thermal bremsstrahlung has decreased by more than three orders of 
magnitude, while the synchrotron radiation has increased by two or more 
orders of magnitude,  to become the dominant emission mechanism 
at all frequencies. 
 
To estimate a ``dynamical time'', 
$\rm \tau_{dyn}$, for the energy deposited by electrons in the CB, 
we forestall that the observed afterglow fluences are the 
ones expected in the CB model. We can then use Eqs.~(\ref{depo}) 
and (\ref{flux}) to deduce a total typical $\rm E_{CB}\sim 3\times 10^{44}$ 
erg and to conclude: 
\begin{equation} 
\rm \tau_{dyn}\equiv \left[{1\over E_{CB}}\,{dE_{CB}\over dt_{CB}}\right]^{-1} 
\sim (8.2\times 10^7\; s)\;\left[{10^3\over\gamma}\right]^2, 
\label{dyn} 
\end{equation} 
for $\rm R=R_{max}=2.2\times 10^{14}$ cm and $\rm n_e=10^{-3}$ cm$^{-3}$. 
 
In the CB's rest frame, the light crossing time for the asymptotic 
radius $\rm R_{max}$ is $\rm\tau_{cr}\sim R_{max}/c\sim 10^4$ s. 
The time for electrons that enter a CB to redistribute their energy 
as they bounce off a few magnetic ``sub-domains'' is a fraction of $\rm\tau_{cr}$; this is much shorter  
than $\rm \tau_{dyn}$, so that a cosmic-ray-like ``source'' 
distribution of electrons (a power law in energy) is steadily generated. 
The synchrotron cooling time of electrons in the CB's rest system is : 
\begin{equation} 
\rm \tau_{syn} \simeq {6\, \pi\, m_e\, c^2 \over \gamma_e\, c\, \sigma_T\, 
B^2} \sim (80\; s)\;\left[{10^3\over\gamma_e}\right]\; 
\left[{100\;\; Gauss\over B}\right]^2. 
\label{tsyn} 
\end{equation} 
In the above equation, we have distinguished 
$\rm\gamma_e$ (the Lorentz factor of an electron in the CB, 
in the CB's rest system) from $\gamma$ (the CB's bulk-motion 
Lorentz factor). Even for $\rm \gamma_e\to 1$, $\rm \tau_{dyn}\gg\tau_{syn}$. 
The Larmor radius of electrons $\rm r_L=p_e/(e\,B)$ in a $\rm B=100$ 
Gauss magnetic field is $\rm r_L\sim (1.5\times 10^5\; cm)(\gamma_e/10^3)$, 
so that even for very high energies, 
the residence time of the electrons in the CB 
is much longer than $\rm\tau_{syn}$. The above inequalities imply that for 
electrons of all energies, a spectrum of 
``Fermi-accelerated'', radiation-loss-modulated electrons is 
steadily generated. 
 
The incoming ISM particles are Fermi-accelerated 
by the turbulent magnetic fields inside the CBs to 
a comoving ``cosmic-ray'' spectral distribution $\rm dn/dE\sim 
E^{-\beta_p}$, with $\rm \beta_p\simeq 2.2$, as indicated 
by simulations (Bendarz \& Ostrowski 1998), analytical estimates 
and the interpretation of the 
observations of cosmic rays (e.g. Dar \& Plaga 1999). 
The acceleration being due to deflections by magnetic 
fields, the spectral shape of the ``source'' distributions of protons 
and electrons ought to be the same: $\rm \beta_e\simeq\beta_p$. 
The index of the equilibrium electron spectrum, modulated by radiation 
losses, is one unit higher: 
$\rm \beta_e\simeq\beta_p+1\approx 3.2$ (Dar \& De R\'ujula 2001). 
The emitted synchrotron radiation has a 
power spectrum with index $\rm\alpha=(\beta_e-1)/2$, so that: 
\begin{eqnarray} 
\rm {dF \over d\nu_{_{CB}}}&\equiv&\rm 
 \nu_{_{CB}}\,{dn_\gamma \over d\nu_{_{CB}}}\propto  
\nu_{_{CB}}^{-\alpha}\nonumber\\ 
\rm \alpha & \approx & \rm 1.1\; . 
\label{spectrum} 
\end{eqnarray} 
The spectrum of Eq.~(\ref{spectrum}) should roughly 
extend between two cutoff frequencies, $\rm \nu_{min}$ 
and $\rm \nu_{max}$, that reflect the energy of the incoming electrons 
and of the maximally accelerated ones. 
The integrated spectrum of Eq.~(\ref{flux}) 
is proportional to $\rm \nu_{min}^{1-\alpha}$ 
and $\rm \nu_{min}\propto \gamma_c^2$ where $\rm \gamma_c$ is the electrons' 
Lorentz factor above which they are in radiative equilibrium. 
Since  the individual frequencies $\nu$ and the limiting frequency 
$\rm\nu_{min}$ all refer to the CB's rest frame, and 
are Doppler-shifted by its motion ($\rm \nu\propto \delta\; \nu_{CB}$) 
as in Eqs.~(\ref{energies}), the non-frequency-integrated version 
of Eq.~(\ref{flux}) ---that is, the 
predicted spectral energy density for a GRB with 
a number $\rm n_{_{CB}}$ of CBs--- is: 
\begin{eqnarray} 
\rm 
F_\nu&\equiv& \rm{dF[\nu,t,\theta]\over dt\,d\nu\,d\Omega} \simeq 
\rm  n_{_{CB}}\,(\alpha-1)\,\pi\, 
[1+z]^{(1-\alpha)}\,m_e\,c^3\nonumber\\ 
&\times&\rm {n_e(x[t])\;[R(t)]^2\,[\gamma(t)]^{2\,\alpha} 
\over 4\,\pi\,[D_L(z)]^2\,\nu_c}\, 
\left[{2\gamma(t)\over 1+[\theta\,\gamma(t)]^2}\right]^{n} 
\,\left[{\nu\over \nu_c}\right]^{-\alpha};\nonumber\\ 
\rm n&\equiv& 3+\alpha\simeq 4.1\nonumber\\ 
\rm\nu_c&\sim&\rm {3 \,e\, B_\infty  \over 4 \pi \, m_e\, c} 
\simeq 0.42\,{B_\infty\over 100\:Gauss} \;GHz, 
\label{fluxdensity} 
\end{eqnarray} 
where we used the explicit form of $\delta$, Eq.~(\ref{doppler}), 
$\rm n_e(x[t])$ is the density along the CB's trajectory 
 $\rm x = \int \gamma \delta c \, dt/(1+z)$, 
and the overall normalization is obtained by assuming 
that relativistic electrons are in radiative equilibrium. 
The predicted normalization for a GRB 
is just an estimate, for  part of the energy 
deposited in the CB by ISM protons, as well as a fraction 
of its magnetic energy, may also be emitted as synchrotron 
radiation\footnote{At very low frequencies, such as those 
corresponding to radio waves in the observer's frame, 
Eq.~(\ref{fluxdensity}) is expected to break down for a variety 
of reasons: a deviation of the low-energy electron spectrum 
from a universal power-law, inverse synchrotron and inverse 
bremsstrhalung self-absorption, plasma frequency cutoff  
and the effect of competing mechanisms other 
than synchrotron radiation which all depend on the exact density profile  
and ionization state of the CB.}. 
 
When a ``typical'' CB, within a minute or two after the end of its GRB, 
reaches its final radius $\rm R_{max}\sim 2\times 10^{14}$ cm, and for 
$\rm n_e=1$ cm$^{-3}$, $\rm z=1$, and $\gamma=\delta=10^3$, 
Eq.~(\ref{fluxdensity}) yields $1.6\, \times 10^{-8}$ erg/(cm$^2$ s), 
of which $6.4\, \times 10^{-10}$ erg/(cm$^2$ s) ($\sim\! 3.7\%$) is in 
the 2--10 keV X-ray range and  $5.6\, \times 10^{-9}$ 
erg/(cm$^2$ s) ($\sim\! 3.2\%$) 
is in the visible range ($\rm 3900$ \AA $\leq \lambda\leq 7600$ 
\AA)  corresponding to a spectral flux density of 2 Jansky (8 
magnitude!) in the R band. 
For the next few hours $\rm\gamma(t)$ 
does not change significantly and the X-ray and optical 
AGs vary as $\rm n_e(x[t])$. This variation should 
in general be a decline, since the CBs are departing from a dense region.   
Such a decline  may 
have been observed both in the optical band in the 
case of GRB 990123 (Akerlof et al.~1999), which rose above $\sim 9$th 
magnitude tens of seconds after the GRB's onset, not far 
from our estimate with ``typical parameters''  (this GRB, whose early 
optical AG we shall discuss in detail, is at 
$\rm z=1.6$, but its initial $\gamma$ and $\delta$ are very large, 
see Table III). 
 
There may be an energy below which the synchrotron cooling time 
of the electrons in the CB is longer than their acceleration time. 
If so, the electron spectrum has an index $\rm \beta_e\simeq 
\beta_p$ and the synchrotron radiation below a certain frequency 
would have an index $\alpha\simeq 0.6$. We have implicitey assumed 
that this frequency is below the smallest optically observed ones, 
an assumption that the data generally  support.  
 
The flux of Eq.~(\ref{fluxdensity}) depends on $\rm \nu_c$
as $\rm \nu_c^{\alpha-1}$, roughly the 10-th root of it.
The actual value of $\rm \nu_c$ is therefore quite irrelevant
to the optical and X-ray AGs discussed here. In our study of
radio AGs (Dado et al. 2002) we find that, in the CB model,
$\rm\nu_c$ is actually the characteristic synchrotron frequency
emitted by the electrons that enter the CB with a Lorentz
factor (in the CB's rest frame) $\rm \gamma_e=\gamma(t)$,
that is $\rm \nu_c\sim 0.22\, \gamma_e^2\,\nu_L$, with 
$\rm \nu_L\propto B\propto \gamma$ the Larmor radius in the 
CB's magnetic field. The spectral index gradually changes from
$\alpha\approx 0.5$ to $\alpha\approx 1.1$ at this frequency.
For a spectrum with this transition in its 
power law, Eqs.~(\ref{fluxdensity}) are to be modified as follows: 
\begin{eqnarray} 
\rm [\gamma(t)]^{2\alpha} &\rightarrow&\rm   [\gamma(t)]^2\nonumber\\ 
\rm (\alpha-1) &\rightarrow&\rm {(\beta_p-2)(3-\beta_p)\over 2}\nonumber\\ 
\rm \alpha & \approx & \rm 0.5~~if~~(1+z) \nu\leq \delta \,\nu_c \; ,\nonumber\\ 
\rm \alpha & \approx & \rm 1.1~~if~~ (1+z)\nu\geq \delta \,\nu_c. 
\label{modif} 
\end{eqnarray} 
The difference between Eqs.~(\ref{fluxdensity}) and (\ref{modif})
is only relevant to the spectral shape of some very early optical AGs, 
and to the ensemble of the radio AGs (Dado et al. 2002).

\subsection{The density of the Inter-Stellar Medium} 
 
The density of ISM protons very close to a GRB progenitor 
plays a role in determining the asymptotic radius, $\rm R_{max}$, 
of a CB, see Eq.~(\ref{Rinfinity}). The density of ISM protons 
along a CB's trajectory controls, as we shall see, the evolution 
in time of the CB's  Lorentz factor $\rm \gamma(t)$. 
This function, and the density of interstellar electrons along 
the CB's trajectory, determine, via Eq.~(\ref{fluxdensity}), the 
AG properties. Clearly, we must discuss these densities in some detail. 
 
An analysis of historical SNe in the Galaxy (e.g. Higdon and 
Lingenfelter, 
1980), of SNe in the LMC (e.g., Dune et al. 2001) and of SNe in 
late-type 
galaxies (Kennicut et al. 1989; van Dyke et al. 1996; 
Higdon et al. 1998) indicates that $85\pm 10$\% of SNe occur in 
{\it superbubbles} (e.g., Lingenfelter et al. 2001). 
These are spaces 
of typical size 0.1 to 0.5 kpc, surrounding star-formation regions, 
that extend all the way into the galactic halo, and 
from which the ISM has been swept away by massive-star winds and 
previous SNe, resulting in an ISM with a low density  ($\rm n\sim 10^{-2}$ 
to $10^{-3}$ cm$^{-3}$) comparable to that in a galactic halo. 
 
At their Wolf-Rayet phase, 
massive stars that finally produce SNeII/Ib/Ic 
emit strong winds with typical velocities, 
$\rm v$,  close to $\rm 10^{3}\, km\, s^{-1}$, at a typical mass-loss rate 
$\rm \dot{M}=$ a few $\rm 10^{-4}\, M_\odot\, yr^{-1}$, over the last 
$\!\sim\! 10^5$ yr before the SN event. The density close to an 
imminent SN is governed by the recent Wolf-Rayet wind and ejections, 
and declines roughly quadratically with distance as:  
\begin{eqnarray} 
\rm n&\!\sim\!& \rm {\dot M \over 4\, \pi\, v\, x^2} \\ 
    & \!\approx\! &\rm(0.18\,cm^{-3})\left[{\dot{M}\over 10^{-4}\,M_\odot\, yr^{-1}} 
\right] 
\left[{10^3\, km\, s^{-1}\over v}\right] \left[{1\,pc\over x}\right]^2\! , 
\nonumber 
\label{WRwind} 
\end{eqnarray} 
till $\rm x\!\sim\! 10$ pc, where the density becomes that of the surrounding 
superbubble (e.g., Ramirez-Ruiz et al. 2000 
and references therein). 
 
\subsection{Complications simplified} 
To predict the explicit 
time-dependence of the AG from a CB, one needs to know $\rm R(t)$, 
$\rm \gamma(t)$ and the ISM density profile, $\rm n_e(x)$, along 
the CB's trajectory. 
Moreover, the various CBs that produce the different pulses in a 
single GRB have slightly different physical parameters (baryon 
number, Lorentz boost) that lead to the differences 
between the individual $\gamma$-ray pulses of a given GRB. 
The large powers of the Lorentz and Doppler factors 
in Eq.~(\ref{fluxdensity}) favour the contribution of CBs with the 
largest $\gamma$.  Given the extremely small fraction of solid angle that 
a CB spans as viewed from the SN centre, we do not expect consecutive CBs to 
hit the SN shell on the same spot (DD2000a). But it is in principle 
possible that the initial expansion and slowing down of the CBs by the SN shell  
and the ISM merges several of them into a single leading CB in the AG phase. 
One seems to be faced with a plethora of parameters and possibilities. 
 
We shall find it sufficient to characterize the 
various ISM densities that the CB encounters 
by two constant densities. One of them is the average proton density very close 
to the parent star, that determines the fast-reached 
asymptotic radius of the CB. For its reference value 
we adopt $\rm n_p^{SN}=1$ cm$^{-3}$. The other is the proton or electron 
density in the superbubble and in the galactic halo, for which we 
adopt as reference $\rm n=10^{-3}$ cm$^{-3}$. 
 
All other putative complications  previously quoted are 
eased by the fact that the times over which AGs extend are 
much longer than the typical intervals between GRB pulses, 
so that the AG light curve is the sum of temporally unresolved 
individual CB afterglows. We can therefore characterize, 
as in Eq.~(\ref{fluxdensity}), the 
AG with the parameters of a single CB, whose actual values 
would represent a weighted average. 
  
\subsection{The slow down of a CB}

A CB ionizing and ploughing through 
an ionized ISM of roughly constant density, would 
lose momentum at a roughly constant rate, independent of whether the ISM 
constituents are rescattered isotropically in the CB's rest frame, or 
their mass is added to that of the CB. 
Energy-momentum conservation for a highly 
relativistic CB of initial mass $\rm M_{CB}\simeq N_{CB}\,m_p$ 
results in the deceleration law (DD2000a): 
\begin{equation} 
\rm d\gamma=-{\pi\,R^2\,n_p\,\gamma^2\over N_{CB}}\, dx \, . 
\label{dgamma} 
\end{equation} 
The element $\rm dx$ of travelled distance (for $\gamma\gg  1$) is, 
according to Eqs.~(\ref{times}), related to the observer's 
time interval $\rm dt$ as: 
\begin{equation} 
\rm dx={c\,\gamma\,\delta\,dt\over(1+z)}\, . 
\label{dxsn} 
\end{equation} 
The law governing the CB's expansion rate is the differential 
version of Eq.~(\ref{tofR}), to wit: 
\begin{equation} 
\rm dR=c\,\beta_{trans}\, 
\sqrt{{R_{max}^3-R^3\over R_{max}^3-R_{trans}^3}}\;dt. 
\label{Roft} 
\end{equation} 
The above set of three equations can be integrated 
numerically for any given $\rm \gamma_0=\gamma(t=0)$, 
$\rm R(t=0)=R_{trans}$ and ISM density along the CB's path 
$\rm n_p(x)$. 
 
We limit our discussion to the case of CBs that, having reached 
their asymptotic radius, are moving 
through a constant-density medium, a case for which there are useful 
analytical expressions for the solution $\rm \gamma(t)$ 
of Eqs.~(\ref{dgamma}), (\ref{dxsn}) and (\ref{Roft}).  
A constant density is a fair approximation, 
for, in a very short observer's time,  the CB reaches the distance 
from the SN at which the density is that of the surrounding 
superbubble. To estimate this brief time, we may use the initial 
Lorentz and Doppler factors to obtain: 
\begin{equation} 
\rm t\sim{(1+z)\over c\,\gamma_0\,\delta_0}\, x\sim (34\, min)\, 
 \left[{x\over 10\;pc}\right]\, , 
\label{tsn} 
\end{equation} 
where we used the typical parameters $\rm z=1$, $\gamma_0=10^3$ and 
$\theta=1/\gamma_0$. 
This constant density approximation is also justified 
a posteriori by the agreement between our predicted AG 
light curves and the observed ones. 
 
We have argued that CBs reach a steady radius $\rm R_{max}$ 
in a few observer's minutes. To ascertain the 
CB's slow-down law for constant radius and constant ISM density, we 
may substitute Eqs.~(\ref{doppler}) and (\ref{dxsn}) into 
Eq.~(\ref{dgamma}) and integrate, to obtain: 
\begin{eqnarray} 
\rm 
&&{1\over\gamma^3}-\rm{1\over\gamma_0^3} 
+3\,\theta^2\,\left[{1\over\gamma}-{1\over\gamma_0}\right]= 
{6\,c\, t\over (1+z)\, x_\infty}\nonumber\\ 
&&\rm x_\infty\equiv{N_{CB}\over\pi\, R_{max}^2\, n_p}\simeq (1.3\;Mpc)\times 
\label{gamoft}\\ 
&&\rm 
\left[{N_{CB}\over 6\!\times\! 10^{50}}\right]^{1\over 3} 
\left[{10^{-3}cm^{-3}\over n_p}\right]\, 
\left[{n_p^{SN}\over 1\,cm^{-3}}\right]^{2\over 3} 
\left[{\gamma_0\over 10^3}{1/(3\sqrt{3})\over\beta_{trans}}\right]^{4\over 3}\!\! , 
\nonumber 
\end{eqnarray} 
where we have distinguished the average density $\rm n_p^{SN}$, close to 
the parent  SN (that determines $\rm R_{max}$) from the density $\rm n_p$ 
in the superbubble or the outer galaxy. 
The function of interest,  $\rm\gamma(t)$, is the real root 
of the above cubic equation, that is: 
\begin{eqnarray} 
\rm \gamma&=&\rm\gamma(\gamma_0,\theta,x_\infty;t) 
=\rm {1\over B} \,\left[\theta^2+C\,\theta^4+{1\over C}\right]\nonumber\\ 
\rm C&\equiv&\rm 
\left[{2\over B^2+2\,\theta^6+B\,\sqrt{B^2+4\,\theta^6}}\right]^{1/3} 
\nonumber\\ 
\rm B&\equiv&\rm 
{1\over \gamma_0^3}+{3\,\theta^2\over\gamma_0}+ 
{6\,c\, t\over  (1+z)\, x_\infty} 
\label{cubic} 
\end{eqnarray} 
 
The distance travelled by the CB 
is given by directly integrating Eq.~(\ref{dgamma}): 
\begin{equation} 
\rm x(\gamma)= x_\infty\, 
\left[{1\over\gamma}-{1\over\gamma_0}\right]\, . 
\label{range} 
\end{equation} 
The characteristic distance over which the Lorentz factor evolves 
from $\gamma_0$ to  $\gamma_0/2$ is $\rm x_\infty/\gamma_0$: 
roughly 1.3 kpc for  $\gamma_0=10^3$ and the reference value of 
$\rm x_\infty$ in Eq.~(\ref{gamoft}).

In Fig.~(\ref{figflux}) we show, 
for $\gamma_0=10^3$ and various viewing angles $\theta$, 
the AG flux predicted by Eq.~(\ref{fluxdensity}), with $\rm n=4.1$ and 
$\rm\gamma(t)$ as in Eq.~(\ref{cubic}) 
with the reference value of $\rm x_\infty$ in Eq.~(\ref{gamoft}). 
For $\theta\neq 0$, and particularly for sufficiently large $\gamma\,\theta_0$, 
the AG curve described by Eqs.~(\ref{fluxdensity}, \ref{cubic}) 
shows a very interesting behaviour. 
Since $\rm \gamma(t)$ is a decreasing function of time, the Doppler 
factor first increases with time, reaches a maximum value 
at $\rm\gamma(t)\,\theta\sim 1$ and then declines. An observer 
initially outside the beaming cone ($\gamma_0\,\theta >1$), 
sees an AG that initially rises with time.  As 
$\gamma$ decreases the cone broadens, and around $\gamma\theta\!\sim\! 1$ 
beaming becomes less efficient, the AG declines. 
 
In Fig.~(\ref{figfluxother}) we show 
the AG flux predicted by Eq.~(\ref{fluxdensity}), with $\rm n=4.1$ and 
$\rm\gamma(t)$ as in Eqs.~(\ref{gamoft}, \ref{cubic}), 
for $\gamma_0=1/\theta=10^3$, $\rm n_p=10^{-3}$ cm$^{-3}$, 
and for various values of the density  $\rm n_p^{SN}$ close 
to the progenitor SN. For the smaller $\rm n_p^{SN}$, the 
limiting CB's radius $\rm R_{max}$ of Eq.~(\ref{Rinfinity}) is 
larger. Consequently the CB, subsequently ploughing through the ISM, 
loses momentum at a faster pace. The figure shows that this may 
extinguish the AG very soon after the GRB, which would make it 
much harder to observe. This effect can also be produced by 
an increase in the ISM density $\rm n_p$, relative to 
the reference choice in Eq.~(\ref{gamoft}), so that the AGs of events not 
occurring in low-density superbubbles would also be hard to observe. 
These may be (along with extinction) the reasons why, in some 50\% 
of cases, GRBs appear not to have afterglows. 
 
For late times, when $\gamma\theta\ll 1$, Eq.~(\ref{gamoft}) implies that 
$\rm\gamma\propto t^{-1/3}$. 
According to Eq.~(\ref{fluxdensity}), then, 
the AG light curve approaches: 
\begin{eqnarray} 
\rm F_\nu (t) &\propto& \rm t^{-\tau};\nonumber\\ 
\tau&=&\rm 1+\alpha\simeq 2.1\, , 
\label{lateAG} 
\end{eqnarray} 
while the $\nu$ dependence stays put at $\nu^{-\alpha}$. This is compatible 
with what is  seen in various late-time AG observations (see e.g. 
GRB 980326: Bloom et al.~1999a; 
GRB 980519: Halpern et al.~1999; 
GRB 990123: Holland et al.~2000a; 
GRB 990510: Stanek et al.~1999, Harrison et al.~1999, Holland et al.~2000a;    
GRB 991208: Castro-Tirado et al.~2001); 
GRB 000301c: Masetti et al.~2000, Jensen et al.~2000; 
GRB 000926: Fynbo et al.~2001, Harrison et al.~2001, Price et al.~2001;  
GRB 010222: Masetti et al~2001, Stanek et al.~2001, Cowsik et al.~2001a).

\section{Comparison with optical observations} 
 
We compare our predictions with raw AG observations, i.e.  observations 
not corrected for extinction. The first step in our procedure 
is to work out what a raw ansatz-standard-candle supernova, SN1998bw, 
would look like at the location of each GRB. For that, we use the 
bare (unextinct) SN1998bw deduced by Galama et al.~(1998a), we 
transport it to the GRB location by way of Eq.~(\ref{bw}) and correct it 
for extinction in the host galaxy and in ours. 
For the extinction in our galaxy 
we use the estimates of Schlegel et al. (1998). 
For the correction at the host (at the emitted 
frequency $\rm (1+z)\,\nu$) we use the wavelength-dependent extinction 
estimated by the observers from the AG spectrum and the colours of the 
host galaxy.  The total extinction of the SN contribution for each 
GRB's optical AG is given in Table II. 
 
The next step in our procedure is to fit the raw data minus the 
SN contribution to a fixed host galaxy luminosity plus the CB's afterglow. 
We do not correct the latter for extinction, which only affects 
its fitted normalization. In the fits, 
the afterglow's spectral energy density is given by Eq.~(\ref{fluxdensity}) 
with $\rm R=R_{max}$ and a constant ISM density, that is: 
\begin{equation} 
\rm 
F_\nu=F \; [\gamma(t)]^{2\alpha}\;[\delta(t)]^{3+\alpha}\, , 
\label{fluxdensity2} 
\end{equation} 
with a normalization factor $\rm F$ which is one of the fitted parameters, 
and with $\rm \gamma(t)$ 
as in Eqs.~(\ref{gamoft})\footnote{Admittedly, our expression for 
the AG light curve is not as simple as Eq.~(\ref{pheno}), 
but it is also analytical. And it is justified.}. The contribution of the 
host galaxy is fixed inside the $\pm 1\sigma$ error range of the photometry 
measurements at the late times when the CB and the SN have 
become sufficiently dim. 
 
The parameters to be fit are $\rm F$ and $\alpha$ in Eq.~(\ref{fluxdensity2}), 
as well as $\rm x_\infty$, $\rm \gamma_0$ and $\theta$ entering the 
expressions for $\delta$ and $\gamma$ as functions of time, Eq.~(\ref{cubic}). 
The fit is done with the program MINUIT, checked over and 
over (in the hunt of false local minima) with different input parameters. 
The values of 
$\gamma_0$, $\rm \alpha$, $\theta$ and $\rm x_\infty$ 
for the different GRB afterglows 
are listed in Table II,  the overall normalization $\rm F$ will be 
 discussed separately. The results of these fits, which are very good, 
are shown in Figs.~(\ref{fig228}) to (\ref{jump508}). Most of them refer to 
R-band observations, which are the most extensive and accurate ones in the 
optical band, and extend to very late times. In order to test that our 
best fitted parameters are independent of frequency we have also fitted 
other bands, when sufficiently accurate data are available, and obtained 
very similar best fitted parameters. This can be seen in Table II 
for GRBs 990510 
and 990712, for which we also present V-band fits. 
Before discussing the results in much more detail in section 13, 
we pay attention to two very peculiar afterglows, and one 
particularly well measured one. 

\subsection{GRB 970508, CBs exiting a superbubble?}

The optical AG of GRB 970508 is the only 
one so far that has been seen to rise and fall very significantly. 
In Fig.~(\ref{1CB508}) we show how 
miserably a fit to this GRB fails, if it is made in the same way 
as all of our other fits.

We have argued in Section 10.1 that GRB progenitors are presumably
located in super-bubbles of 0.1 to 0.5 kpc size. There may be instances
in which the jet of CBs, after travelling for such a distance, does not
continue onwards to a similarly low-density halo region, but encounters
a higher-density domain. To test whether this may explain the very
peculiar shape of this AG, we have made a fit with two 
values of the ISM particle-number density, instead of one,
and a time (or distance from the progenitor) at which the
transition occurs.  The result is shown in Fig.~(\ref{jump508})
and it is fairly satisfactory. 

The fit parameters correspond to a density increasing by a factor 
of $\sim 2.2$ at $\rm t\sim 1.1$ day after burst,
at which point the CBs have travelled some $\sim 0.24$ kpc,
a very reasonable radius for a superbubble. The remaining
parameters are in the usual range, but for $\theta$, which,
at $\sim 3.5$ mrad, is on the large side. Given this large value
and $\gamma_0\sim 1123$, the time at which 
$\rm\gamma(t)\,\theta=1$ is reached exceptionally late; this
explains the rise and fall of the theoretical curve; see Dar
and De R\'ujula (2000a) for an earlier version of this result.
The relatively large $\theta$ is also in accordance with the
fact that, in spite of a relatively large
$\gamma$, the equivalent spherical energy of this GRB is
particularly low, see Tables I and II.

In an earlier version of this paper, we attributed the shape
of the AG of GRB 970508 to the effects of gravitational lensing
by an intervening star or binary, of mass $\rm\sim 2\,M_\odot$. 
That was an error. The required mass for an object placed mid-way
to the GRB location is almost three orders of magnitude bigger.
An effect of the observed size and shape
 could also be due to an even heavier object 
(such as a globular cluster) kiloparsecs away from the source,
but the chance probability for that is negligible. The possibility
of lensing by stars ---which has a few percent probability and
would produce amplification effects typically lasting $\sim 1$ hour--- 
is still interesting. We discuss it in Appendix III.

\subsection{GRB 980425, a very special case?} 
 
As reported in Table I, this GRB is by far the closest and yet, its measured 
$\gamma$-ray fluence is not large. We have argued (DD2000a) that 
in the CB model this is simply due to the fact that it is observed at a very 
large $\theta$: the GRB fluence has the same angular dependence 
as Eq.~(\ref{fluence}). The association of this GRB with SN1998bw is clear. 
In fact, the ``afterglow'' is dominated by the SN. Only the very last 
measured AG point is significantly above the $\rm ^{56}Co$ decay 
trend of the SN ejecta and would be due to the proper afterglow: that 
of the CBs (see Fig. 4  of D2000a). With only one point above the SN 
``background'' we cannot make, in the case of GRB 980425, 
a detailed fit of the sort we have made for the other GRB afterglows. 
But, as we shall see after we gain confidence on the success of the 
CB model in describing X-ray afterglows, we can use the X-ray 
measurements for this GRB to determine its parameters and to show 
that, after all, it is not at all a very special case. 
 
\subsection{GRB 990123, the early optical data} 
 
In the case of this GRB, there are good optical data  starting 
exceptionally early: during the $\gamma$-ray activity at $\rm t=22.18$ s after 
its detected beginning (Akerlof et al. 1999).  
The AG rises abruptly to a second point at $\rm t=47.38$ s, 
 and decreases thereafter. 
At the earlier stage, the CB is still hot and fully ionized, its  
thermal bremsstrahlung (free-free) self absorption  
is very large and fast decreasing, resulting in a fast rising AG 
that turns into a declining light curve 
when the CBs become transparent to short radio waves (corresponding to 
optical light in the observer frame). It is possible to explain the 
initial rise in detail, but the scarcity of the data and the surplus 
of parameters make the exercise moot.  
We choose to describe this AG from $\rm t=47.38$ s $\sim 5.5\times 10^{-4}$ 
d onwards, 
the first point shown of the measured decline and of Fig.~(\ref{early123}).  
We can use the relation between local distance and observer's 
time, Eq.~(\ref{dxsn}), 
and the specific values of $\rm z$, $\gamma_0$ and $\delta_0$ 
reported in the Tables for this GRB, to conclude that at the start of the 
optical AG data the CBs are a mere 0.46 pc away from the progenitor star. 
This is precisely in the domain where the density profile ought 
to be that of Eq.~(\ref{WRwind}), $\rm n\propto r^{-2}$, induced 
by the parent-star's wind and ejecta. Since 
at these early times the CB's deceleration is negligible, an 
$\rm r^{-2}$ density profile translates directly into an 
optical AG that declines as $\rm t^{-2}$, see Eq.~(\ref{fluxdensity}).  
 
In Fig.~(\ref{early123}) we report the result of a fit to the AG 
that includes a term proportional to $\rm t^{-2}$. The parameters 
of the late ($\rm t\!>\! 1$ d) AG remain essentially unchanged 
relative to the ones used before in constructing Fig.~(\ref{fig123}). 
The normalization of the fitted $\rm t^{-2}$ term is very close to 
that implied by Eqs.~(\ref{fluxdensity}) and (\ref{WRwind}): 
the first point of the data is exactly reproduced for a density 
$\rm n=0.54 \,cm^{-3}$ at that point, at which $\rm x\simeq 0.46$ pc. 
Thus, the CB model also succeeds in describing this very early 
AG in magnitude and shape\footnote{Even the most adamant defenders 
of fireballs admit that, in their scenarios, the absence of ``windy'' AGs is  
a problem, see, e.g. Piran (2001).}. 
 
\section{Comparison with X-ray observations}

The CBs enter their AG phase when they become transparent to radiation. 
As we have seen, their X-ray AG is dominated by thermal bremsstrahlung 
and first declines with time as $\rm \sim t^{-5}$, as described by 
Eqs.~(\ref{Tbrem}, \ref{fbrem}). In a matter of observer's minutes, for typical 
parameters, the CB's radius reaches its limiting value and 
the synchrotron radiation of Eq.~(\ref{fluxdensity}),  
which is proportional to $\rm n_p\, R^2$, becomes the dominant AG 
source in both the X-ray and optical domains: the corresponding lightcurves 
are approximately proportional. Once again for typical parameters, a CB 
reaches the galactic halo in just a few hours. The light curves before that  
time would be hard to model in minute detail, 
since the ISM density is no doubt changing rapidly. 
Coincidentally, there are not enough continuous X-ray measurements during 
the first few hours after the GRBs to reliably extract a density 
profile. 
 
The previous considerations justify a very simple 
description of the X-ray light curves. Let $\rm  W$ 
be the (constant) ratio of X-ray to R-band optical AGs. We expect: 
\begin{equation} 
\rm   F_X(t) \simeq F_X(t_{trans})\, \left[{t_{trans}\over t}\right]^5 
+ W\, F_R(t) \, . 
\label{Xdensity} 
\end{equation} 
where $\rm t$ is the observer's time since the ejection of the (last) CB, and 
 $\rm F_R(t)$ is the spectral energy density in the R-band, 
i.e., Eq.~(\ref{fluxdensity}), that we have fitted to the R-band AG 
observations.  The two parameters we fit to X-ray light curves 
are $\rm F_X(t_{trans})\times [t_{trans}]^5$ and 
$\rm W$, resulting in a very good description 
of  the measured X-ray afterglows of the GRBs 
of known redshift: 970228, 970828, 971214, 980613, 990123, 
990510, 000926 and 010222. The cases of GRBs 970508 
and 991216 require a non-constant density profile. 
The fits to the X-ray AGs are shown in 
Figs.~(\ref{X228}-\ref{X222}).  
GRB 980425 is discussed separately below. 
 
During a GRB the X-ray luminosity fluctuates as the $\gamma$-ray 
luminosity does, changing abruptly at the end of the GRB  into 
a very fast decline. This is expected in the CB model, in which the two 
behaviours have slightly different origins: thermal bremsstrahlung from 
the various CBs' surfaces during the GRB, thermal bremsstrahlung from the 
rapidly cooling CBs' volumes as they become transparent to the radiation 
they enclose, in a short time $\rm t_{trans}$ ---of the order of 
seconds--- at the end of each individual CB. In 
Figs.~(\ref{X228}-\ref{X222}) we have therefore 
shown our fits to data beginning 
at the start of the sharp X-ray decline.

The use in Eq.~(\ref{Xdensity}) of a more careful treatment of the 
bremsstrahlung contribution (evolving from $\rm t^{-5}$ to $\rm t^{-3}$) 
is unwarranted: the assumption of a constant ISM density 
should be inappropriate between 
$ \rm \sim\! 2\times 10^{-3}$ and $\sim\! 0.2$ days, and there are no 
data in that domain except for GRB 991216 and perhaps 970508 that, like the 
early optical AG of GRB 990123, suggest an initial density variation  
$\rm\propto 1/r^2$, resulting in an observed $\rm\sim t^{-2}$ decline. 
In these cases, we have fitted the data with 
Eq.~(\ref{Xdensity}) for a $\rm \sim 1/r^2$ plus constant density along the 
CB trajectory. For the particularly interesting case of GRB 970508,  
shown in Fig.~(\ref{X508}), we have also subtracted from the data 
the contribution of the X-ray line observed before 0.8 d.  
This reduces the two points observed  between 
0.2 and 1 day by a factor $\sim 0.39$. 
The overall result is compatible with an effect that, at late times, 
is achromatic, since both the late optical and X-ray AGs are
proportional to $\rm n_e$; see Fig.~(\ref{jump508}) 
for the optical counterpart.
 
In what follows we refer to the initially rapidly-falling part of 
an X-ray AG, as ``early'' and to the subsequent much flatter 
AG as ``late''. 
 
We have chosen to write the prediction for X-rays 
as in Eq.~(\ref{Xdensity}) to better expose the expected achromaticity 
of the late AG. Moreover, the data for late X-ray AGs is generally more 
sparse than for optical light; it is therefore advisable to exploit 
the fact that the CB's parameters are better fit from the R-band data, 
and to write the late X-ray AG as a rescaling of that data.  
But the effects of absorption are much less severe 
for X rays and there is one parameter, 
the overall normalization, for which it is preferable to use the 
X-ray fluence in testing the model. The values of 
$\alpha$, $\theta$, $\gamma_0$, and $\rm x_\infty$, 
fit for each GRB to the R-band AG, are sufficient to deduce,  
via Eq.~(\ref{fluxdensity}), the shape ---but not the normalization--- of the 
expected {\it late} X-ray flux in the 2-10 keV domain.  
This flux depends, via $\rm n_e\,R_{max}^2$ at fixed $\rm x_\infty$, 
on the number of CBs and their individual baryon number, as 
$\rm F\propto n_{_{CB}}\,N_{CB}=N_{jet}$ (the dependence on 
$\rm B_\perp$, via $\nu_0^{\alpha-1}$, is extremely weak). 
Let $\rm q$, for a given GRB, be the ratio of the observed flux to the one  
expected for our canonical $\rm N_{CB}=6\times 10^{50}$, and  
for one dominant (largest $\gamma$) cannonball: $\rm n_{_{CB}}=1$. 
The values of $\rm q$ are reported in Table III. They range from $\sim 1/3$ to 
$\sim 3$, indicating that the CB model satisfactorily explains the  
{\it late} X-ray AG normalizations 
and that the total baryon number of the  
ensemble of CBs that dominate the AG appears to be quite constant.

One can see in Figs.~(\ref{X228}) to (\ref{X222}), that the 
X-ray fluences at the start of the X-ray decline are 
$\rm \sim 10^{-8}-10^{-7}\; erg\, cm^{-2}\, s^{-1}$. These figures  
for the beginning of the early X-ray AG compare 
quite favourably with the typical prediction quoted after Eq.~(\ref{fbrem}): 
$\rm 4.35\times 10^{-9}\; erg\, cm^{-2}\, s^{-1}$ for a single CB. 
Once again, the parameters extracted from fitting the optical AGs 
are not sufficient to fix case by case the overall {\it early} X-ray normalization in 
Eq.~(\ref{fbrem}), which also depends directly on other parameters 
(notably $\rm N_{CB}$, $\rm n_{CB}$ and the radii of the still-growing CBs).  
Using these degrees of freedom we could fit not only the late, but also the 
early absolute X-ray fluences. Suffice it to say that the magnitude of the  
{\it early} X-ray afterglow is also the one expected in the CB model.

The fitted values of $\rm W$ in Eq.~(\ref{Xdensity}) 
are in fair agreement with the absorption-dependent expectation 
$\rm  [A(\nu_X)/A(\nu_O)]\,(\nu_X/ \nu_O)^{-\alpha_{_{\!OX}}}$ 
from the spectrum of Eq.~(\ref{spectrum}). 
Some examples of values of $\rm\alpha_{_{\!OX}}$ 
extracted this way are 
$1.06\pm 0.12$  for GRB 970228 (Frontera et al., 1998), 
$1.12 \pm 0.07$  for GRB 970508 (Galama et al. 1998b), 
$0.95\pm 0.1$ for GRB 971214 (Dal Fiume et al. 2000), 
$>1$ for GRB 980703 (Vreeswijk et al. 1999) 
$0.96 \pm 0.26$  for GRB 990510 (Pian et al. 2001), 
0.9 to 1.1 for GRB 000926 (Piro et al. 2001) and 
$0.97 \pm 0.05$  for GRB 010222 (In `t Zand et al. 2001). 
However, the inferred values of $\rm \alpha_{OX}$ are affected by large 
and very uncertain extinctions in the CB and host galaxies. Evidence for 
large extinction of optical AGs of GRBs in their host galaxies is provided 
by the large column densities ($\rm N_H > 10^{22}\, cm^{-1}$) extracted 
from the X-ray observations of some GRBs (e.g., GRBs 970228, 970508, 
970828, 971214, 980329, 980519: Owens et al. 1998;  GRB 980703: Vreeswijk 
et al. 1999) and from the absorbed spectra of the optical AG of other GRBs 
(e.g. GRB 990712: Vreeswijk et al. 2000; GRB 991216: Halpern et al. 2000a; 
GRB 000131: Andersen et al. 2000). In fact, the failure to detect the 
optical AG of many long duration GRBs with well localized X-ray AG ---like 
GRB 970111, GRB 970616, GRB 970815, and 970828--- may be due to strong 
extinction of their optical AG in the host galaxy (Djorgovski et al. 2001 
and references therein).

\subsection{GRB 980425 is not exceptional} 
 
\begin{figure} 
\begin{center} 
\vspace*{.003cm} 
\hspace*{-0cm} 
\epsfig{file=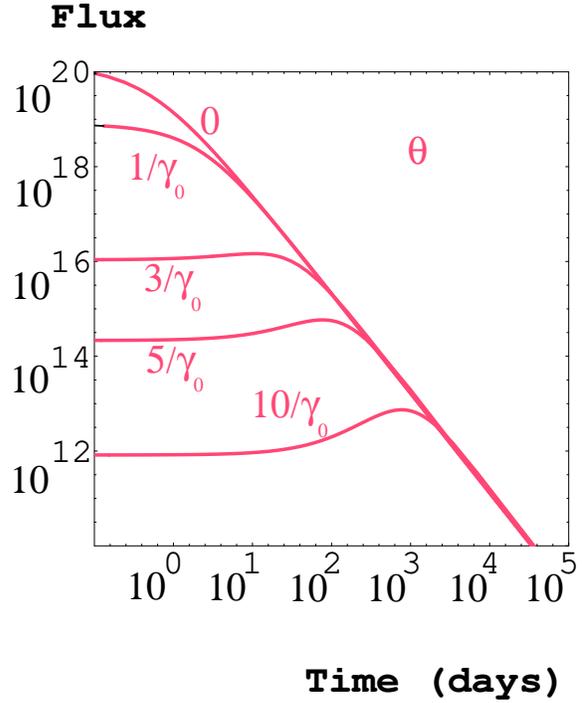,width=7.5cm} 
\caption{Afterglow flux, in arbitrary units, as a function of observer's 
time, for $\gamma_0=10^3$ and various viewing angles $\theta$, 
as given by Eqs.~(\ref{fluxdensity}) with $\rm n=4.1$ and 
$\rm\gamma(t)$ as in Eq.~(\ref{cubic}) 
with the reference value of $\rm x_\infty$ in Eq.~(\ref{gamoft}).} 
\vspace*{-0.5cm} 
\label{figflux} 
\end{center}  
\end{figure}

\begin{figure} 
\begin{center} 
\vspace*{.003cm} 
\hspace*{-0cm} 
\epsfig{file=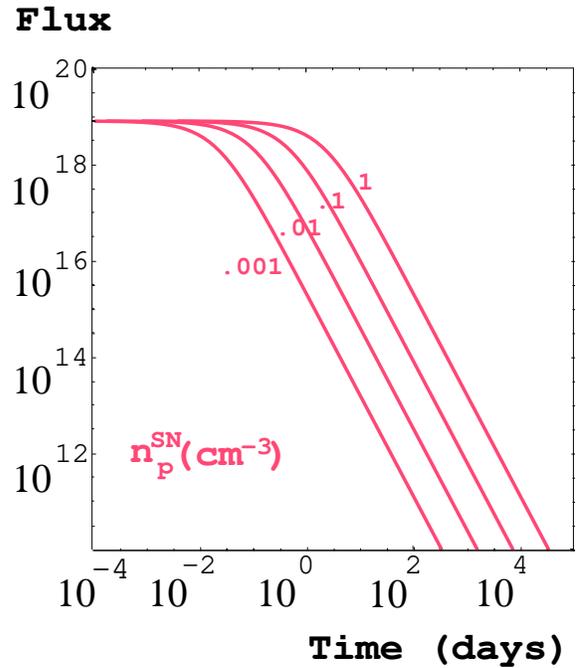,width=7.5cm} 
\caption{ 
Afterglow flux, in arbitrary units, as a function of observer's 
time, for $\gamma_0=1/\theta=10^3$ and various values 
of the (constant) ISM density, $\rm n^{SN}$, close to the GRB progenitor, 
as given by Eqs.~(\ref{fluxdensity}), with $\rm n=4.1$ and 
$\rm\gamma(t)$ as in Eq.~(\ref{cubic}).} 
\vspace*{-0.5cm} 
\label{figfluxother} 
\end{center}  
\end{figure}

The $\gamma$-ray fluence of this GRB is not atypical but 
its redshift, $\rm z=0.0085$, is extremely small. 
If it is not intrinsically exceptional, its CBs must be viewed from 
an atypically large angle (DD2000a). 
For large $\theta$ the CBs' afterglow is strongly 
reduced, as can be seen from Fig.~(\ref{figflux}), allowing for the possibility 
that the AG is dominated by the SN. 
Consequently, the CB parameters cannot be derived 
from the optical light curve of the blended SN 1998bw/GRB 980425 system. 
But they can be deduced from the X-ray emission of the system if we 
assume, unlike the observers do (e.g., Pian et al. 1999;  Pian et al. 2001) 
that it was produced by the CBs and {\it not} by the conventional 
quasi-spherical SN ejecta\footnote{We are indebted to E. Pian for 
discussions on this point.}. Indeed,  
significant X-ray emission from SNe has been detected only 
at much later times after the event. Moreover, the exceptionally slow 
decline of the X-ray AG in this GRB is what is expected from the 
large viewing angle interpretation, see Fig.~(\ref{figflux}). 
 
Given all of the above, for this GRB 
we have ``reversed'' our procedure by first fitting the X-ray AG of the 
SN 1998bw/GRB 980425 system. The fit is shown in 
the upper part of Fig.~(\ref{425}) and the 
fitted parameters are listed in Table II. The data are not very 
precise and the fit (for which we assumed $\alpha=1.1$, 
as fitted for all other GRBs) is not one of our best,  
but it inescapably requires a very large viewing angle $\theta$. 
The best fitted angle is $\theta\sim 8.3$ mrad, 
corresponding, for the fitted $\gamma\sim 750$, 
to an initial Doppler factor  $\delta\sim 37$.  
If the CBs of GRB 980425 had been viewed from a typical viewing 
angle, $\theta\!\leq\! 1/\gamma_0 $, the equivalent isotropic energy 
would have been in the range $ 7.3\times 10^{51}$ 
to $5.8 \times 10^{52}$ erg, like that of all other GRBs. 
 
If we assume that for GRB 980425 
the extinction, ISM density, and CB 
radius were the same as for other GRBs well measured in X-rays, 
such as GRB 990510 or GRB 010222, we can use Eq.~(\ref{fluxdensity}) 
to derive the expected intensity of the X-ray AG plateau of GRB 980425, 
see Fig.~(\ref{425}), and its caption. 
The results are $\rm F_X[425] = 0.32\, F_X[510]$, and  $\rm F_X[425] = 0.15\, 
F_X[222] $, both yielding $\rm F_X[425] 
\!\sim\! 4\times 10^{-13}\, erg\, cm^{-2}\, s^{-1}$, in 
agreement with the BeppoSAX observations (Pian, 1999; Pian et al. 2000). 
The double success in deducing a ``normal''  GRB equivalent 
isotropic energy, and the 
intensity of the X-ray AG, constitutes a very strong support for 
the alleged association of SN1998bw with (a not exceptional) GRB 980425. 
 
The  fitted parameters of the X-ray AG can be used to predict the 
magnitude and shape of the 
optical AG of the blended SN 1998bw/GRB 980425 system, 
if we assume the same  V/X extinction ratio as in GRBs 990510, 000926 and 
010222. This we do in the lower part of Fig.~(\ref{425}). The 
CBs' contribution dominates at very late time and, 
remarkably, it is in perfect agreement with the HST 
observation (Fynbo et al. 2000) on day 778 after the GRB.    
 
\section{Discussion of the results on optical AGs} 
 
\begin{figure}[t] 
\begin{tabular}{cc} 
\hskip 2truecm 
\vspace*{2cm} 
\hspace*{-1.7cm} 
\epsfig{file=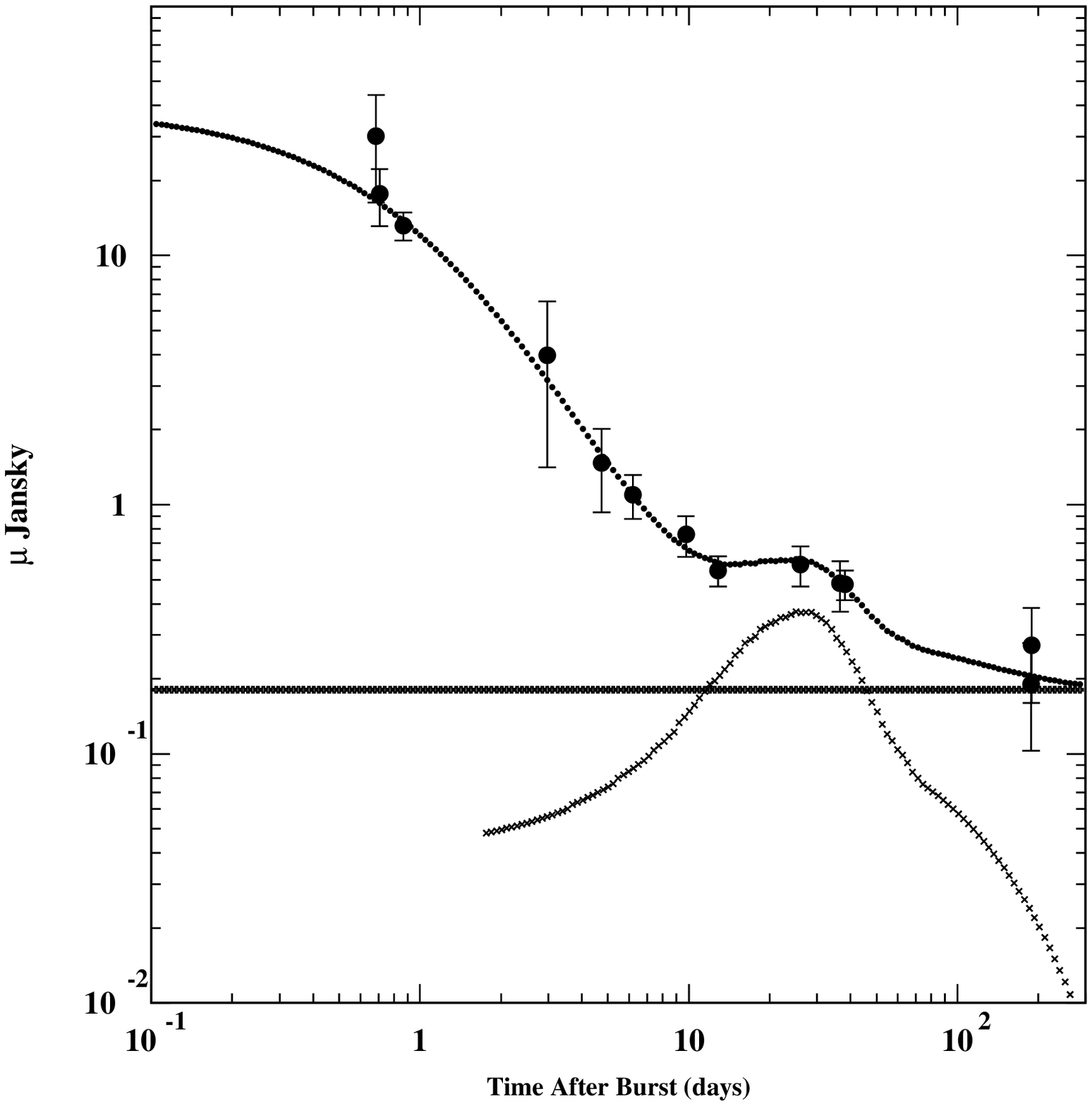, width=8cm} \\ 
\hspace*{.5cm} 
\epsfig{file=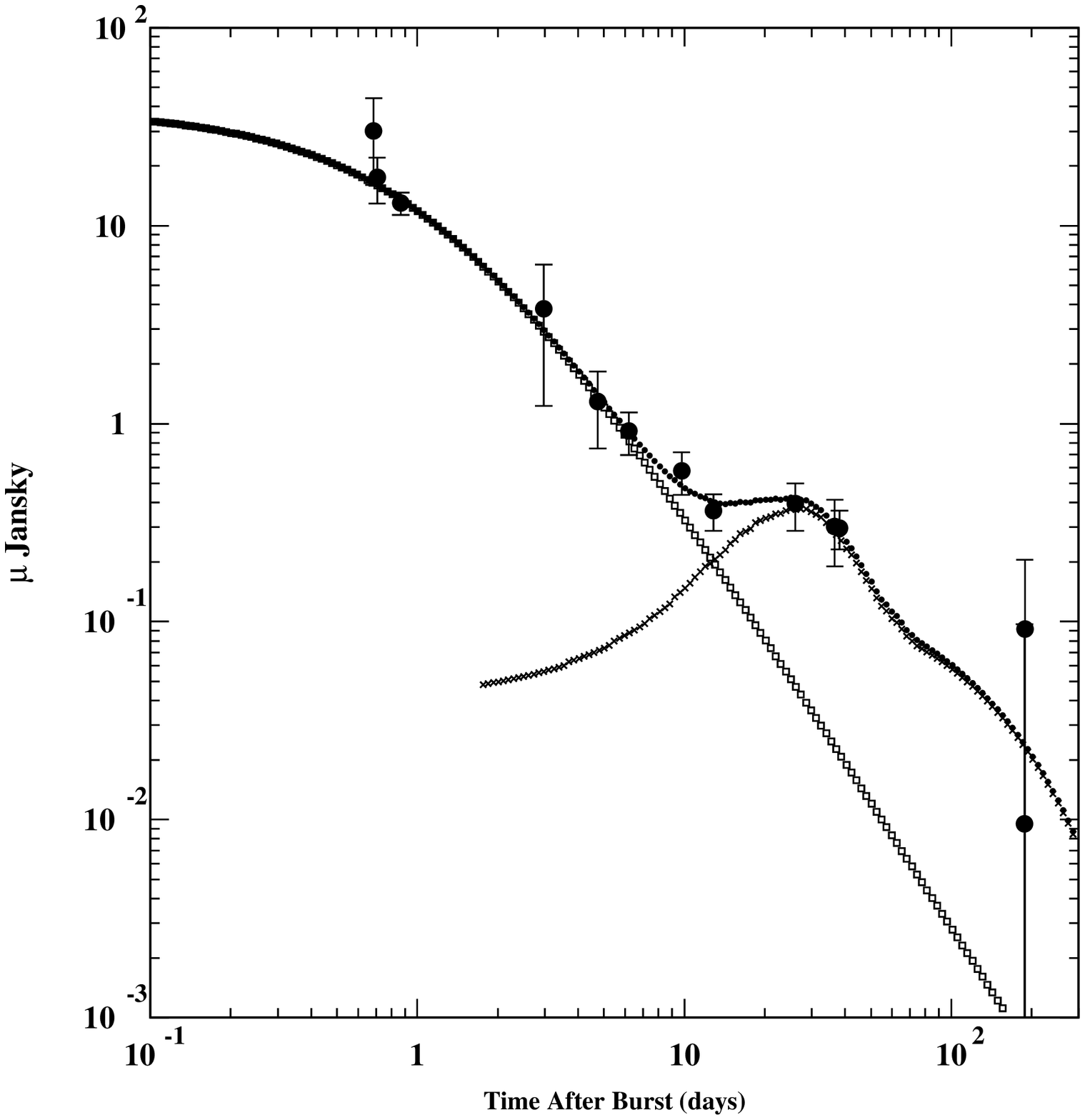, width=8cm} 
\end{tabular} 
\caption{Comparisons between our fitted R-band afterglow 
(upper curves) and the observations listed in Table II, 
not corrected for extinction, 
for GRB 970228, at $\rm z=0.695$. 
Upper panel: Without subtraction of the host 
galaxy's contribution (the straight line). 
Lower panel: With the host galaxy subtracted, and the 
CB's AG (the line of squares) given by Eqs.~(\ref{fluxdensity}) and 
(\ref{cubic}). The contribution 
from a 1998bw-like supernova placed at the GRB's 
redshift, Eq.~(\ref{bw}), corrected for  extinction, 
is indicated in both panels by a line of crosses. 
The SN bump is clearly discernible.} 
\label{fig228} 
\end{figure}

\begin{figure}[t] 
\begin{tabular}{cc} 
\hskip 2truecm 
\vspace*{2cm} 
\hspace*{-1.7cm} 
\epsfig{file=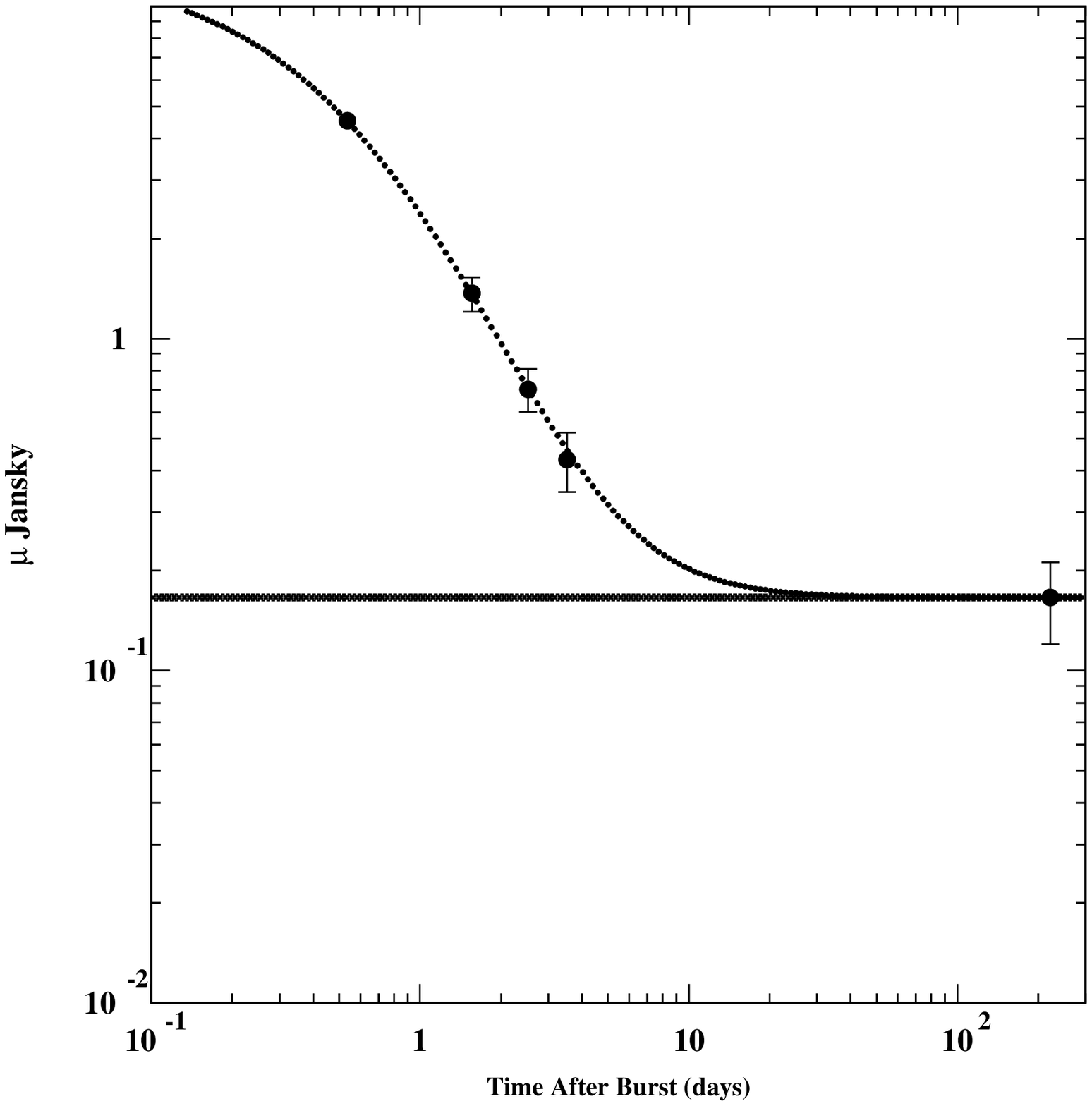, width=8cm} \\ 
\hspace*{.5cm} 
\epsfig{file=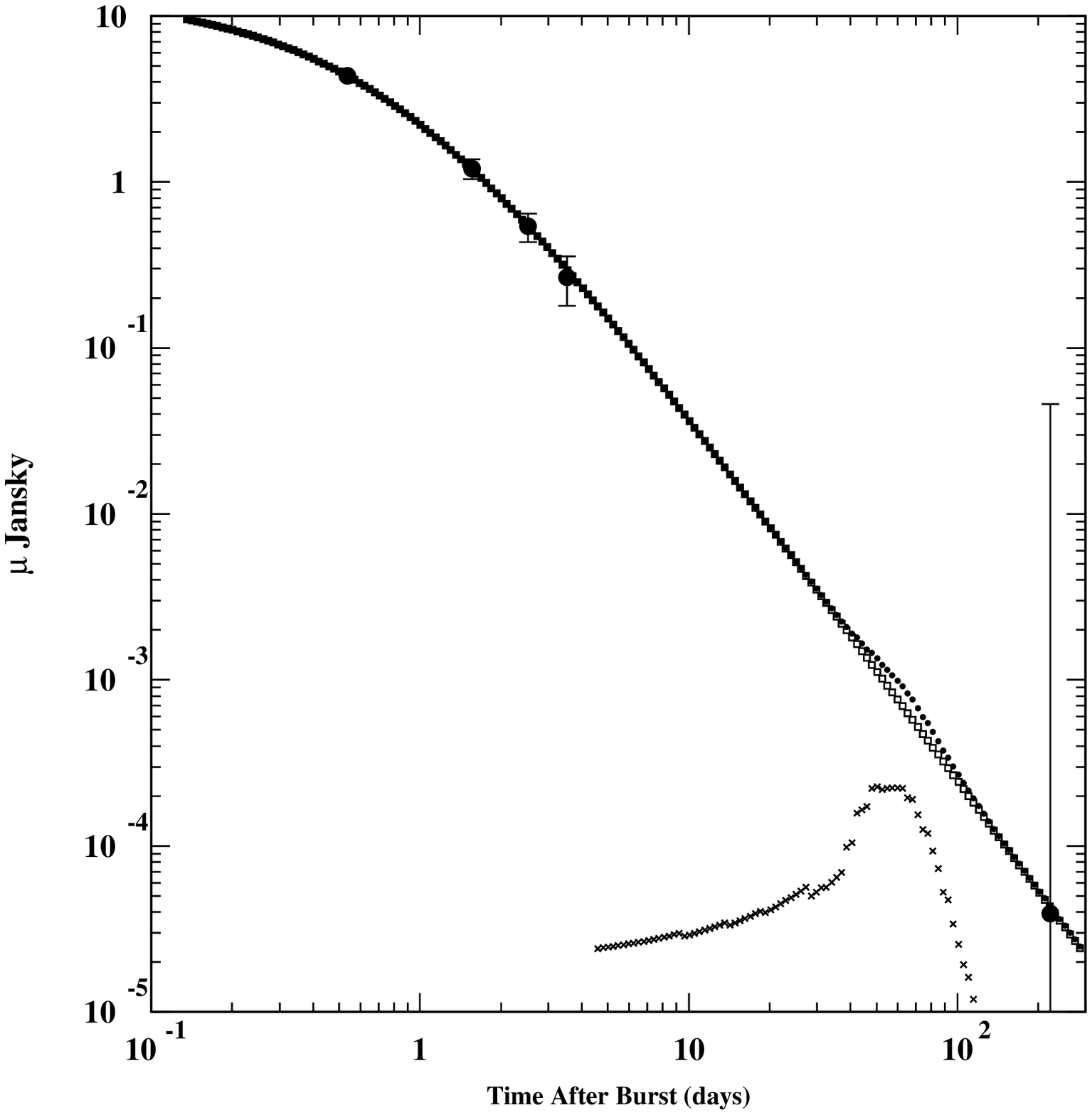, width=8cm} 
\end{tabular} 
\caption{Comparisons between our fitted R-band afterglow 
(upper curves) and the observations listed in Table II, 
not corrected for  extinction, 
for GRB 971214, at $\rm z=3.418$. 
Upper panel: Without subtraction of the host 
galaxy's contribution (the straight line). 
Lower panel: With the host galaxy subtracted, and the 
CB's AG (the line of squares) given by Eqs.~(\ref{fluxdensity}) and 
(\ref{cubic}). The contribution 
from a 1998bw-like supernova placed at the GRB's 
redshift, Eq.~(\ref{bw}), corrected for  extinction, 
is indicated by a line of crosses. The expected supernova 
contribution is too weak to be observed.} 
\label{fig214} 
\end{figure}

\begin{figure}[t] 
\begin{tabular}{cc} 
\hskip 2truecm 
\vspace*{2cm} 
\hspace*{-1.7cm} 
\epsfig{file=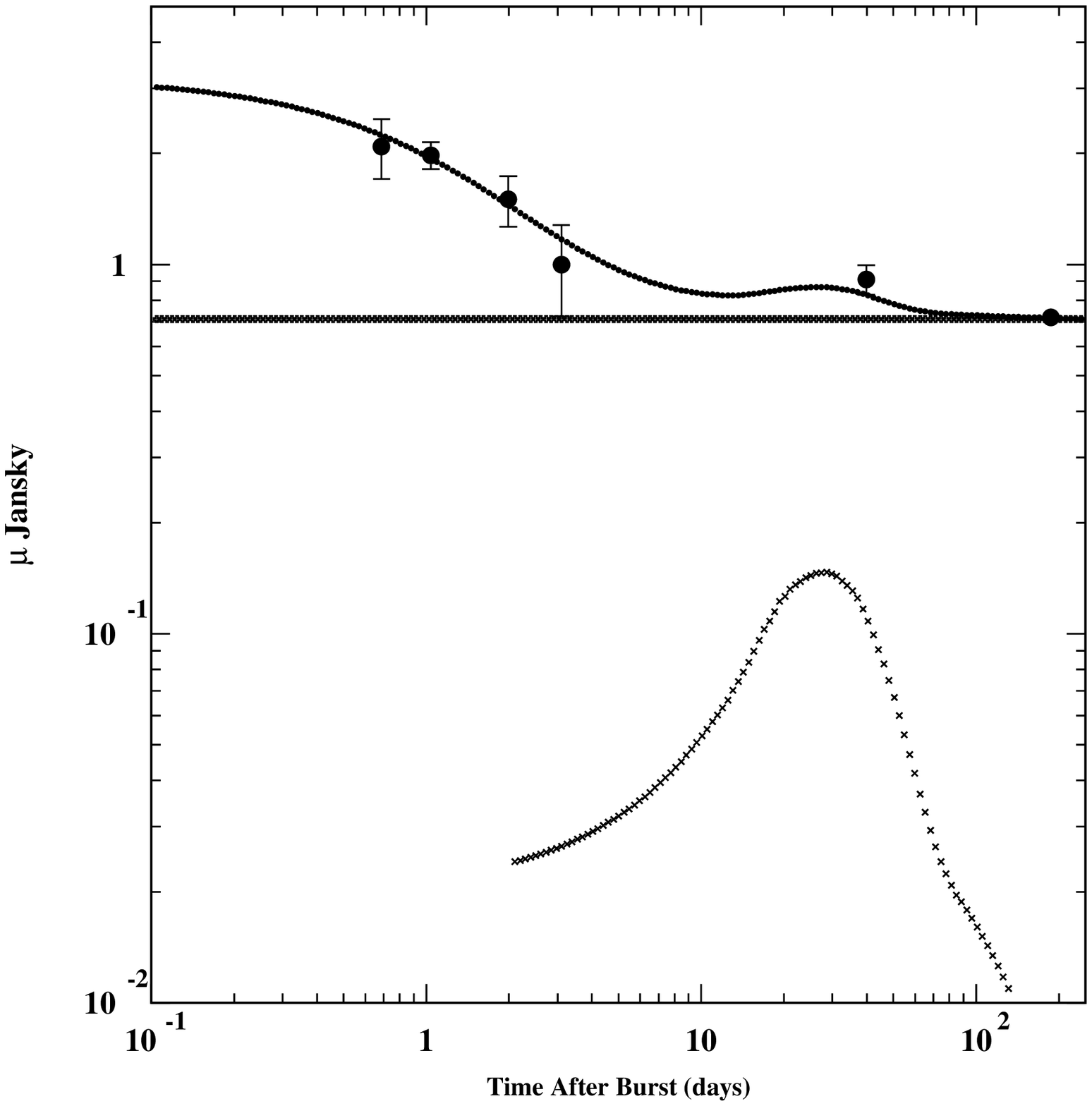, width=8cm} \\ 
\hspace*{.5cm} 
\epsfig{file=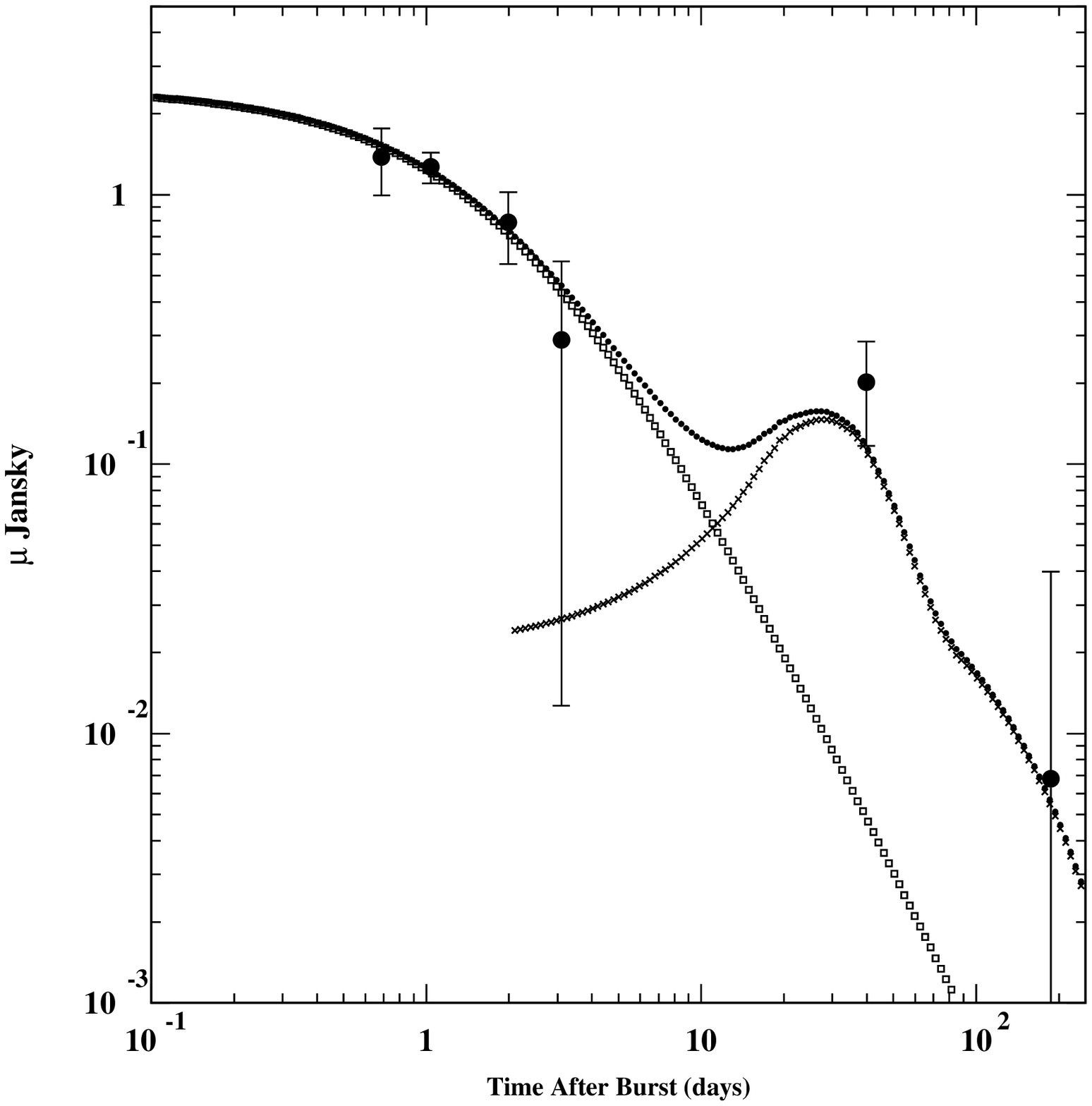, width=8cm} 
\end{tabular} 
\caption{Comparisons between our fitted R-band afterglow 
(upper curves) and the observations listed in Table II, 
not corrected for  extinction, 
for GRB 980613, at $\rm z=1.096$. 
Upper panel: Without subtraction of the host 
galaxy's contribution (the straight line). 
Lower panel: With the host galaxy subtracted, and the 
CB's AG (the line of squares) given by Eqs.~(\ref{fluxdensity}) and 
(\ref{cubic}). The contribution 
from a 1998bw-like supernova placed at the GRB's 
redshift, Eq.~(\ref{bw}), corrected for  extinction, 
is indicated in both panels by a line of crosses. 
A SN 1998bw-like contribution, though based on 
only one significant point, appears to be required.} 
\label{fig613} 
\end{figure}

\begin{figure}[t] 
\begin{tabular}{cc} 
\hskip 2truecm 
\vspace*{2cm} 
\hspace*{-1.7cm} 
\epsfig{file=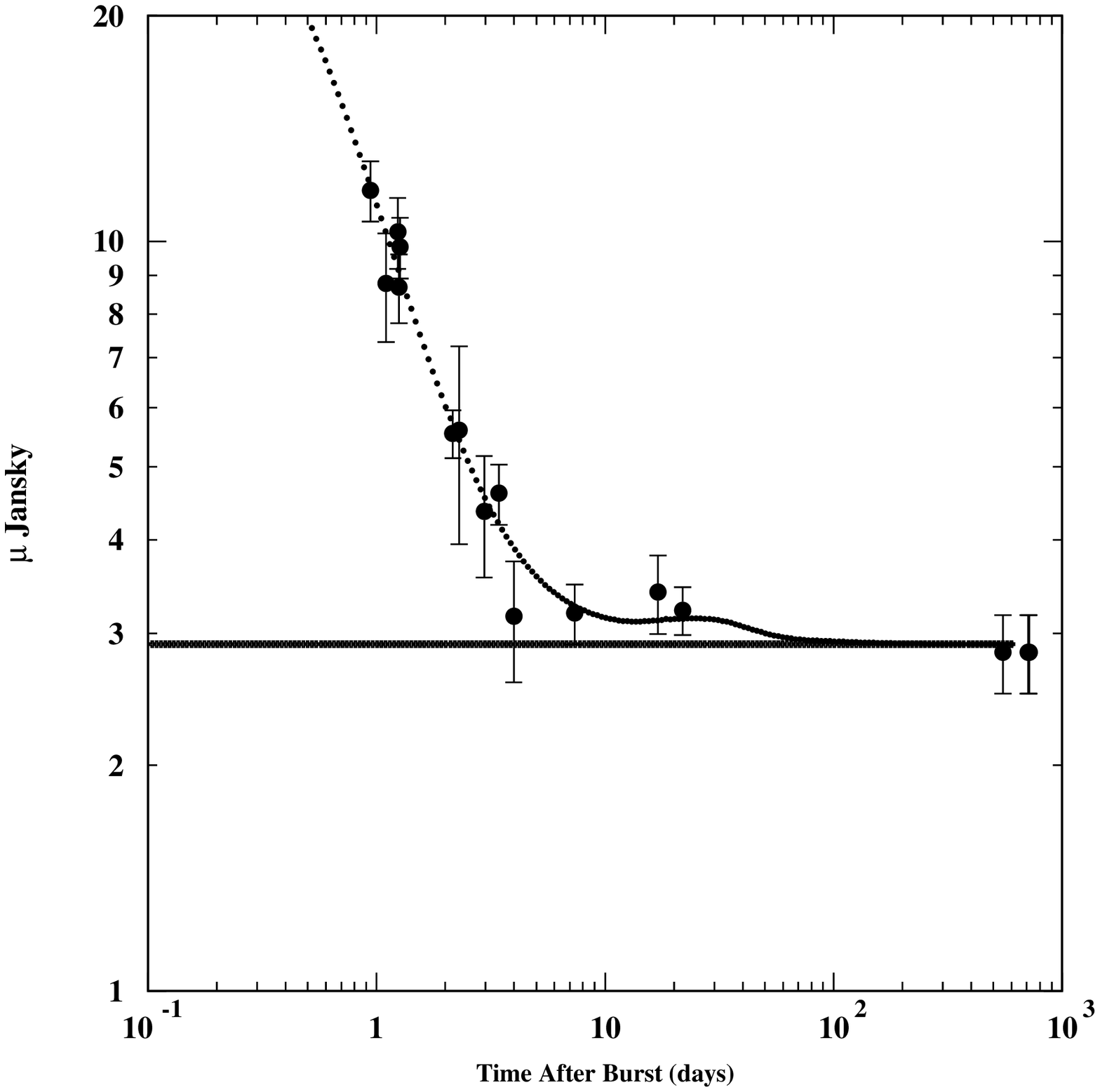, width=8cm} \\ 
\hspace*{.5cm} 
\epsfig{file=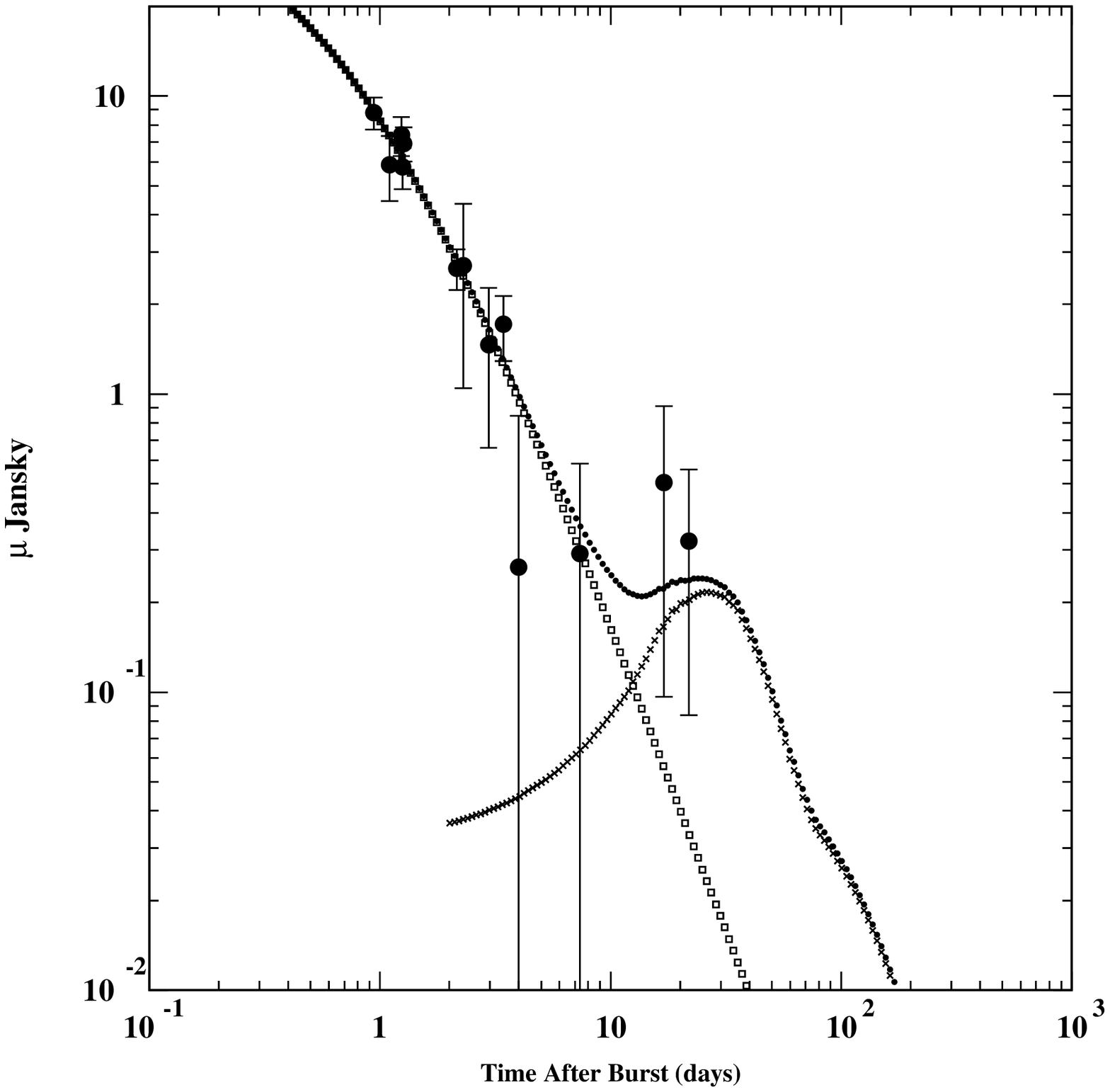, width=8cm} 
\end{tabular} 
\caption{ Comparisons between our fitted R-band afterglow 
(upper curves) and the observations listed in Table II, 
not corrected for  extinction, 
for GRB 980703, at $\rm z=0.966$. 
Upper panel: Without subtraction of the host 
galaxy's contribution (the straight line). 
Lower panel: With the host galaxy subtracted, and the 
CB's AG (the line of squares) given by Eqs.~(\ref{fluxdensity}) and 
(\ref{cubic}). The contribution 
from a 1998bw-like supernova placed at the GRB's 
redshift, Eq.~(\ref{bw}), corrected for  extinction, 
is indicated  by a line of crosses. 
A SN 1998bw-like contribution, though the errors 
are large, appears to be required.} 
\label{fig703} 
\end{figure}

\begin{figure}[t] 
\begin{tabular}{cc} 
\hskip 2truecm 
\vspace*{2cm} 
\hspace*{-1.7cm} 
\epsfig{file=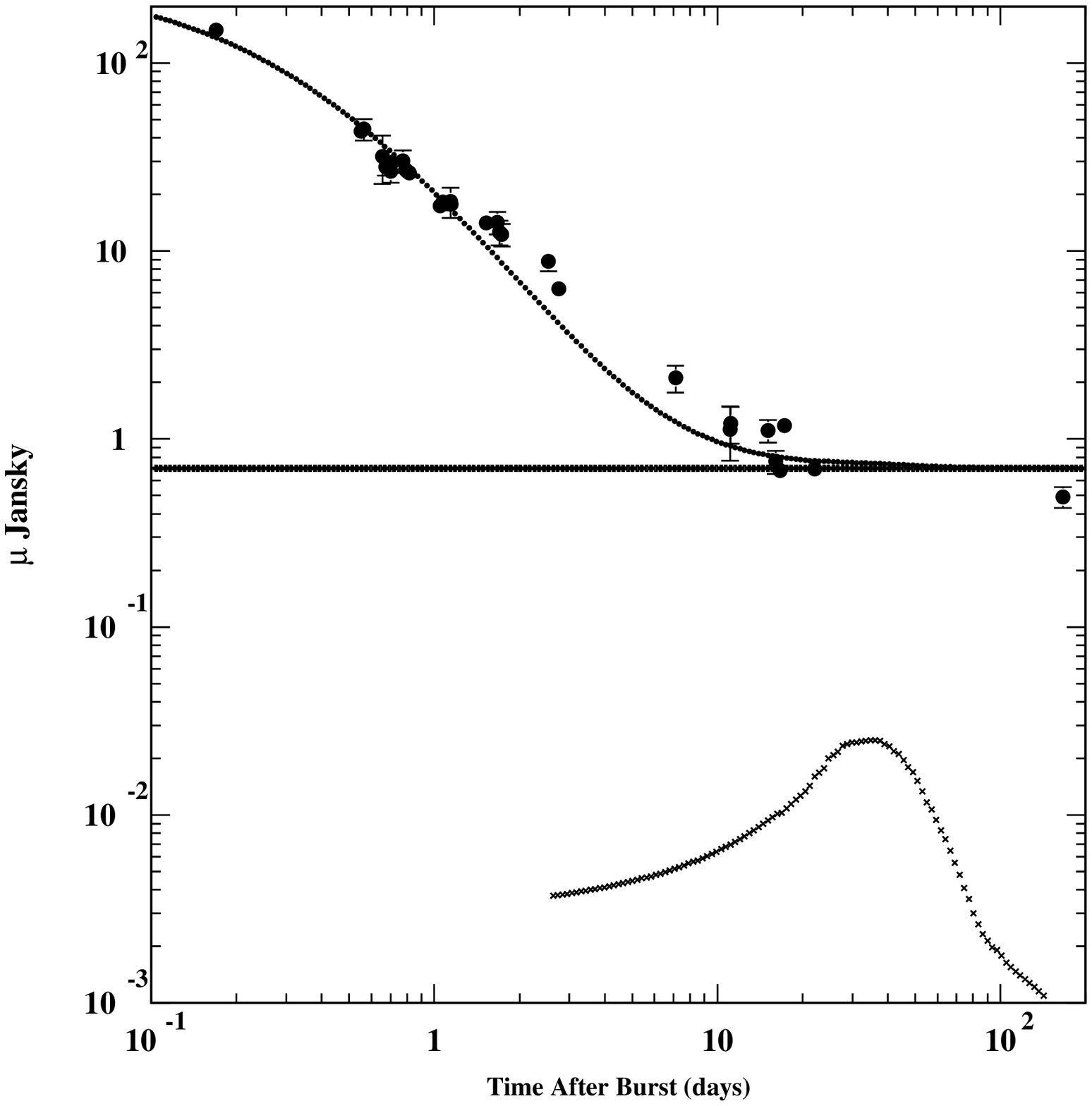, width=8cm} \\ 
\hspace*{.5cm} 
\epsfig{file=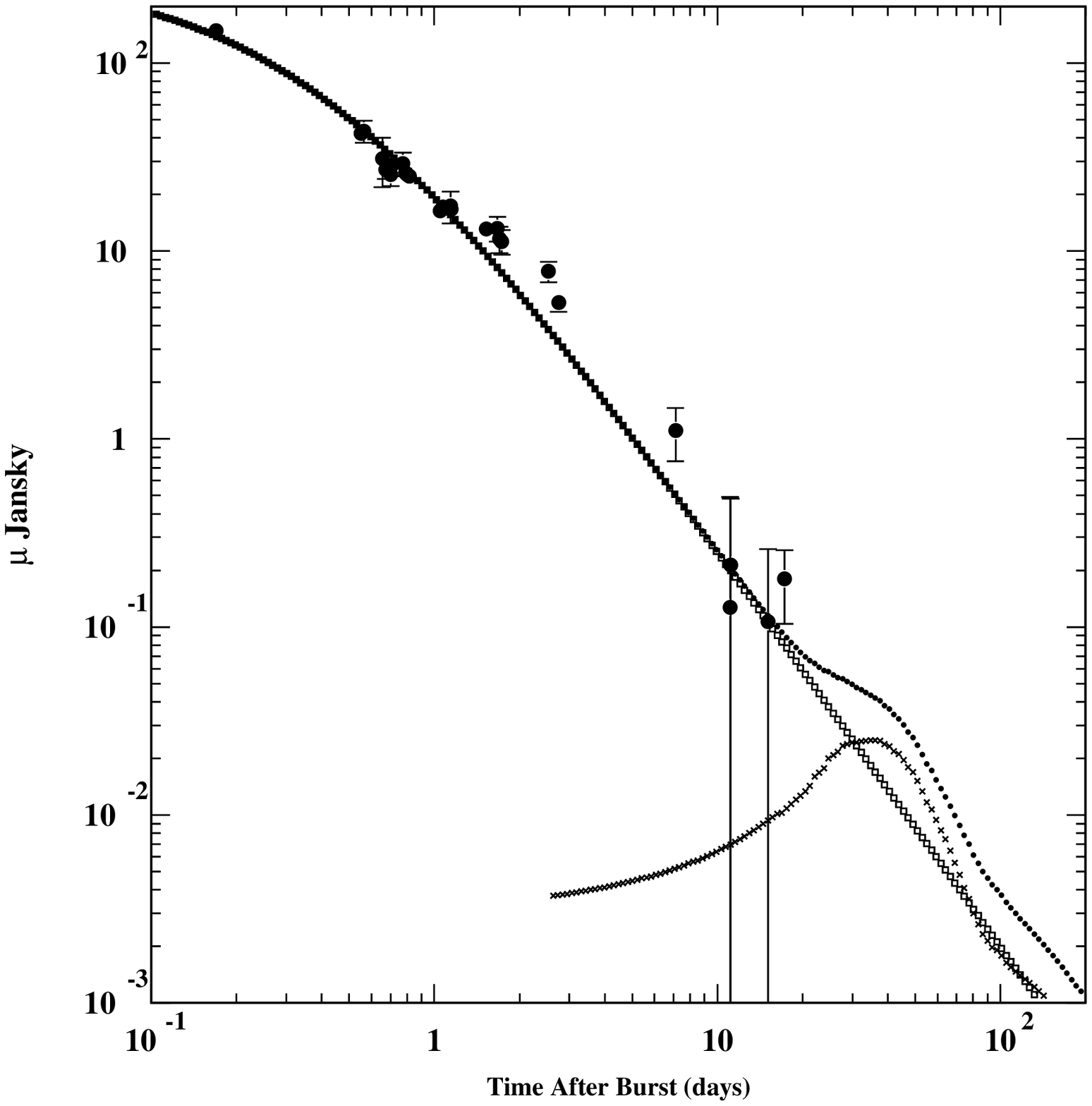, width=8cm} 
\end{tabular} 
\caption{Comparisons between our fitted R-band afterglow 
(upper curves) and the observations listed in Table II, 
not corrected for  extinction, 
for GRB 990123, at $\rm z=1.6$. 
Upper panel: Without subtraction of the host 
galaxy's contribution (the straight line). 
Lower panel: With the host galaxy subtracted, and the 
CB's AG (the line of squares) given by Eqs.~(\ref{fluxdensity}) and 
(\ref{cubic}). The contribution 
from a 1998bw-like supernova placed at the GRB's 
redshift, Eq.~(\ref{bw}), corrected for  extinction, 
is indicated by a line of crosses. The fit is not very good. 
There are no observations at the time when a SN1998bw-like 
contribution may have been seen.} 
\label{fig123} 
\end{figure}

\begin{figure}[t] 
\begin{tabular}{cc} 
\hskip 2truecm 
\vspace*{2cm} 
\hspace*{-1.7cm} 
\epsfig{file=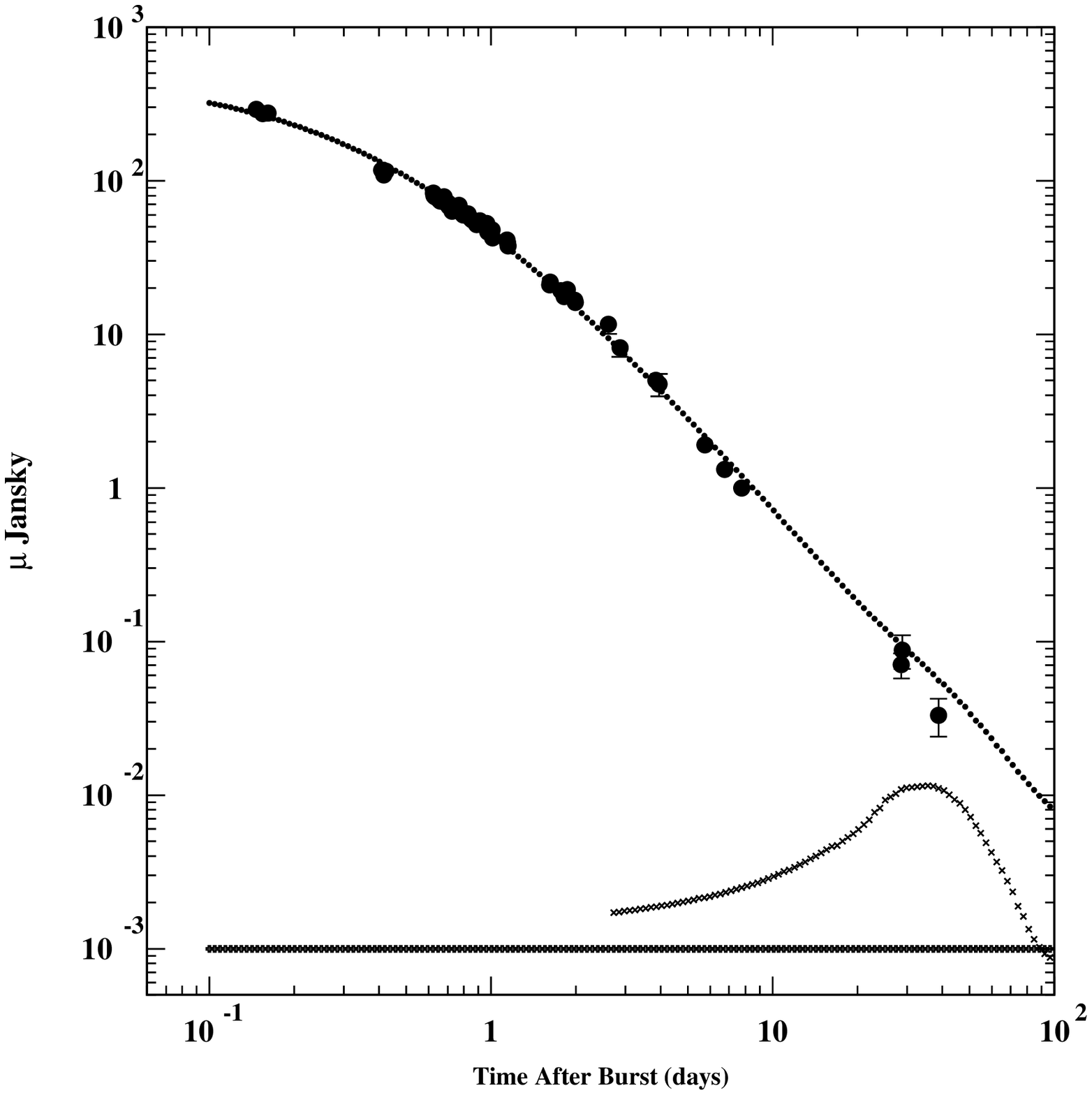, width=8cm} \\ 
\hspace*{.5cm} 
\epsfig{file=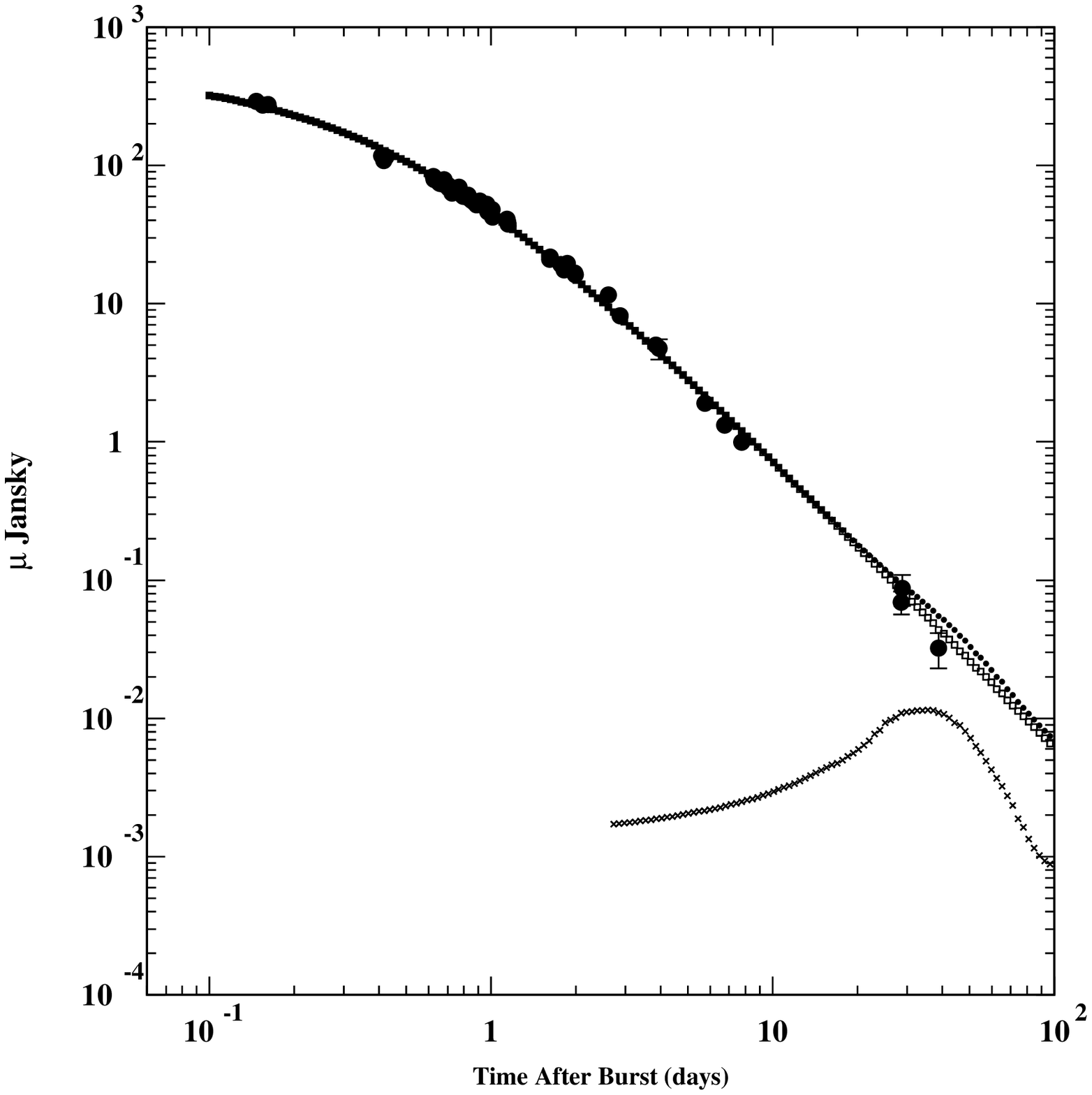, width=8cm} 
\end{tabular} 
\caption{Comparisons between our fitted R-band afterglow 
(upper curves) and the observations listed in Table II, 
not corrected for  extinction, 
for GRB 990510, at $\rm z=1.619$. 
Upper panel: Without subtraction of the host 
galaxy's contribution (the straight line). 
Lower panel: With the host galaxy subtracted, and the 
CB's AG (the line of squares) given by Eqs.~(\ref{fluxdensity}) and 
(\ref{cubic}). The contribution 
from a 1998bw-like supernova placed at the GRB's 
redshift, Eq.~(\ref{bw}), corrected for  extinction, 
is indicated in both panels by a line of crosses. 
The expected supernova contribution is too weak to be observed.} 
\label{fig510} 
\end{figure}

\begin{figure}[t] 
\begin{tabular}{cc} 
\hskip 2truecm 
\vspace*{2cm} 
\hspace*{-1.7cm} 
\epsfig{file=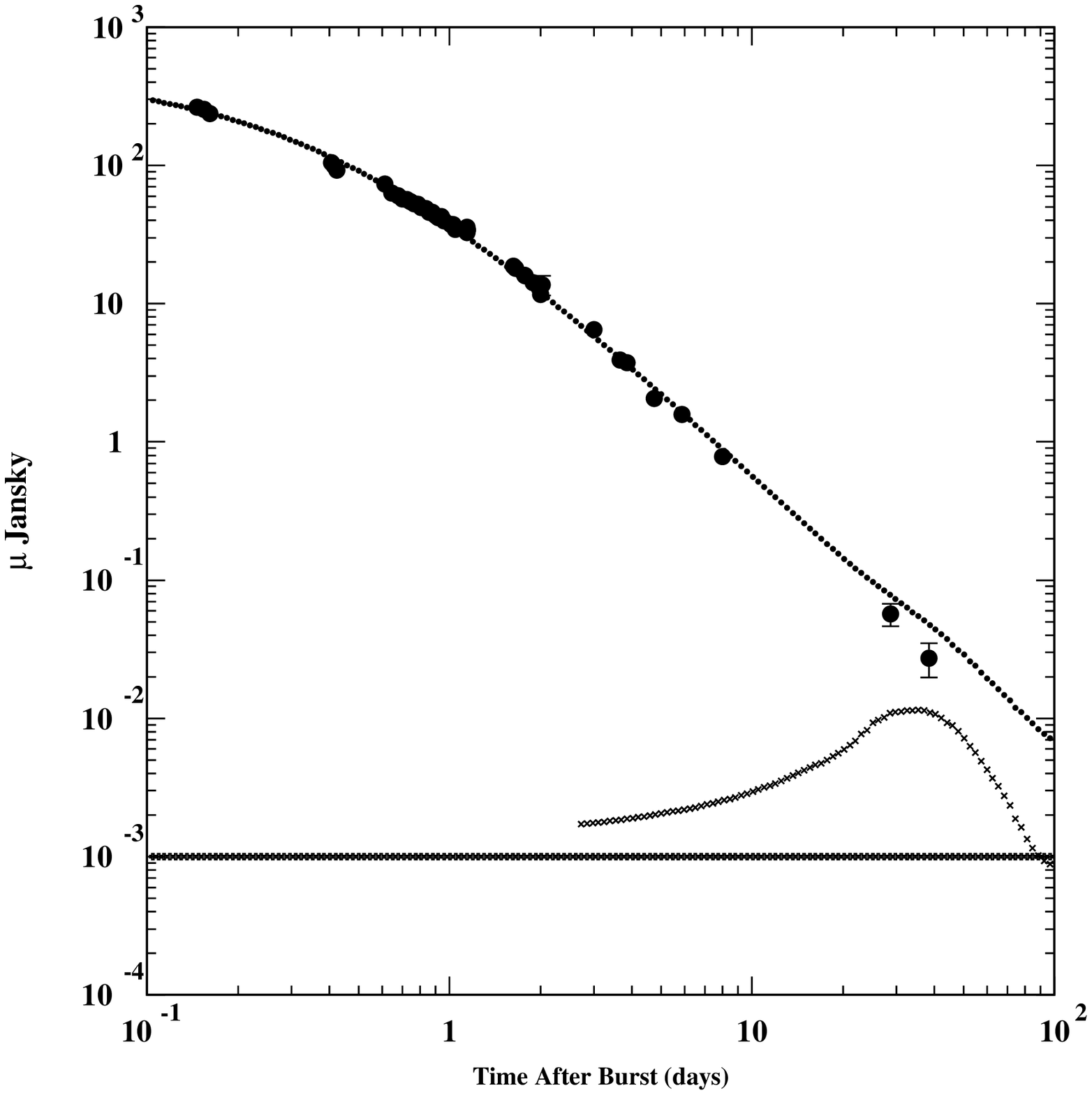, width=8cm} \\ 
\hspace*{.5cm} 
\epsfig{file=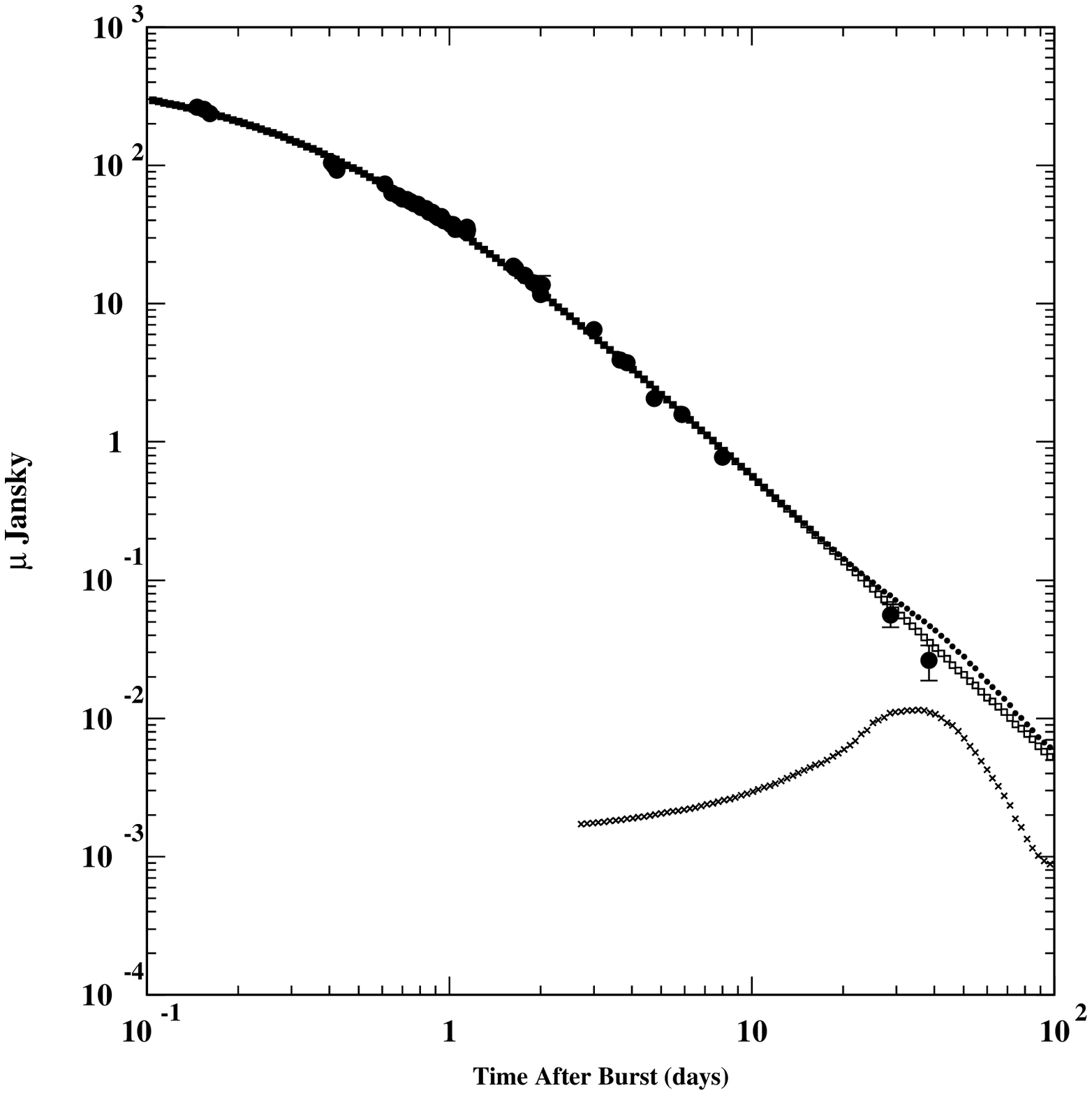, width=8cm} 
\end{tabular} 
\caption{Comparisons between our fitted V-band afterglow 
(upper curves) and the observations listed in Table II, 
not corrected for  extinction, 
for GRB 990510, at $\rm z=1.619$. 
Upper panel: Without subtraction of the host 
galaxy's contribution (the straight line). 
Lower panel: With the host galaxy subtracted, and the 
CB's AG (the line of squares) given by Eqs.~(\ref{fluxdensity}) and 
(\ref{cubic}). The contribution 
from a 1998bw-like supernova placed at the GRB's 
redshift, Eq.~(\ref{bw}), corrected for  extinction, 
is indicated in both panels by a line of crosses. 
The expected supernova contribution is too weak to be observed.} 
\label{fig510V} 
\end{figure}

\begin{figure}[t] 
\begin{tabular}{cc} 
\hskip 2truecm 
\vspace*{2cm} 
\hspace*{-1.7cm} 
\epsfig{file=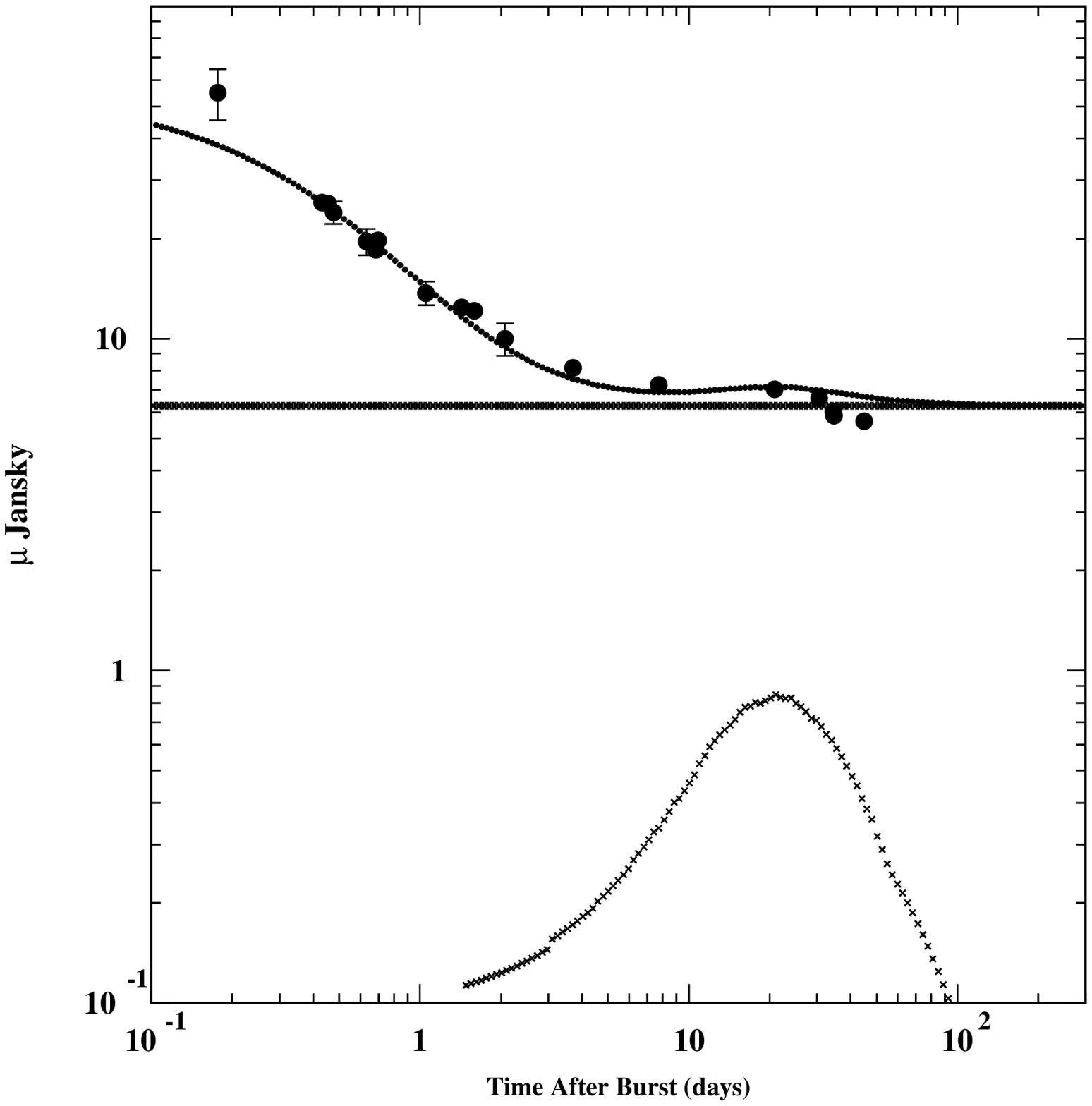, width=8cm} \\ 
\hspace*{.5cm} 
\epsfig{file=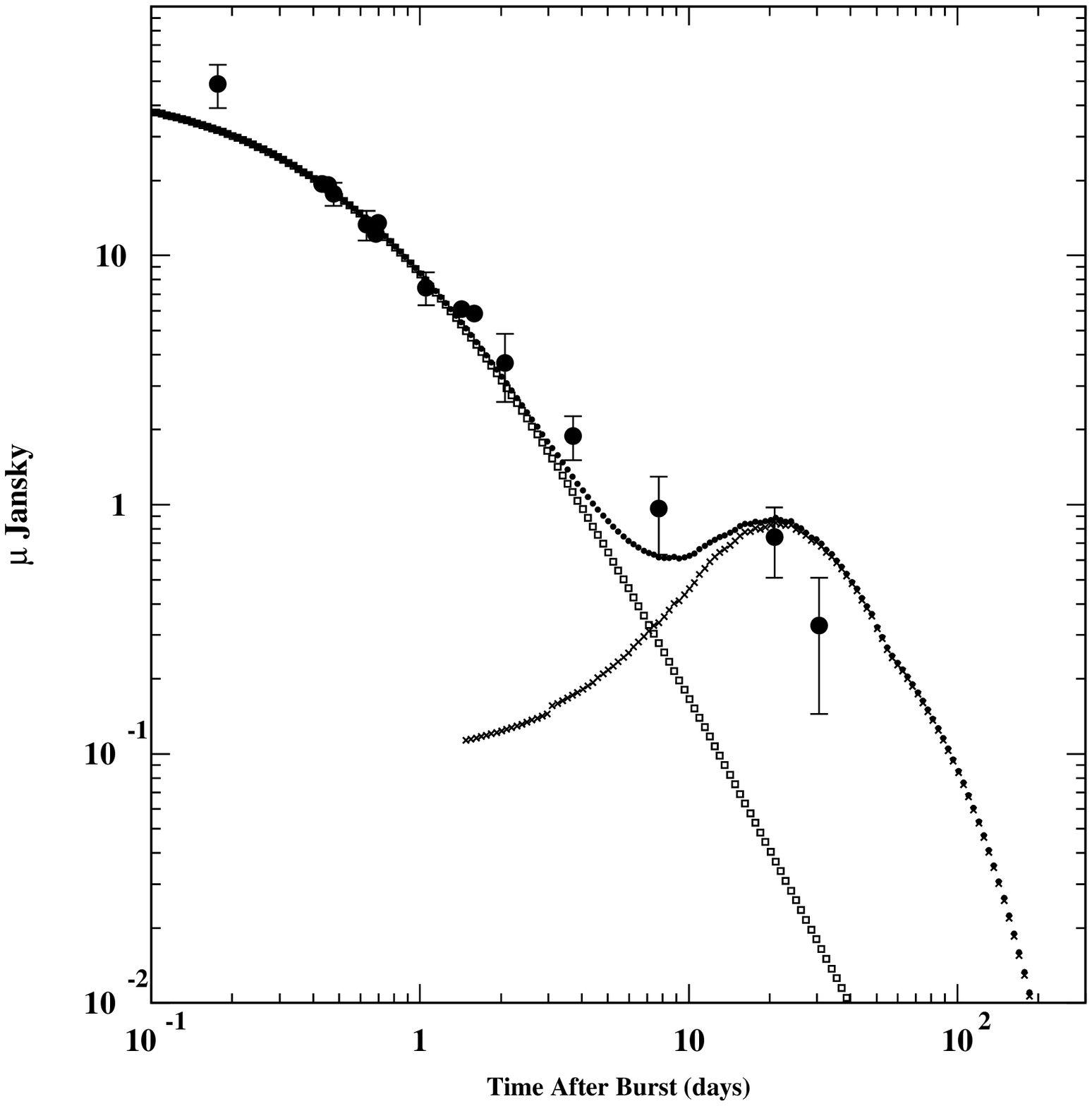, width=8cm} 
\end{tabular} 
\caption{Comparisons between our fitted R-band afterglow 
(upper curves) and the observations listed in Table II, 
not corrected for  extinction, 
for GRB 990712, at $\rm z=0.434$. 
Upper panel: Without subtraction of the host 
galaxy's contribution (the straight line). 
Lower panel: With the host galaxy subtracted, and the 
CB's AG (the line of squares) given by Eqs.~(\ref{fluxdensity}) and 
(\ref{cubic}). The contribution 
from a 1998bw-like supernova placed at the GRB's 
redshift, Eq.~(\ref{bw}), corrected for  extinction, 
is indicated in both panels by a line of crosses. 
The SN contribution is clearly discernible, but a bump at slightly 
earlier times than that of our standard-candle 
SN1998bw would provide a better description.} 
\label{fig712} 
\end{figure}

\begin{figure}[t] 
\begin{tabular}{cc} 
\hskip 2truecm 
\vspace*{2cm} 
\hspace*{-1.7cm} 
\epsfig{file=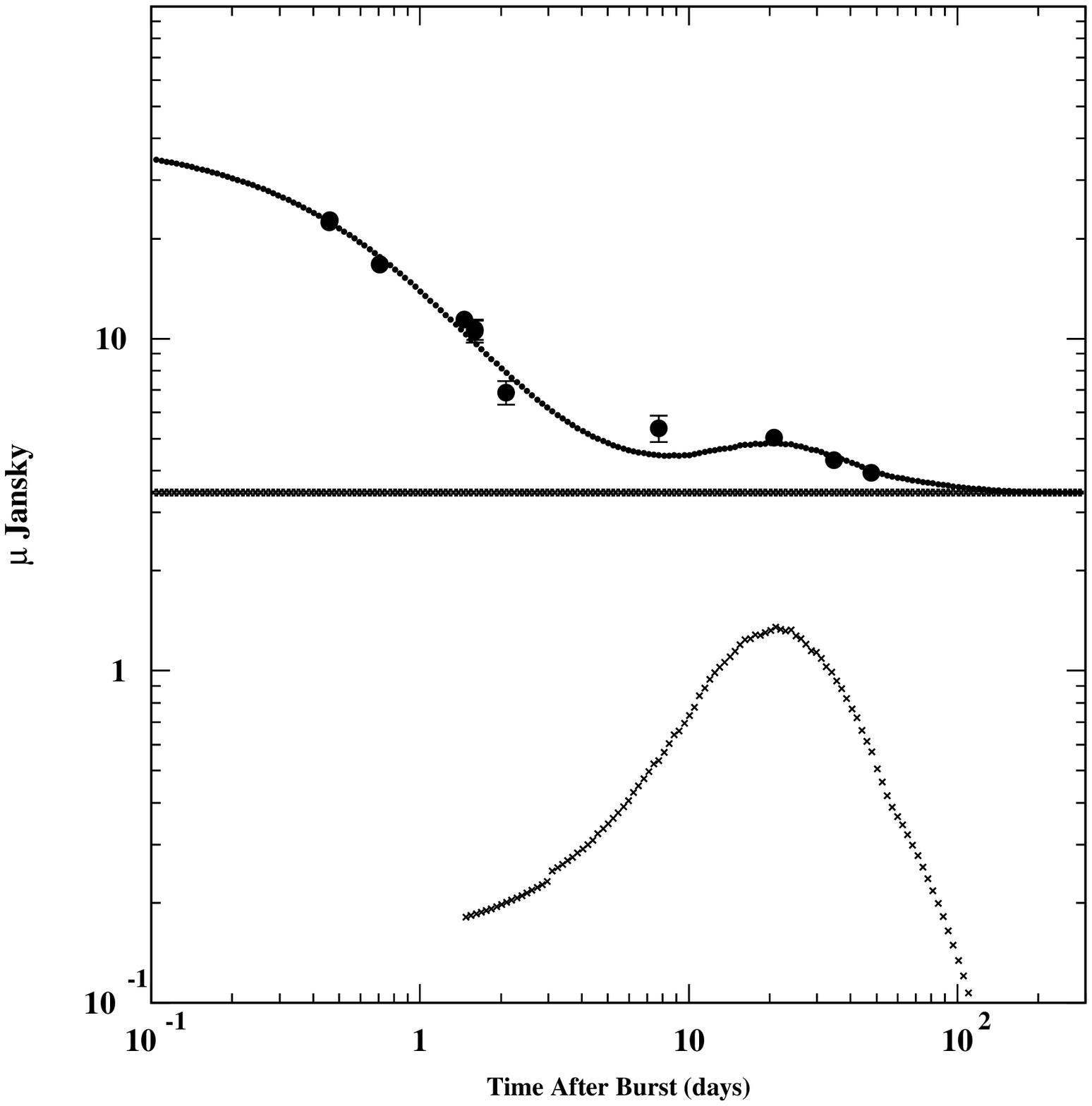, width=8cm} \\ 
\hspace*{.5cm} 
\epsfig{file=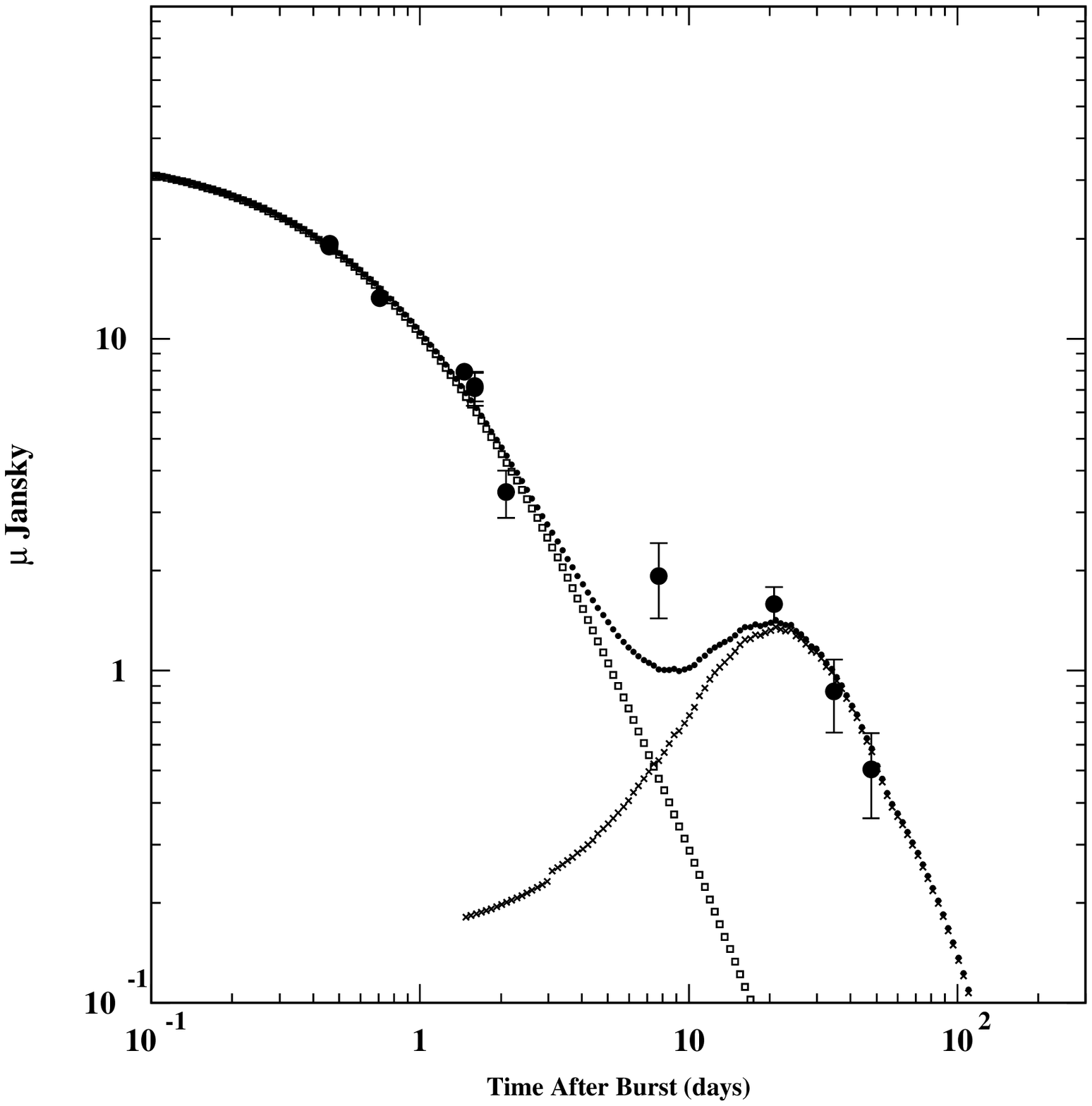, width=8cm} 
\end{tabular} 
\caption{Comparisons between our fitted V-band afterglow 
(upper curves) and the observations listed in Table II, 
not corrected for  extinction, 
for GRB 990712, at $\rm z=0.434$. 
Upper panel: Without subtraction of the host 
galaxy's contribution (the straight line). 
Lower panel: With the host galaxy subtracted, and the 
CB's AG (the line of squares) given by Eqs.~(\ref{fluxdensity}) and 
(\ref{cubic}). The contribution 
from a 1998bw-like supernova placed at the GRB's 
redshift, Eq.~(\ref{bw}), corrected for  extinction, 
is indicated in both panels by a line of crosses. 
The SN  is clearly discernible, but a bump at slightly 
earlier times than that of our standard-candle 
SN1998bw would provide a better description.} 
\label{fig712V} 
\end{figure}

\begin{figure}[t] 
\begin{tabular}{cc} 
\hskip 2truecm 
\vspace*{2cm} 
\hspace*{-1.7cm} 
\epsfig{file=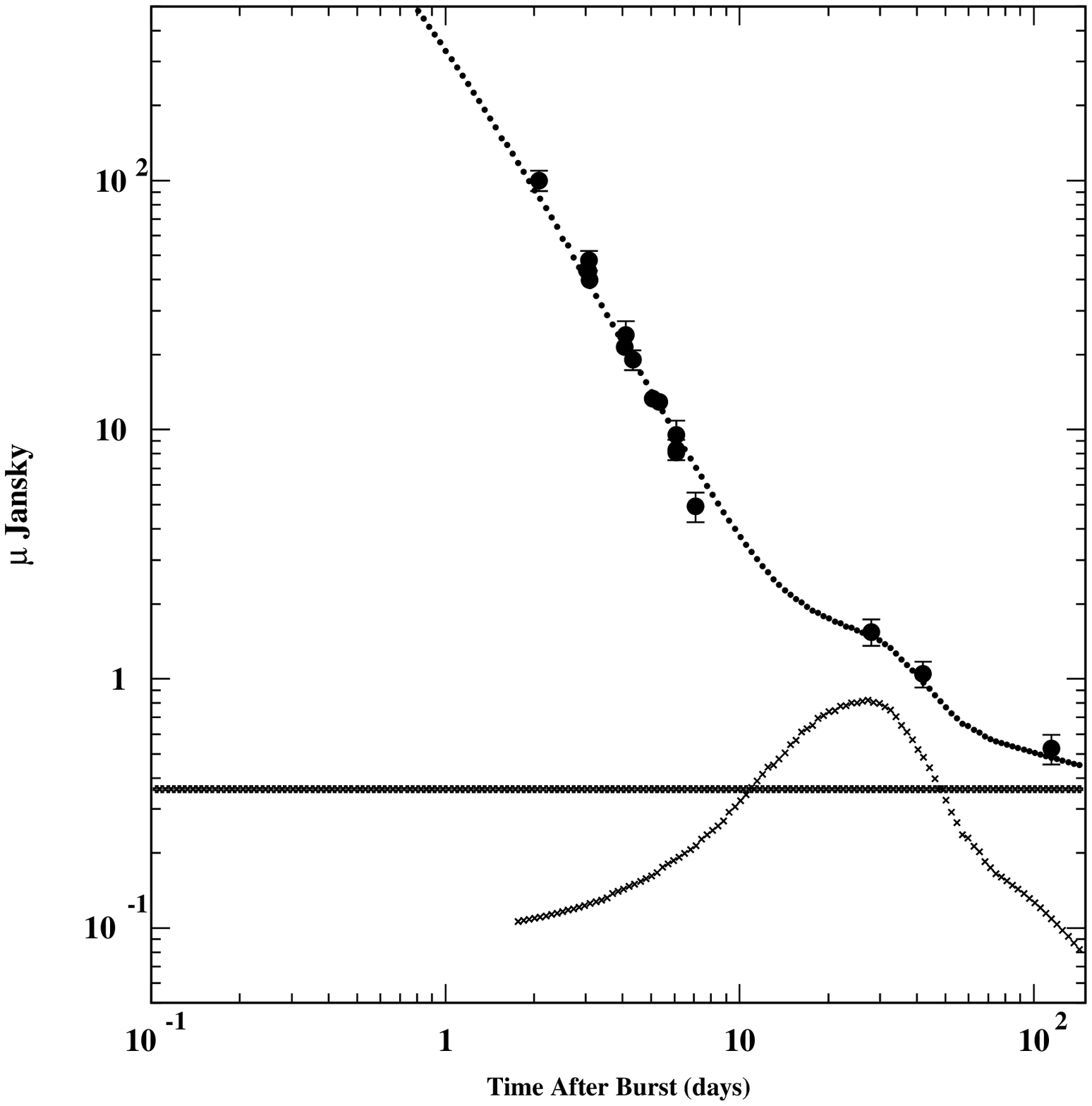, width=8cm} \\ 
\hspace*{.5cm} 
\epsfig{file=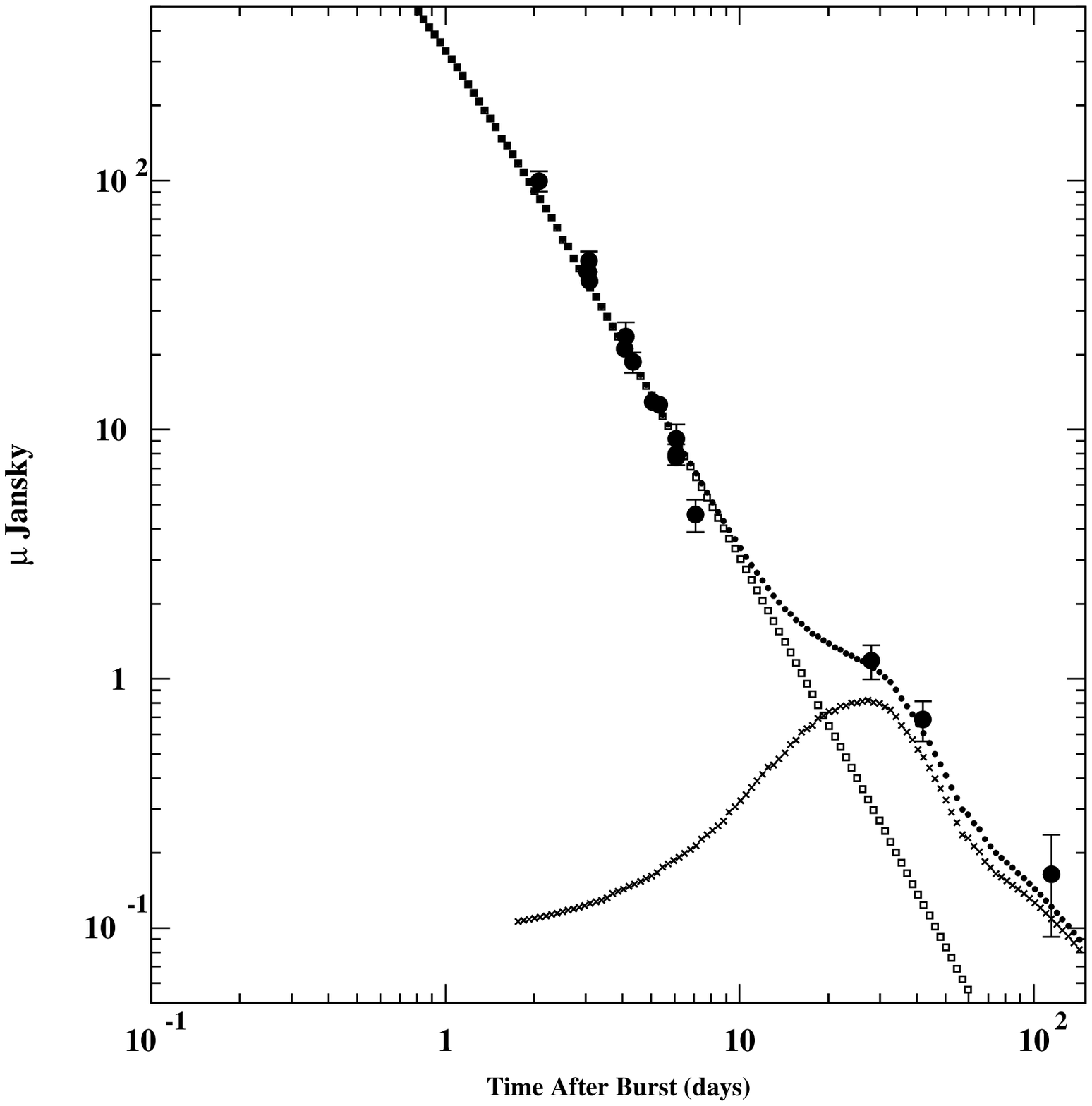, width=8cm} 
\end{tabular} 
\caption{Comparisons between our fitted R-band afterglow 
(upper curves) and the observations listed in Table II, 
not corrected for  extinction, 
for GRB 991208, at $\rm z=0.706$. 
Upper panel: Without subtraction of the host 
galaxy's contribution (the straight line). 
Lower panel: With the host galaxy subtracted, and the 
CB's AG (the line of squares) given by Eqs.~(\ref{fluxdensity}) and 
(\ref{cubic}). The contribution 
from a 1998bw-like supernova placed at the GRB's 
redshift, Eq.~(\ref{bw}), corrected for  extinction, 
is indicated in both panels by a line of crosses. 
The SN contribution is clearly discernible.} 
\label{fig208} 
\end{figure}

\begin{figure}[t] 
\begin{tabular}{cc} 
\hskip 2truecm 
\vspace*{2cm} 
\hspace*{-1.7cm} 
\epsfig{file=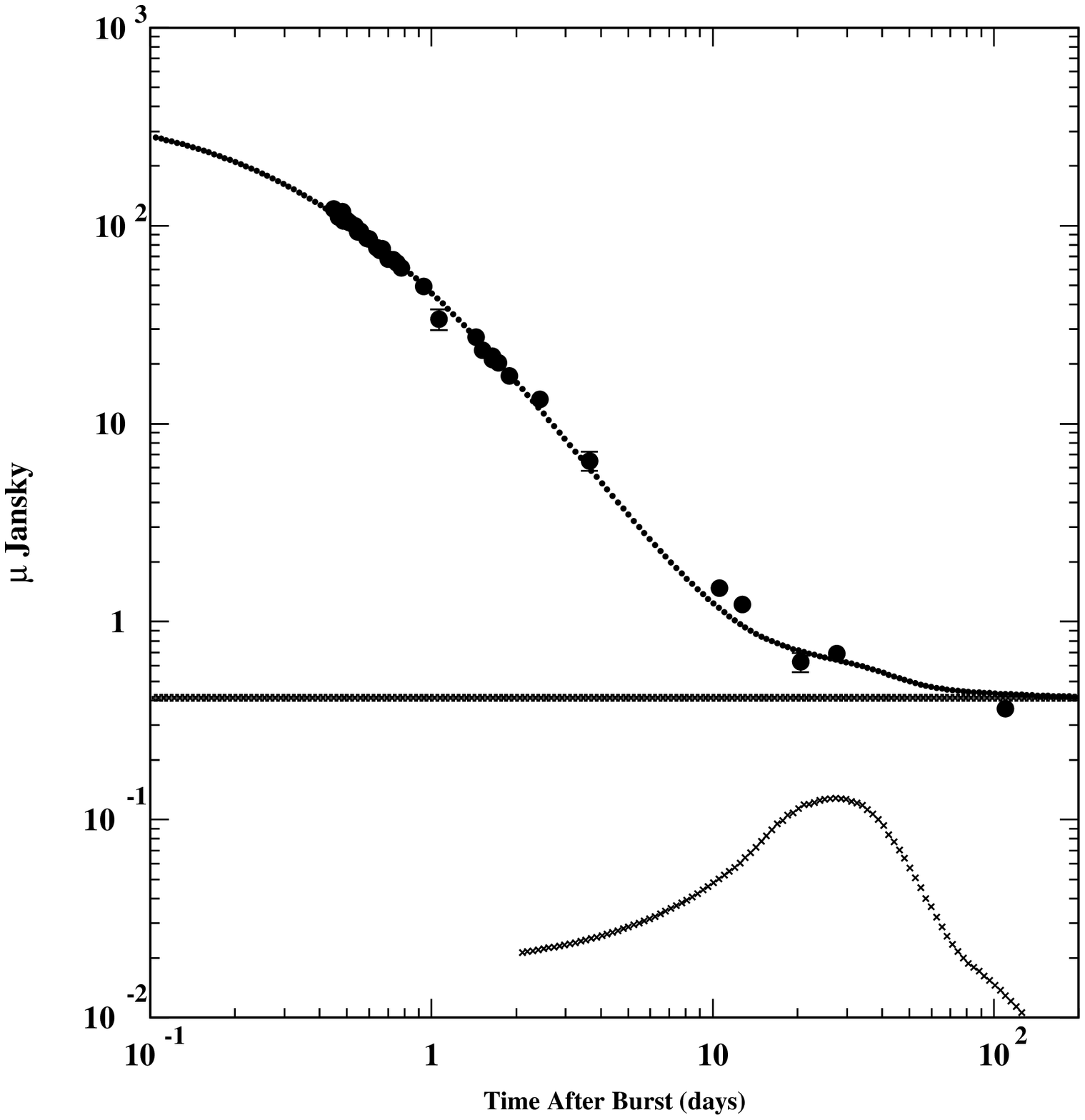, width=8cm} \\ 
\hspace*{.5cm} 
\epsfig{file=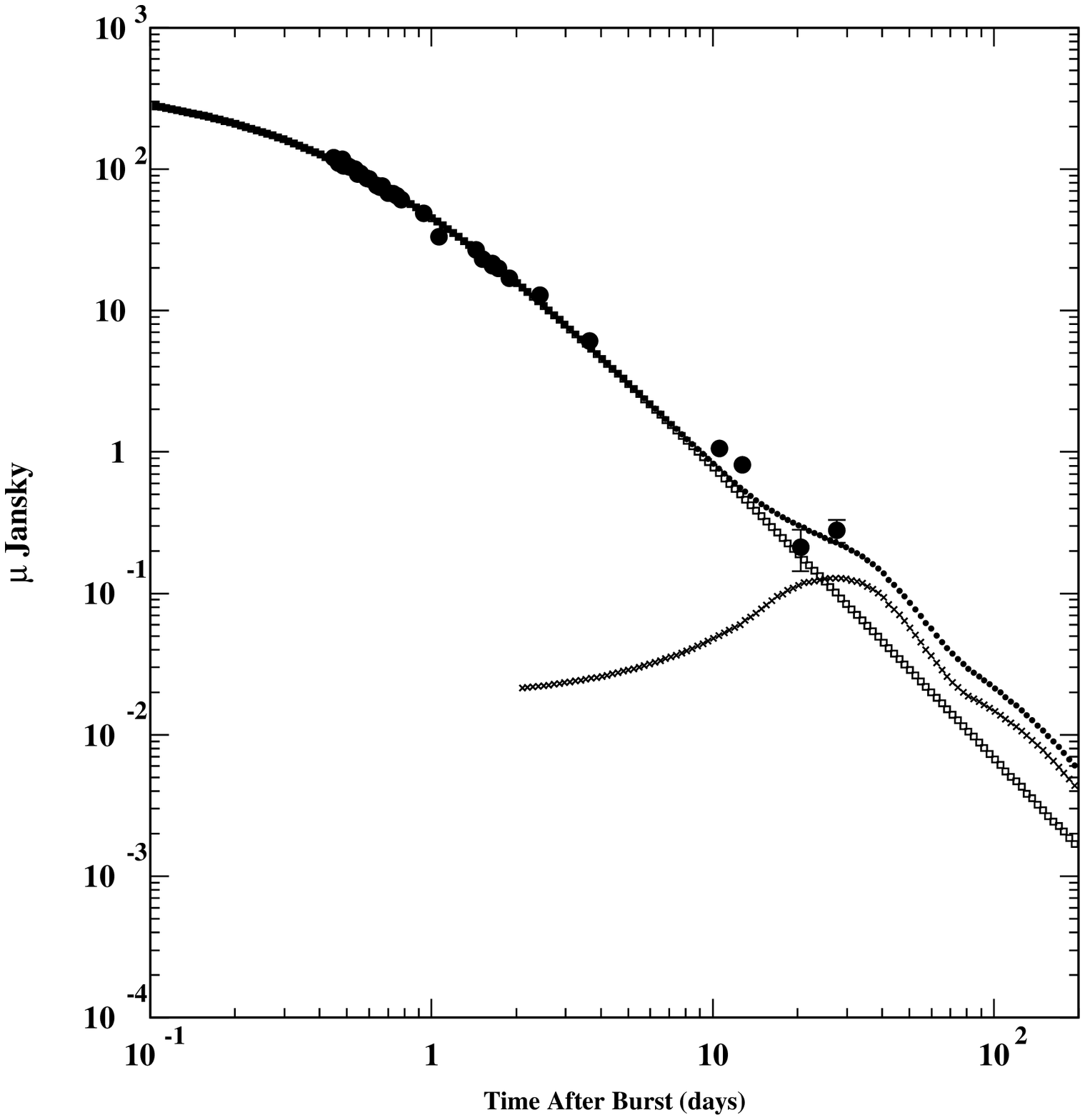, width=8cm} 
\end{tabular} 
\caption{Comparisons between our fitted R-band afterglow 
(upper curves) and the observations listed in Table II, 
not corrected for  extinction, 
for GRB 991216, at $\rm z=1.020$. 
Upper panel: Without subtraction of the host 
galaxy's contribution (the straight line). 
Lower panel: With the host galaxy subtracted, and the 
CB's AG (the line of squares) given by Eqs.~(\ref{fluxdensity}) and 
(\ref{cubic}). The contribution 
from a 1998bw-like supernova placed at the GRB's 
redshift, Eq.~(\ref{bw}), corrected for  extinction, 
is indicated in both panels by a line of crosses. 
The data show possible evidence for a SN bump.} 
\label{fig216} 
\end{figure}

With the exception of the AG of GRB 970508, which has the sharp ``break up'' 
that we have explained via a sudden change in density, 
a look at Figs.~(\ref{fig228}) to (\ref{jump508})  clearly reveals that the 
observed AGs have absolutely no ``breaks''. 
In the CB model, the gradual evolution of the proper 
afterglow (that of the CBs) is simply a consequence of 
the gradual decrease of the Lorentz factor $\rm \gamma(t)$. 
We give in Table II the list of the parameters resulting from 
our fits ($\alpha$, $\gamma_0$, $\theta$ and $\rm x_\infty$) 
to optical R-band afterglows (and in two cases, to 
the V-band afterglow as well). 
 
The fits to the CB model are satisfactory, particularly since the best-fit 
parameters turn out to be precisely in the expected ranges. 
On close inspection one notices that our curves occasionally 
undershoot or overshoot some points by a small factor, as in 
GRBs 990123 and 000301c\footnote{The feature at $\rm t\!\sim\! 4$ days in 
GRB 000301c has been interpreted by Garnavich et al. (2000b) 
as due to gravitational lensing.}. 
This is not a surprise: the AG fluences 
are proportional to the ISM number density, which we do not 
expect to be exactly constant for kpc distances, even 
in the halo of galaxies. If such ``defects'' were not present 
in our fits, we would have concluded 
that the data had been over-parametrized. For the same reasons, 
and because of the systematic errors in the data, 
the values of the parameters we extract from our fits should not 
be taken entirely at face value, even though the minimization procedure 
---which attributes to the errors a counterfactual purely statistical origin--- 
results in tiny 1 $\sigma$ spreads for the fitted parameters, and in 
$\chi^2 $ values that are in most cases extremely satisfactory. 
 
All the  figures (\ref{fig228}) to (\ref{jump508}) refer to optical data 
for $\rm t\!\geq\! 0.1$ days, for which it is reasonable 
to approximate the ISM density by a constant value, 
describing the density of the superbubble and/or the galactic halo. 
We have already discussed the early observations of 
GRB 990123, for which this approximation breaks down. 
 
\subsection{The distribution of fitted parameters} 
 
In the CB model, the parameter $\alpha$ of 
Eqs.~(\ref{spectrum}, \ref{fluxdensity}) is 
the only one for which we have no reason to expect 
a range of different values. It is therefore extremely 
satisfactory  that the fitted values of $\alpha$ are, within errors, 
compatible with {\it all} of the GRBs having a universal 
behaviour with the theoretically predicted value: $\alpha\approx 1.1$, 
Eq.~(\ref{spectrum}). The narrow distribution of best fitted $\alpha$ 
values is shown in Fig.~(\ref{alphadist}). 
In the CB model, we have extracted the 
values of $\alpha$ from 
the temporal shape of the AG and ---adding consistency 
to the picture--- they agree  well with the values 
obtained from spectral observations, either in X-rays (with spectra 
modified by a best-fit hydrogen column density) or in the 
optical domain (with a galactic colour-dependent extinction). 
Some examples are:\\ 
GRB 970228: $\rm\alpha_X=1.06\pm 0.12$  (Costa et al.~1997).\\ 
GRB 970508:  $\rm\alpha_{O}=1.12\pm 0.07$ (Galama et al.~1998c);\\ 
 $\rm\alpha_X=1.11\pm 0.06$ (Galama et al.~1998c).\\ 
GRB 990123: $\rm\alpha_{O}=1.29\pm 0.23$ (Holland et al.  1999a).\\ 
GRB 990510:  $\rm\alpha_X=0.96 \pm 0.26 $ (Pian et al.~2001);\\  
 $\rm\alpha_{O}=1.26\pm 0.15$ (Stanek et al.~1999).\\ 
GRB 991208: $\rm\alpha_{O}=1.05\pm 0.09$ (Castro-Tirado et al. 2000).\\ 
GRB 991208: $\rm\alpha_{O}\approx 1.1$ (Takeshima et al. 1999).\\ 
GRB 000301c: $\rm\alpha_{O}=1.15\pm 0.26$  (Jensen et al.~2000).\\ 
GRB 000926:  $\rm\alpha_X=1.2\pm 0.3$    (Piro et al.~2001);\\  
 $\rm\alpha_{O}=1.02\pm 0.02$ (Sagar et al.~2001a).\\ 
GRB 010222:  $\rm\alpha_X=0.97 \pm 0.15$ (In `t Zand et al.~2001);\\ 
 $\rm\alpha_{O}=1.07\pm 0.09$  (Stanek  et al.~2001). 
 
There are also cases for which the reported value of $\alpha$ 
differs significantly from 1.1. One notable instance is GRB 990510, 
for which Beuerman et al. (1999) report $\alpha\sim 0.55$ and 
Stanek et al. (1999) find $\alpha=0.61\pm 0.12$. To extract this value 
the authors extrapolate the measured extinction: $\rm E(B-V)=0.20$ 
(Schlegel et al. 1998). 
If this measured extinction is used to correct {\it only} the measured  
$\rm B-V$ 
values for GRB 990510: $0.57\pm 0.02$ (Beuerman et al. 1999), 
and $0.56\pm 0.03$ (Stanek et al. 1999), 
one obtains $\rm \alpha_O=1.08\pm 0.12$. 
The uncertainties entailed by absorption corrections are the 
reason why we have chosen to de-emphasize results that are sensitive 
to them, whether they do, or do not, agree with our expectations. 
 
The distribution in initial Lorentz factors, $\gamma_0$, shown in 
Fig.~(\ref{gammadist}), agrees snugly with our expectation, 
$\gamma_0\sim 10^3$, extracted from independent 
information: the fluence and the individual-photon energies of GRBs 
(DD2000a,b) and the energies of X-ray lines in their afterglow 
(DD2001a). Notice how surprisingly narrow this distribution is. 
 
The distribution of viewing angles $\theta$ is shown 
in Fig.~(\ref{thetadist}). 
The AG data for GRB 000131 consist in only three points, 
while for GRBs 991208  and 000301c the measurements start rather late. 
The sensitivity to $\theta$ in our fit to these GRBs is not good. We reflect 
this fact in Fig.~(\ref{thetadist}) by having the corresponding 
entries unshaded. The distribution is compatible with the 
expectation that the limited experimental sensitivity to GRBs introduces 
a sharp cutoff as $\theta$ increases; see the steep fluence function, Eq.~(\ref{fluence}). 
 
The parameter $\rm x_\infty$ of Eq.~(\ref{gamoft}) is the only 
one for which we expect a rather broad distribution. 
Indeed, it depends on the densities close to the GRB 
progenitor, which ought to be 
quite variable, and in the region where 
the CB light is emitted; see Eq.~(\ref{gamoft}). 
The values of $\rm x_\infty$ reported in Table II show 
a spread of a bit over one order of magnitude, 
supporting the expectation. 
In Fig.~(\ref{xdist}) we show the distribution of 
$\rm Log_{10}[x_\infty(Mpc)]$, which peaks at 
the reference value of Eq.~(\ref{gamoft}) and extends 
to smaller values, as expected if the average density close 
to the progenitor is sometimes much smaller than our rather large 
reference value: $\rm n_p^{SN}=1\;cm^{-3}$; and/or the density 
of the ISM is bigger than our rather low reference value: 
$\rm n_p=10^{-3}\;cm^{-3}$. 
 
The values of $\alpha$, $\theta$, $\gamma_0$ and $\rm x_\infty$ 
are not sufficient to predict the overall normalization 
of an AG: $\rm F$ in Eq.~(\ref{fluxdensity2}), whose approximate 
value is given by the absolutely normalized 
 Eq.~(\ref{fluxdensity}). Indeed, 
$\rm F$ is proportional to the product of the number of CBs 
 and their baryon number. To skirt absorption 
corrections we have discussed in section 12 the values of $\rm F$ 
in connection with late X-ray AGs. There, we compared 
the data and the naive expectation for  a single (highest-$\gamma$)  
dominant CB ($\rm n_{_{CB}}=1$) and 
our canonical $\rm N_{CB}$, to work out the ratio $\rm q$ 
of fitted to predicted values of $\rm F$ in the X-ray band.  
The same exercise can be carried along for the R-band AGs, 
with the result that the optical values of $\rm q$ are not 
within a factor of three, but within an order of magnitude 
of $\rm q=1$. It is tempting to conclude that this may 
be due to poorly-understood absorption. It is in any case clear that 
the AG magnitude, in the CB model, is not a problem\footnote{In the 
fireball model and its descendants, the 
efficiency of conversion of kinetic energy to photons is claimed to be 
high, and both the GRBs and their AGs are due to the same 
mechanism: synchrotron radiation. It is therefore difficult to explain why, 
at the end of the GRB, the radiation rate suddenly drops by two orders of 
magnitude or more, and why there is more energy in the GRB than in the 
afterglow (see, e.g. Burenin et al.~1999; Pian et al.~2001, In `t Zand et 
al.~2001).}. 
 
The GRB consists of the photons emitted by the hot CBs as they 
exit the SN shell. In the rest system of a CB, the individual energies, as in 
Eq.~(\ref{energies}), are a fraction $\rm (1+z)/\delta_0$ of 
the observed energies, with $\delta_0=\delta[\gamma_0,\theta]$. 
The total energy emitted by a GRB, in the rest system of its CBs, is 
in the form of isotropically distributed photons that appear to us as the collimated $\gamma$-ray burst. This comoving total energy 
 is related to the observed fluence 
$\rm F_\gamma$ by: 
\begin{equation} 
\rm E^{CB}_\gamma = {4\,\pi\,D^2_L\,F_\gamma\over(1+z)\,\delta_0^3}\, . 
\label{ECBrest} 
\end{equation} 
For the GRBs all of whose parameters are well determined, 
we list in Table III the values of $\rm E^{CB}_\gamma$. 
Interestingly, their distribution ---shown in 
Fig~(\ref{Edist})---  is also quite narrow. 
 
We have not discussed in this subsection the parameters of 
GRB 980425, listed in Table II. 
They are obtained from a fit to the X-ray ---as opposed to optical--- AG, 
and they are imprecise. The deduced value of $\rm E_\gamma^{CB}$ 
is $0.16 \times 10^{44}$ erg, a bit lower than those listed 
in Table II. This is to be expected, the small $\delta$ of this GRB 
makes its GRB softer, and less prominent within the BATSE energy window. 
 
To summarize, the distributions of parameters are in extremely good 
agreement with the expectations of the CB model and, if anything, 
they are astonishingly close to what they would be 
for ``standard candle'' GRBs. 
 
\subsection{The GRB/SN association in view of our results} 
 
It is useful to discuss the evidence for a SN component in 
the GRB optical AGs in order of decreasing redshift. 
The fact that we have a consistent and successful description 
of optical afterglows strengthens the interpretation of this 
putative evidence.

Examining Figs.~(\ref{fig228}) to (\ref{jump508}), we draw the 
following conclusions. 
In the six more distant GRBs, ranging from GRB 000131 
at $\rm z=4.5$ to GRB 010222 at $\rm z\!>\!1.474$, there is 
no evidence for {\it or against} a SN 1998bw-like component. 
In GRB 000418, at $\rm z=1.119$, there is an indication of an excess, compatible 
with the SN. In GRB 980613, at $\rm z=1.096$, the evidence, though based 
mainly on just one point, is very strong. In GRB 991216, at $\rm z=1.02$, 
there is a clear indication of a late excess over the CB's afterglow, though a 
SN1998bw-like contribution does not describe it very well 
(a slightly earlier bump would do a very good job, indicating 
that the standard-candle hypothesis for the SN is good, but not perfect). 
In GRB 980703, at $\rm z=0.966$, the SN excess is visible and well fitted, 
but the errors are large. For GRB 970828, at $\rm z=0.957$, there 
are no AG observations at optical wavelengths. In the peculiar
case of GRB 970508, at $\rm z=0.835$, 
there is in the data a clear excess at late times that is very 
well fitted by our SN ansatz. For the next three closer GRBs 
(991208, 970228, 990712), at $\rm z=0.706$ to 0.434, the evidence 
is completely convincing that a SN1998bw-like contribution 
is required to fit the data. In the case of  GRB 990712, once again, 
a SN peak occurring slightly earlier than that of a redshift-corrected 
SN1998bw would provide a better description of the AG. 
Finally, GRB 980425, at $\rm z=0.0085$ is indeed associated to a SN: 
our fairly satisfactory standard-candle choice. 
 
A clear trend is apparent in the last paragraph. The closer a GRB 
is, the better the evidence for its association with a SN. The trend 
is entirely consistent with the fact that, for the more distant GRBs, 
a SN contribution to the AG could not be seen, even if it was there. 
In all cases where the SN could be seen, it was seen, with the evidence 
gaining in significance as the distance diminishes. 
The temptation to conclude that all long-duration GRBs are 
associated with SNe appears to us to be irresistible, even if an 
irrefutable proof will never be possible. 
 
\subsection{Do cannonballs deserve their name?} 
 
We have argued that CBs should reach an asymptotic radius in 
a very short time and travel thereafter as literal, i.e. non-expanding, 
``cannonballs''. Can this statement be contrasted with the data? 
To answer this question we have analyzed the AGs produced by 
CBs whose radius inertially increases at a fixed speed, 
$\rm \beta_{exp}\, c$, in their rest system. The details are 
given in Appendix II. The result is that the late AGs behave in this 
case as $\rm F_\nu\propto t^{-\tau}$, with $\tau=9\,(1+\alpha)/5$. 
For $\alpha\simeq 1.1$, as we have argued, $\tau\simeq 3.8$, 
which completely disagrees with the AG light curves. For $\tau=2.1$, 
in agreement with the latter, $\alpha\simeq 0.17$, which completely 
disagrees with the measured spectra. Thus, the cannonballs 
of the CB model do deserve their name.

\section{Summary and conclusions} 
 
We have previously argued that the cannonball 
model offers a successful and simple explanation 
of the fluence, energy spectrum, and temporal behaviour of 
the prompt $\gamma$-rays of a GRB (DD2000b).
>From these considerations we extracted a CB's 
typical Lorentz factor, $\gamma_0\sim 10^3$, and typical 
baryon number $\rm N_{CB}$: in the 
vicinity of $6\times 10^{50}$, or ${\cal{O}}(10)$ times 
as much for a jet of CBs in a multipulse GRB 
(DD2000b, DD2001b). Using only these CB-related input parameters 
and a reasonable initial CB expansion velocity, we have explicitly 
worked out all of the properties of X-ray and optical afterglows. 
 
As an intermediate result, we 
derived the temporal behaviour of the radius of 
a CB and showed that, faithful to their name, CB radii tend to a constant 
$\rm R_{max}$ in mere minutes of observer's time. The 
value of $\rm R_{max}$ depends 
on the ISM density close to the GRB progenitor. 
GRBs whose CBs have large radii have afterglows whose temporal 
decline is very fast. This is also the case for GRBs whose CBs travel through a 
relatively dense ISM. 
These may be the reasons, along with strong extinction in the host galaxy, 
why the search for AGs is not always successful. 
 
We have shown how well the CB model describes 
all the properties of the X-ray and optical afterglows of GRBs. 
Our results do not ---and could not--- take into account possible variations 
of the ISM density along a CB's path; they are in this sense 
``descriptions'' rather than fits. In spite of this, the descriptions 
are excellent and the consistency of the results is impressive. 
 
The observed behaviour of both X-ray and optical AGs is the 
predicted one. All the parameters extracted from the fits have 
values or distributions close to the expected  ones. 
This is the case for the integrated fluences in  
the early and late X-ray and optical 
bands, for the values of the Lorentz factor 
$\gamma$, for the distribution of observation angles $\theta$, 
for the spectral index $\alpha$ (that we determine, 
not from the spectra themselves, but 
from the time-dependence of the late afterglow), 
and for the parameter $\rm x_\infty$ of Eq.~(\ref{gamoft}), 
that governs the pace 
of the CB's slowdown, and is a combination 
of the ISM densities, and the CB's baryon number and radius. 
Throughout, we have chosen to de-emphasize the results that are most  
sensitive to systematic errors, such as absorption corrections. Thus, 
we have not systematically extracted parameters from  
flux normalizations, nor reported any $\chi^2$ tests of the quality of the fits 
(which would be misleadingly good) or statistical-error estimates on the fitted 
parameters (which would be misleadingly small).

The distribution of the derived 
quantity $\rm E_\gamma^{CB}$ (the energy 
of the GRB photons in the CBs' rest system) 
has a range of less than a factor of four, 
making long-duration GRBs not far from standard candles. 
We have demonstrated that GRB afterglows can be 
understood in detail. Perhaps this will pave the way to the 
use of GRBs as cosmological beacons. 
 
GRB 980425 is at first sight a special case: its $\gamma$-rays 
are rather soft, it is extremely near by cosmological standards, 
and very clearly associated with a SN. In the CB model it is not 
exceptional, only seen at a relatively large angle. 
Its X-ray afterglow is normal and emitted 
by its CBs, not by the isotropic ejecta of SN1998bw. 
Its optical AG is dominated by the SN for almost two years, 
but its last measured point is due to the CBs, and agrees with 
the expectation. 
 
Very early optical AG data are only available for GRB 990123. 
The CB model is also capable of describing them naturally: 
their magnitude and time-dependence are those expected 
for CBs moving through the density profile produced by winds in 
the Wolf-Rayet phase of the progenitor. Future early observations 
should test this feature of the CB model: a linear dependence 
of the AG fluence on the varying local-density profile. 
 
For GRB 970508 we can provide a good fit 
to its peculiar AG only if we assume that its CBs 
encountered a sudden change in ISM density, 
as they would if they are exiting a superbubble into
a higher-density region.
 
The very early X-ray and optical AGs are not achromatic: 
they are dominated by two mechanisms with different 
time-dependence: thermal bremsstrahlung 
and synchrotron radiation. 
But in a matter of a few hours all of the AG is dominated by 
synchrotron radiation, and the light curves
should be achromatic, as observed. 
 
Our descriptions of the optical AGs indicate or require 
the contribution to the light curve of a supernova akin 
to SN1998bw, in all eight out of sixteen cases for which 
the event occurred close enough for such a contribution 
to be observable. The conclusion that there is a roughly one-to-one 
association between core-collapse SNe (or perhaps just type Ib and Ic SNe) 
and long-duration GRBs is very tempting. The enormous beaming 
factor of the CB model makes this conclusion tenable and consistent. 
 
We are currently completing the study of the predictions 
of the CB model for radio afterglows, for which self-absorption in the CBs
is relevant, and requires a careful analysis. The results, soon to 
be announced (Dado et al. 2002),
are excellent. We also plan to discuss elsewhere  
the interesting implications of the CB model 
for cosmic-ray physics,  and for other 
accreting compact objects that eject relativistic jets, such as 
radiogalaxies, AGNs, quasars, microquasars, blazars and microblazars. 
 
It would be interesting to compare our results 
 to those of conically ejected shocked fireballs. 
We have argued that, for this, it would be 
more convincing not to have the observed firetrumpets 
pointing precisely to planet Earth. For the time being, 
we have proved that the CB model ---based on simple 
and definite hypothesis--- makes predictions that 
are univocal (as opposed to multiple-choice), explicit, 
analytical in fair approximations, quite simple, 
very complete,  and very successful. 
 
\vskip .5cm 
\noindent 
{\bf Acknowledgements:} 
We are very indebted to Rainer Plaga for numerous fruitful discussions. 
We thank Elena Pian and Jean In `t Zand 
for kindly providing us with the detailed data on the 
BeppoSAX X-ray observations of GRBs 990510 and 010222; 
Donald Smith and Atsunama Yoshida for the data on GRB 970828; 
and Luigi Piro for the data on GRB 970508. 
We are indebted to an anonymous referee for many 
sensible suggestions and questions. 
This research was supported in part  
by the Helen Asher Fund for Space Reseach and by the  
V.P.R. Fund - Steiner Research Fund at the Technion. 
 
\vskip .5cm 
\noindent 
\section*{Appendix I: Subdominant AG mechanisms} 
 
Three mechanisms declining less fast than thermal bremsstrahlung 
---but typically subdominant relative to synchrotron radiation--- 
contribute to GRB afterglows: 
relativistic bremsstrahlung  from Coulomb collisions 
of CB electrons with the ISM constituents, 
atomic transitions of CB atoms excited or ionized by these 
same collisions, and inverse Compton scattering of 
the CBs' electrons on the cosmic background radiation. 
We discuss them here to substantiate our assertion that 
thermal bremsstrahlung and synchrotron radiation 
 dominate the AGs. 
 
\subsection*{Relativistic bremsstrahlung} 
 
The electrons of the CB, in their highly relativistic collisions 
with the ISM nuclei, emit non-thermal bremsstrahlung 
radiation. 
The total power per unit area received by the observer 
from this source  is: 
\begin{equation} 
\rm {dF_{RB}\over dt\, d\Omega} \approx { 45.7\, \alpha\, r_e^2\, N_{CB}\, 
n_p\, 
m_e\, c^3\, \gamma^2\,\delta^4  \over 4\, \pi\, D_L^2}\, , 
\label{RBflux} 
\end{equation} 
where $\rm r_e=2.18\, fm$ is the classical radius of the electron 
and $\alpha\approx 1/137$ is the fine structure constant. 
For $\rm N_{CB}= 6\times 10^{50}$, $\gamma=\delta=10^3$, 
$\rm z=1$ and $\rm n_p=1\, cm^{-3}$, the above expression yields 
$3.8\times 10^{-11}$ erg/(cm$^2$ s). 
This radiation has the same dependence on $\gamma$ and $\delta$ 
as the synchrotron radiation of Eq.~(\ref{flux}), but it is some two orders 
of magnitude smaller. 
This spectrum extends to $\rm\!\sim\! m_e\,\delta/(1+z)\sim 250$ 
MeV; it may be a contribution to the very high energy 
$\gamma$-rays observed by EGRET and COMPTEL in some GRBs. 
 
\subsection*{Atomic transitions} 
 
The recombination and de-excitation of the CB atoms 
excited and dissociated by Coulomb collisions with 
the ISM particles contributes to the AG with a power: 
\begin{equation} \rm 
{dF_C\over dt\, d\Omega} \approx {8\, \pi \, r_e^2\, n_p\, m_e\, c^3\, 
N_{CB}\, ln[\Lambda] \, \gamma\,\delta^4 \over 4\, \pi\, D_L^2}\, , 
\label{Cflux} 
\end{equation} 
where $\rm ln[\Lambda]$ is the Coulomb logarithm, 
\begin{equation} \rm 
ln[\Lambda]= ln \left[ {\sqrt{2\, m_e\, c^2\, \gamma^2\, T_{max}} \over I} 
\right ]\, , 
\label{Lambda} 
\end{equation} 
$\rm I=13.6$ eV is the 
ionization energy of hydrogen, and $\rm T_{max}$ is the maximum kinetic 
energy that can be imparted to a stationary electron in a single collision 
with a relativistic particle of mass $\rm M$: 
\begin{equation} 
\rm T_{max}= {2\, m_e\, c^2\, \gamma^2 \over 1+2\, \gamma^2\, m_e/M + 
             (m_e/M)^2}\, . 
\label{Tmax} 
\end{equation} 
For $\rm N_{CB}= 6\times 10^{50}$, $\gamma=\delta=10^3$, 
$\rm n_p=1$ cm$^{-3}$ and 
$\rm z=1$,  Eq.~(\ref{Cflux})  yields $7.15\times 10^{-11}$ erg/(cm$^2$ s), 
 which 
is negligible with respect to the synchrotron radiation of Eq.~(\ref{flux}). 
This power is radiated mostly as synchrotron emission from the 
knocked-on electrons and by line emission  from hydrogen recombination 
(boosted to X-ray energies).

\subsection*{Inverse Compton scattering} 
 
The scattering of the CBs' electrons 
on ambient starlight and the cosmic 
background radiation (CBR) produces a radiated power: 
\begin{equation} 
\rm {dF_{CBR} \over dt\, d\Omega} \approx {32\, \pi \, r_e^2\, 
\rho_\gamma\, 
c\, N_{CB}\, \gamma^2\,\delta^4  \over 36\, \pi\, D_L^2}\, , 
\label{ICSflux} 
\end{equation} 
where $\rho_\gamma$ is the energy density of the radiation field 
($\rm \rho_\gamma= 0.24\,(1+z)^4\, eV\, cm^{-3}$ for 
the CBR). For $\rm N_{CB}= 6\times 10^{50}$, $\gamma=\delta=10^3$ 
and $\rm z=1$, the above expression 
yields $1.5\times 10^{-14}$ erg/(cm$^2$ s) for the CBR contribution, 
which is negligible relative to synchrotron radiation 
up to a very late afterglow phase, 
when $\rm n_e$ may be very small. The contribution of 
Compton scattering to a polarization of the signals, 
on the other hand, may not be negligible (Shaviv and Dar, 1995). 
 
Inverse Compton scattering of the CBs' high energy electrons, with 
a proper spectrum $\rm dn_e/dE \sim E_e^{-3.2}$, on their self-produced 
synchrotron and thermal bremsstrahlung radiation produces very 
high energy photons with a spectrum $\rm dn_\gamma/dE\sim E^{-2.1}$. 
For distant GRBs, this 
spectrum is cutoff at sub TeV energies by pair 
production on the infrared background radiation. But, for very nearby 
GRBs, it may be observable up to extremely high energies, larger than those 
observed from blazars. 
 
\section*{Appendix II: Ever expanding ``cannonballs''} 
 
It is very instructive to study the possibility that CBs, 
instead of reaching an asymptotic radius, would 
continue to expand significantly. 
To find $\rm\gamma(t)$ in this case, Eq.~(\ref{Roft}) 
is to be substituted by: 
\begin{equation} 
\rm dR=\beta_{exp}\;{dx\over\gamma}, 
\label{Rofx} 
\end{equation} 
 while Eqs.~(\ref{dgamma},  \ref{dxsn}) remain unchanged. 
Insert Eq.~(\ref{Rofx}) into Eq.~(\ref{dgamma}) and integrate, to 
obtain, for constant $\rm n_p$: 
\begin{eqnarray} 
\rm R^3(\gamma)&=&\rm R^3_{trans}+\widehat R^3_{max}, 
\left[{1\over\gamma^2}-{1\over\gamma_0^2}\right]\nonumber\\ 
\rm \widehat R^3_{max}&\equiv&\rm {3\,N_{CB}\,\beta_{exp}\over  
2\,\pi\,n_p}\, . \label{Rhat} 
\end{eqnarray} 
The CBs reach an asymptotic radius as $\gamma\to 1$, which bears 
some resemblance to that of Eq.~(\ref{Rinfinity}), but it is reached much later 
and it is much larger (e.g. $\rm \widehat R\sim 3.8\times 10^{17}$ cm for 
$\rm n_p=10^{-3}$ cm$^{-3}$, $\rm\beta_{trans}=1/(3\sqrt{3})$, and 
 $\rm N_{CB}=6\times 10^{50}$). 
 
Upon insertion of Eq.~(\ref{Rhat}) into Eq.~(\ref{dgamma}), we obtain 
the analogue of Eq.~(\ref{gamoft}): 
\begin{eqnarray} 
-\rm\int_\gamma^{\gamma_0} &\rm d\gamma&\rm 
{1+\theta^2\gamma^2\over \gamma^{8/3}} 
\left[1-{\gamma^2\over\gamma_0^2}\right]^{-2/3} 
={2\,c\, t\over 3\, (1+z)\, \widehat x_\infty} 
\nonumber\\ 
\rm\widehat x_\infty&\equiv& 
\rm{N_{CB}\over\pi\,\widehat R_{max}^2\, n_p} 
\label{gamoftbis} 
\end{eqnarray} 
where we have neglected $\rm R^3_{trans}/\widehat R^3_{max}$. 
The integral can be done analytically, but is not compact. 
 
For late times, when $\gamma\theta\ll 1$, Eq.~(\ref{gamoftbis}) implies that 
$\rm\gamma\propto t^{-3/5}$.   According to Eq.~(\ref{fluxdensity}), then, 
the AG light curve approaches: 
\begin{eqnarray} 
\rm F_\nu (t) &\propto& \rm t^{-\hat\tau};\nonumber\\ 
\widehat\tau&=&\rm{9\over 5}\;(1+\alpha)\simeq 3.8. 
\label{lateAGbis} 
\end{eqnarray} 
This behaviour cannot be reconciled with the data, 
as explained in the text. 

\section*{Appendix III: gravitational lensing of moving CBs} 

The phenomenon of gravitational lensing is well known. 
A lensing object of mass $\rm M$ has a Schwarzschild radius 
$\rm R_S=2\,G_N\,M/c^2$. If $\rm D_A(z)$ is the angular distance 
from source to observer and $\rm x$ is the fractional distance to the 
lens, the Einstein radius of the system is $\rm R_E=[2\,R_S\,D_A\,x(1-x)]^{1/2}$. 
As the lensing object crosses close to the line of sight (or, in our case, as 
the line of sight to the fast-moving CB passes close to the lensing object) 
the amplification $\rm A$ is: 
\begin{eqnarray} 
\rm A &=& \rm {2+u^2\over u\,\sqrt{u+u^2}}\, ,\nonumber\\ 
\rm u(t)&\equiv&\rm \left[u_{min}^2+ 
{[t-t_{max}]^2\over \tau^2}\right]^{1/2}\, ,\nonumber\\ 
\rm \tau(t)&\equiv&\rm {R_E\over v_\perp(t)}\, ,\nonumber\\ 
\rm v_\perp(t)&\simeq& \rm 
c\,{\theta\,\gamma(t)\,\delta(t)\over 1+z}\, , 
\label{Gravamp} 
\end{eqnarray} 
where $\rm u_{min}$ is the minimum distance to the lens, in Einstein radii, 
of the line of sight to the CB during its motion. 
 
Gravitational lensing of a moving CB is peculiar in two ways: 
the apparent velocity is superluminal, and 
the time ``width''  of the effect, $\rm \tau(t)$, is itself  
time dependent, since the CB is decelerating as the lensing occurs: 
\begin{equation} 
\rm \tau(t)=\tau(0)\;{v_\perp(0)\over v_\perp(t)}=\tau(0)\; 
\left[{\gamma_0\over\gamma(t)}\right]^2\; 
{1+[\theta\,\gamma(t)]^2\over 1+[\theta\,\gamma_0]^2}\, . 
\label{tauoft} 
\end{equation} 

For a solar-mass star placed halfway to a GRB at $\rm z=1$, the
typical duration of a lensing event is $\tau(0)\sim 1$ hour.
The average Einstein radius of a solar-mass star placed somewhere
on the way to such a location is 
$\rm R_E[\odot]\!=\!R_E\,(M_\odot/M)^{1/2}$ 
$\rm 2\, \langle [x(1-x)]^{1/2}\rangle 
\!\sim\! 1860$ AU. What is the optical depth (or apriori probability), $\epsilon$, 
for an observable lensing by such an object? 
Consider lensing during the first 10 days of an AG, when its fluence
is relatively high and during which, for typical parameters,
the CBs travelled a (local) distance $\rm x\!\simeq\! 2$ kpc. 
The apparent transverse distance is 
$\rm x_\perp\!=\! x\,\theta/(1+z)\!\simeq\! 1$ pc. 
The average luminosity density of the local Universe is 
$\rm \rho\sim (1.8\pm 0.2)\,h\times10^8\,L_\odot/Mpc^3$ and the mass to luminosity 
ratio of star populations is $\rm M/L\sim 5$ to 10 in solar units, 
so that the number density of ``typical'' solar-mass stars is 
$\rm n_\odot\!\sim\! \rho\,M/L\!\sim\! 8.8\times 10^8/Mpc^3$ for 
$\rm h\!\sim\! 0.65$. 
The optical depth is: 
\begin{equation} 
\rm \epsilon=x_\perp\, R_\odot\,D_A\,n_\odot\,\langle (1+z)^3 \rangle\, . 
\label{epsilon} 
\end{equation} 
In the interval extending to $\rm z=1$, and for 
our adopted cosmological parameters, the volume average 
$\langle (1+z)^3 \rangle$  is $\sim 5$. 
Thus we obtain $\epsilon\sim (4\pm 2)$\%, which makes
the lensing effects hopefully visible. 
 
A rough estimate of $\epsilon$ taking into account that stars gather in 
galaxies gives a similar result. Let the surface-number density of stars in 
a galaxy, as a function of distance to the centre, be approximated 
by $\rm \Sigma_*(r)\!=\!\Sigma_*(0)\, e^{-r/h}$, with $\rm h\!\sim\! 5$ kpc. 
For a reference galaxy with $\rm N_*\!=\!10^{11}$ stars $\Sigma(0)\!\simeq\! 640$  
pc$^{-2}$. Define a galaxy's effective lensing radius so that 
$\rm \Sigma_*(r_{eff})\,x_\perp\,R_E\!\sim\! 1$. For the quoted values of 
$\rm x_\perp$ and $\rm R_E$ this means $\rm r_{eff}\!\sim\! 9$ kpc. 
Approximate the surface density of galaxies at $\rm z\!<\! 1$ by 
the observed value for galaxies with $\rm R$-magnitude below 25: 
$\rm\Sigma_G\!\sim\! 4.6\times 10^8$ rad$^{-2}$ (Casertano et al. 2000). 
The lensing probability at an angular distance $\rm D_A$ is then 
$\rm \epsilon\!\sim\!\pi\,r_{eff}^2\,\Sigma_G/D_A^2\!\sim\! 4$\%.

\newpage 
\vspace{.5  cm} 
 
{\bf 
\noindent 
Table I - Gamma-ray bursts of known  redshift} 
\begin{table}[h] 
\hspace{0.3cm} 
\begin{tabular}{|l|c|c|c|c|c|c|} 
\hline 
 
\hline 
\hline 
GRB   &z & D$_{\rm L}$  & ${\rm F_\gamma}$ 
&${\rm E_\gamma}$ &  ${\rm R[HG]}$\\ 
\hline 
970228$^1$      &0.695  &4.55  &1.1   & 0.22      & 25.2$^{18}$  \\ 
970508$^2$      &0.835  &5.70  &0.49  & 0.10      & 25.0$^{19}$  \\ 
970828$^3$      &0.957  &6.74  &9.6   & 2.06      & 24.5$^{20}$  \\ 
971214$^4$      &3.418  &32.0  &0.94  & 2.11      & 25.6$^{21}$  \\ 
980425$^5$      &.0085  &.039  &0.44  &8.1E-6     & 14.3$^{22}$  \\ 
980613$^6$      &1.096  &7.98  &0.17  & 0.61      & 24.0$^{23}$  \\ 
980703$^7$      &0.966  &6.82  &2.26  & 1.05      & 22.6$^{24}$  \\ 
990123$^8$      &1.600  &12.7  &26.8  &19.80      & 23.9$^{25}$  \\ 
990510$^9$      &1.619  &12.9  &6.55  & 5.00      & 27.0$^{26}$  \\ 
990712$^{10}$   &0.434  &2.55  &6.5   & 0.53      & 21.8$^{27}$  \\ 
991208$^{11}$   &0.70   &4.64  &10.0  & 1.51      & 24.4$^{28}$  \\ 
991216$^{12}$   &1.020  &7.30  &19.4  & 5.35      & 24.8$^{29}$  \\ 
000131$^{13}$   &4.500  &44.4  & 4.2  &11.60      & 27.8$^{30}$  \\ 
000301c$^{14}$  &2.040  &17.2  &0.41  & 0.46      & 28.0$^{31}$  \\ 
000418$^{15}$   &1.119  &8.18  &2.0   & 0.82      & 23.9$^{32}$  \\ 
000926$^{16}$   &2.066  &17.4  &2.20  & 2.60      & 25.6$^{33}$  \\ 
010222$^{17}$   &1.474  &11.5  &12.0  & 7.80      & 25.9$^{34}$  \\ 
 
\hline 
\hline 
\end{tabular} 
\end{table} 
\vskip -0.3 true cm 
\noindent 
{\bf Comments:} $\rm z$: Redshift. $\rm D_L$: Luminosity distance in Gpc, 
for $\rm \Omega_m=0.3, 
\; \Omega_\Lambda=0.7$ and ${\rm H_0=65\, km\, s^{-1}\,Mpc^{-1}}$. 
$\rm F_\gamma$: BATSE $\gamma$-ray fluences in units of 
$10^{-5}$ erg cm$^{-2}$. $\rm E_\gamma$: (Equivalent spherical) energy in 
units of  $10^{53}$ ergs. 
$\rm R[HG]$: R-magnitude of the host galaxy, except for GRB 990510, for 
which the V-magnitude is given, corrected for galactic extinction\\ 
\noindent 
{\bf References}: 
1: Djorgovski et al.~1999a; 
2: Metzger et al.~1997; 
3: Djorgovski et al.~2000; 
4: Kulkarni et al.~1998b; 
5: Tinney et al.~1998; 
6: Djorgovski et al.~1998a; 
7: Djorgovski et al.~1998b; 
8: Kelson et al.~1999; 
9: Vreeswijk et al.~1999a; 
10: Hjorth et al.~1999; 
12: Vreeswijk et al.~1999b; 
13: Andersen et al.~2000; 
14: Feng et al.~2000; 
15: Bloom et al.~2001; 
16: Fynbo et al.~2001; 
17: Jha et al.~2001; Fruchter et al.~2001; 
18: Fruchter et al.~1999b; 
19: Pian 2001; 
20: Djorgovskyet al.~2001 
21: Odewahn et al.~1998; 
22: Galama et al.~1998a; 
23: Djorgovski et al.~2000; 
24: Bloom et al.~1998b; 
25: Bloom et al.~1999b; 
26: Pian 2001; 
27: Hjorth et al.~1999; 
28: Diercks et al.~2000; 
29: Djorgovski et al.~1999b; 
30: Andersen et al.~2000; 
31: Smette et al.~2000; 
32: Bloom et al.~2000; 
33: Fynbo et al.~2001; 
34: Fruchter et al.~2001. 
\newpage

{\bf 
\noindent 
Table II -  The CB, host-galaxy, and extinction parameters} 
\begin{table}[h] 
\hspace{+.1cm} 
\begin{tabular}{|l|c|c|c|c|c|c|c|l|} 
\hline 
\hline 
GRB   &$\gamma_0 $ &$\alpha$& $\theta\, $ &$\rm x_\infty\, $  
&   R[HG]  & $\rm A_{SN}$ \\ 
\hline 
970228        & 540   &1.10  &1.686    & 0.155    & 25.55   &0.50 \\ 
971214        & 999   &1.20  &0.708    & 0.373    & 25.69   &0.94 \\ 
980613        & 509   &1.09  &1.619    & 0.241    & 24.07   &0.82 \\ 
980703        & 779   &1.08  &0.953    & 0.344    & 22.54   &0.88 \\ 
990123        &1325   &1.09  &0.420    & 0.954    & 23.90   &0.96 \\ 
990510        & 991   &1.10  &0.261    & 0.777    & 27.80   &0.50 \\ 
990510$^a$    & 907   &1.08  &0.318    & 0.504    & 27.80   &0.40 \\ 
990712        & 948   &1.09  &0.750    & 1.191    & 21.93   &0.50 \\ 
990712$^a$    & 957   &1.08  &0.863    & 1.319    & 22.57   &0.40 \\ 
991208        &1034   &1.26  &0.100    & 1.357    & 24.81   &0.80 \\ 
991216        & 972   &1.09  &0.375    & 0.953    & 24.64   &0.80 \\ 
000131        &1200   &1.26  &0.100    & 0.793    & 27.80   &0.93 \\ 
000301c       &1040   &1.19  &2.223    & 0.141    & 28.00   &0.90 \\ 
000418        &1017   &1.17  &0.970    & 1.961    & 23.74   &0.92 \\ 
000926        & 760   &1.20  &0.740    & 0.133    & 25.63   &0.96 \\ 
010222        & 1178  &1.10  &0.465    & 1.026    & 25.76   &0.94 \\ 
              &       &      &         &          &         &     \\ 
970508$^b$    & 1123   &1.10   & 3.51   & 0.293     & 24.69   &0.94 \\ 
980425$^b$    & 769   &1.10   & 8.30   & 0.252    & 14.30   &0.93 \\ 
\hline 
\hline 
 
\end{tabular} 
\end{table} 
\vskip -0.3 true cm 
\noindent 
{\bf Comments:} $\gamma_0$: Initial Lorentz factor. 
$\theta$: Viewing angle relative to the CB line of motion, 
 in milliradians. 
 $\rm x_\infty$: CB slow-down parameter, in Mpc 
($\gamma=\gamma_0/2$ at $\rm x= x_\infty/\gamma_0$). 
$\rm R[HG]$: Fitted value of the 
R-magnitude of the host galaxy (except for GRB 990712, for 
which also the V-magnitude is given) not 
corrected for galactic extinction. 
$\rm A_{SN}$: Attenuation of the SN1998bw-like contribution due to  
galactic extinction. 
$a$: V-band afterglow parameters.  
$b$: Two GRBs are special: GRB 970508 is fit with two constant
ISM densities, the  $\rm x_\infty$ quoted value corresponds to the
initial one; in GRB 980425 
the SN outshines the CBs in the optical, the fit is to the X-ray AG, 
$\alpha=1.1$ was assumed, and the parameter determinations 
are very imprecise.

\noindent 
{\bf References}:\\ 
{\bf GRB 970228}: 
Castander, et al.~1999a; 
Castander, et al.~1999b; 
Djorgovski et al.~1999a; 
Fruchter et al.~1997a; 
Galama et al.~1997; 
Galama et al.~2000; 
Garcia et al.~1998; 
Guarnieri, et al.~1997; 
Metzger et al.~1997a; 
Pedichini et al.~1997; 
Sahu et al.~1997a; 
Sahu et al.~1997b; 
van Paradijs et al.~1998. \\ 
{\bf GRB 970508}: 
Bloom et al.~1998b; 
Castro-Tirado et al.~1998b; 
Chevalier \& Ilovaisky~1997; 
Djorgovski et al.~1997; 
Fruchter et al.~1997b; 
Fruchter et al.~2000; 
Galama et al.~1998b; 
Metzger et al.~1997b; 
Pedersen et al.~1998a; 
Schaefer et al.~1997; 
Sokolov et al.~1997; 
Sokolov et al.~1998; 
Zharikov et al.~1998. \\ 
{\bf GRB 971214}: 
Diercks et al.~1998; 
Halpern et al.~1998b; 
Kulkarni et al.~1998b.\\ 
{\bf GRB 980613}: 
Djorgovski et al.~1998a; 
Djorgovski et al.~2000; 
Halpern et al.~1998a; 
Hjorth et al.~1998.\\ 
{\bf GRB 980703}: 
Bloom et al.~1998b; 
Castro-Tirado et al.~1999a; 
Holland et al.~2000b; 
Holland et al.~2001; 
Pedersen et al.~1998b; 
Sokolov et al.~1998; 
Vreeswijk et al.~1999c; 
Zapatero Osorio et al.~1998.\\ 
{\bf GRB 990123}: 
Castro-Tirado et al.~1999b; 
Fruchter et al.~1999a; 
Galama et al.~1999a; 
Holland et al.~2000a; 
Kulkarni et al.~1999. \\ 
{\bf GRB 990510}: 
Beuermann et al.~1999; 
Covino et al.~1999; 
Fruchter et al.~1999c; 
Galama et al.~1999b; 
Harrison et al.~1999; 
Holland et al.~2000a; 
Marconi et al.~1999a,b; 
Pietrzy{\'n}ski \& Udalski~1999a,b,c; 
Stanek et al.~1999. \\ 
{\bf GRB 990712}: 
Hjorth et al.~2000a; 
Sahu, et al.~2000\\ 
{\bf GRB 991208}: 
Castro-Tirado et al.~2001; 
Sagar et al.~2000a.\\ 
{\bf GRB 991216}: 
Djorgovski et al.~1999b; 
Garnavich et al.~2000a; 
Halpern et al.~2000a; 
Sagar et al.~2000a.\\ 
{\bf GRB 000131}: 
Andersen et al.~2000.\\ 
{\bf GRB 000301c}: 
Jensen et al.~2001; 
Halpern et al.~2000b; 
Garnavich et al.~2000b; 
Masetti et al.~2000; 
Rhoads \& Fruchter~2001; 
Sagar et al.~2000b; 
Veillet 2000a.\\ 
{\bf GRB 000418}: 
Berger et al.~2000; 
Henden et al.~2000; 
Klose et al.~2000;  
Metzger \& Fruchter~2000.\\ 
{\bf GRB 000926}: 
Fynbo et al.~2001; 
Halpern et al.~2000C 
Harrison et al.~2001; 
Hjorth et al.~2000b; 
Price et al.~2001; 
Sagar et al.~2001a; 
Veillet~2000b. \\ 
{\bf GRB 010222}: 
Cowsik et al.~2001; 
Jha et al.~2001; 
Fruchter et al.~2001; 
Garnavich et al.~2001; 
Masetti et al.~2001; 
Oksanen et al.~2001; 
Orosz et al.~2001; 
Price et al.~2001; 
Sagar et al.~2001b; 
Stanek et al.~2001; 
Valentini et al.~2001; 
Watanabe et al.~2001; 
Veillet~2001.\\ 
 

\vskip 0.3 true cm 
 
{\bf 
\noindent 
Table III - The rest frame GRB energy and X-ray AG of GRBs with  
measured redshift and X-ray AG 
from their observed $\gamma$-ray fluence and the optical AG parameters} 
\begin{table}[h] 
\hspace{-.1cm} 
\begin{tabular}{|l|c|c|c|c|c|c|c|c|c|} 
 
\hline 
\hline 
GRB   &z &$\rm D_L$  &${\rm F_\gamma}$  
&$\gamma_0$&$\delta_0$&$\rm q $&$\rm E_\gamma^{CB}\,$  \\ 
\hline 
970228   &0.695    &4.55  &1.10   & 540  & 591  & 0.86   &0.78 \\ 
970508   &0.835    &5.70  &1.10   & 1123  & 137  & 1.26   &1.47 \\ 
970828   &0.957    &6.74  &1.10   &1153  &1160  & 0.77   &1.34 \\ 
971214   &3.418    &32.0  &0.94   & 999  &1331  & 1.28   &1.11 \\ 
980613   &1.096    &7.98  &0.17   & 509  & 606  & 1.34   &1.74 \\ 
990123   &1.600    &12.7  &26.8   &1325  &2023  & 1.45   &1.84 \\ 
990510   &1.619    &12.9  &6.55   & 991  &1858  & 2.60   &0.78 \\ 
991216   &1.020    &7.30  &19.4   & 972  &1716  & 0.38   &1.54 \\ 
000926   &2.066    &17.4  &2.20   & 761  &1115  & 2.12   &0.60 \\ 
010222   &1.474    &11.5  &12.0   &1109  &1812  & 1.43   &1.31 \\ 
 
\hline 
\hline 
\end{tabular} 
\end{table} 
\vskip -0.3 true cm 
\noindent 
{\bf Comments:} $\rm D_L$: Luminosity distance, for $\rm \Omega_m=0.3, 
\; \Omega_\Lambda=0.7$ and ${\rm H_0=65\, km\, s^{-1}\,Mpc^{-1}}$, in Gpc.\\ 
$\rm F_\gamma$: BATSE/BeppoSAX  $\gamma$-ray fluences in units of 
$10^{-5}$ erg cm$^{-2}$. $\rm \delta_0$: Initial Doppler factor. 
$\rm q$: The ratio between observed and predicted late-time X-ray fluxes in   
the 2-10 keV band, for a single unextinct standard CB with  
$\rm N_{CB}=6\times 10^{50}\, .$  
$\rm E_\gamma^{CB}$: Energy radiated by the ensemble of CBs 
in its rest frame, in units of $\rm 10^{44}\,erg\,.$  \\ 
 
{}

\newpage

\clearpage

\clearpage 

\clearpage 
 
\clearpage 
 
%
 
 
 
\clearpage 
 
\clearpage 
 
\clearpage 
 
\begin{figure}[t] 
\begin{tabular}{cc} 
\hskip 2truecm 
\vspace*{2cm} 
\hspace*{-1.7cm} 
\epsfig{file=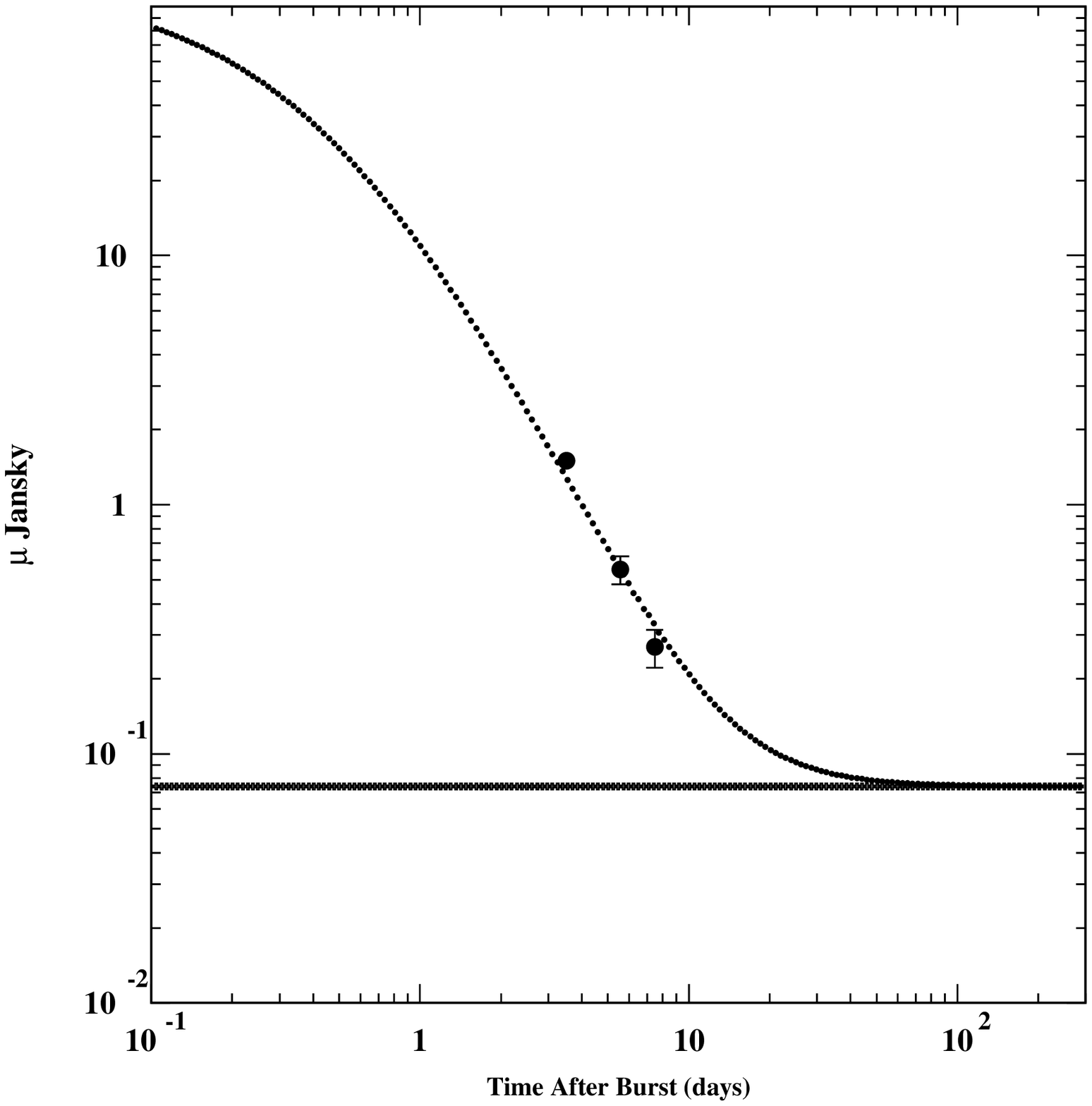, width=8cm} \\ 
\hspace*{.5cm} 
\epsfig{file=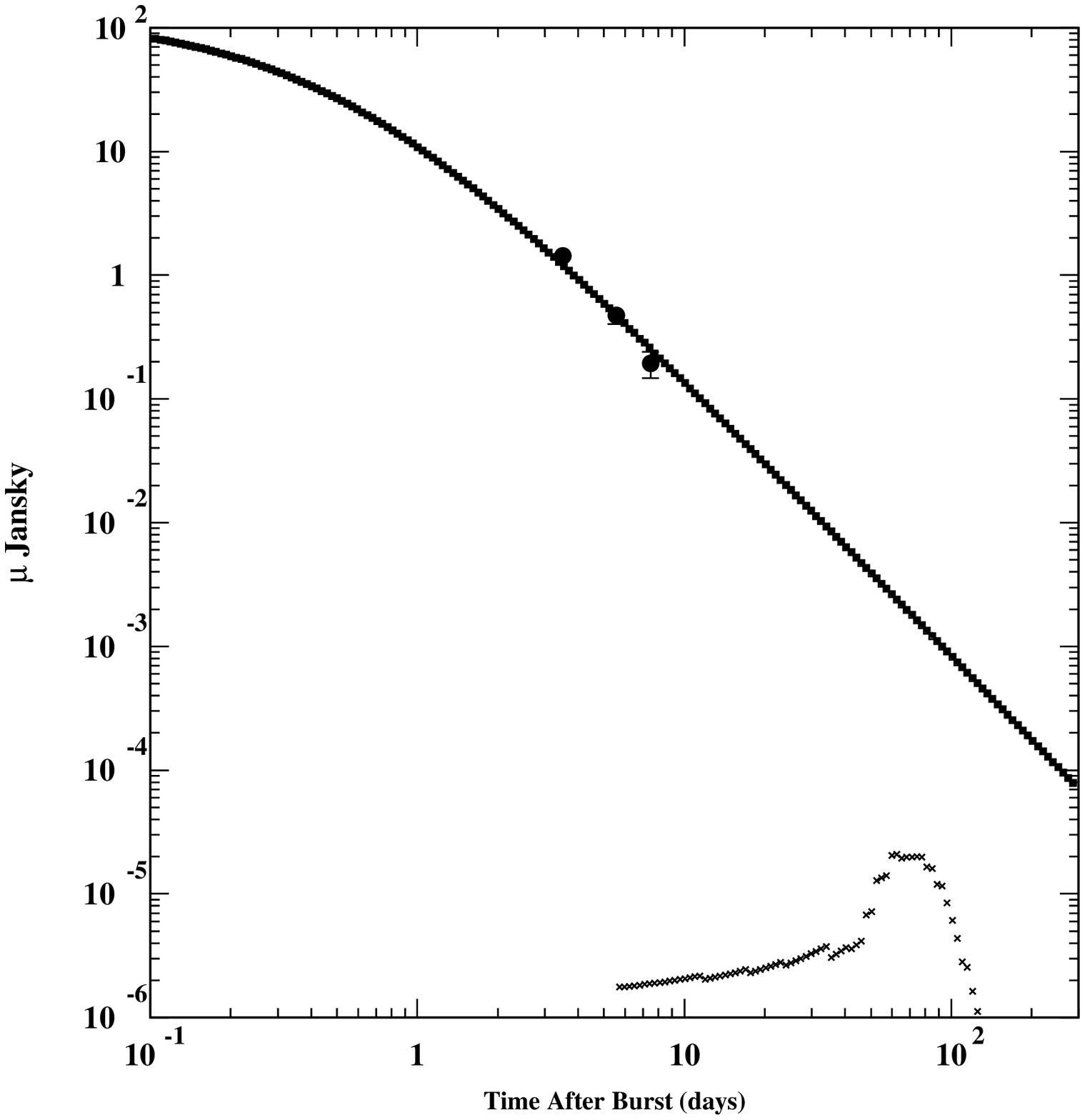, width=8cm} 
\end{tabular} 
\caption{Comparisons between our fitted R-band afterglow 
(upper curves) and the observations listed in Table II, 
not corrected for  extinction, 
for GRB 000131, at $\rm z=4.5$. 
Upper panel: Without subtraction of the host 
galaxy's contribution (the straight line). 
Lower panel: With the host galaxy subtracted, and the 
CB's AG (the line of squares) given by Eqs.~(\ref{fluxdensity}) and 
(\ref{cubic}). The contribution 
from a 1998bw-like supernova placed at the GRB's 
redshift, Eq.~(\ref{bw}), corrected for  extinction, 
is indicated by a line of crosses. 
The data are too scarce for conclusions to be drawn.} 
\end{figure} 
\clearpage 
 
\begin{figure}[t] 
\begin{tabular}{cc} 
\hskip 2truecm 
\vspace*{2cm} 
\hspace*{-1.7cm} 
\epsfig{file=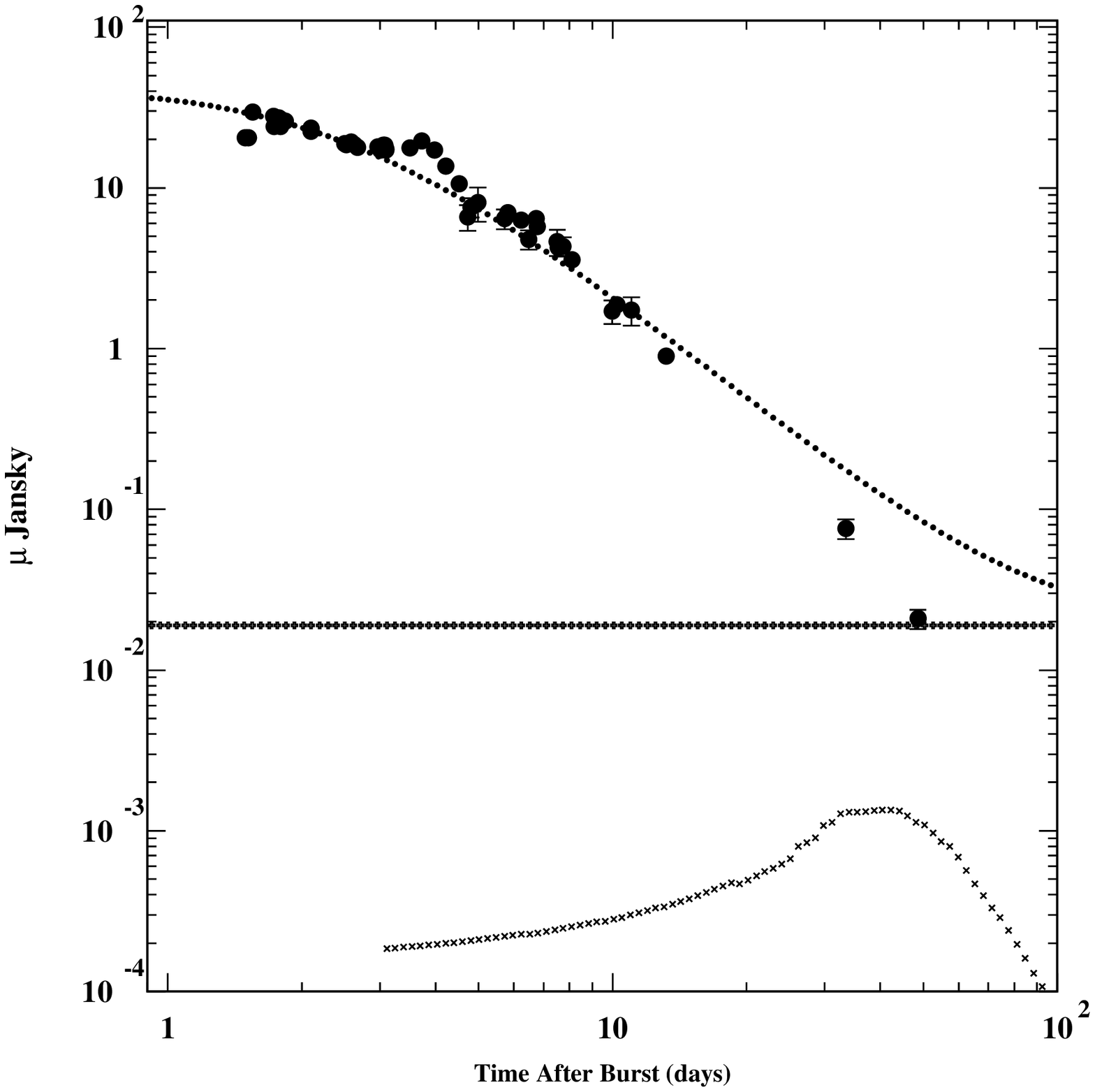, width=8cm} \\ 
\hspace*{.5cm} 
\epsfig{file=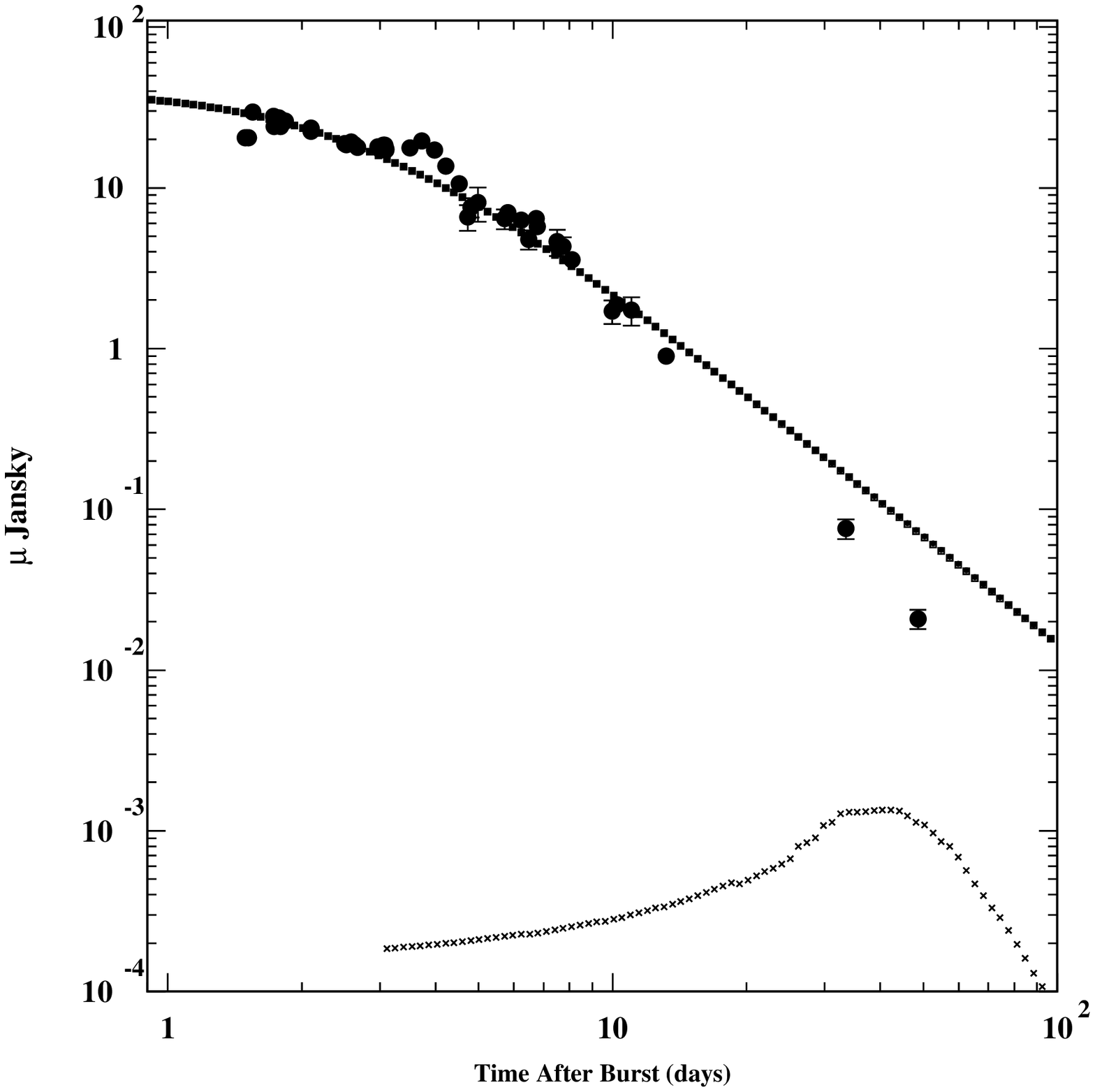, width=8cm} 
\end{tabular} 
\caption{Comparisons between our fitted R-band afterglow 
(upper curves) and the observations listed in Table II, 
not corrected for  extinction, 
for GRB 000301c, at $\rm z=2.040$. 
Upper panel: Without subtraction of the host 
galaxy's contribution (the straight line). 
Lower panel: With the host galaxy subtracted, and the 
CB's AG (the line of squares) given by Eqs.~(\ref{fluxdensity}) and 
(\ref{cubic}). The contribution 
from a 1998bw-like supernova placed at the GRB's 
redshift, Eq.~(\ref{bw}), corrected for  extinction, 
is indicated by a line of crosses. 
The expected SN contribution is too weak to be 
observable.} 
\label{fig301} 
\end{figure} 
 
 
 
\begin{figure}[t] 
\begin{tabular}{cc} 
\hskip 2truecm 
\vspace*{2cm} 
\hspace*{-1.7cm} 
\epsfig{file=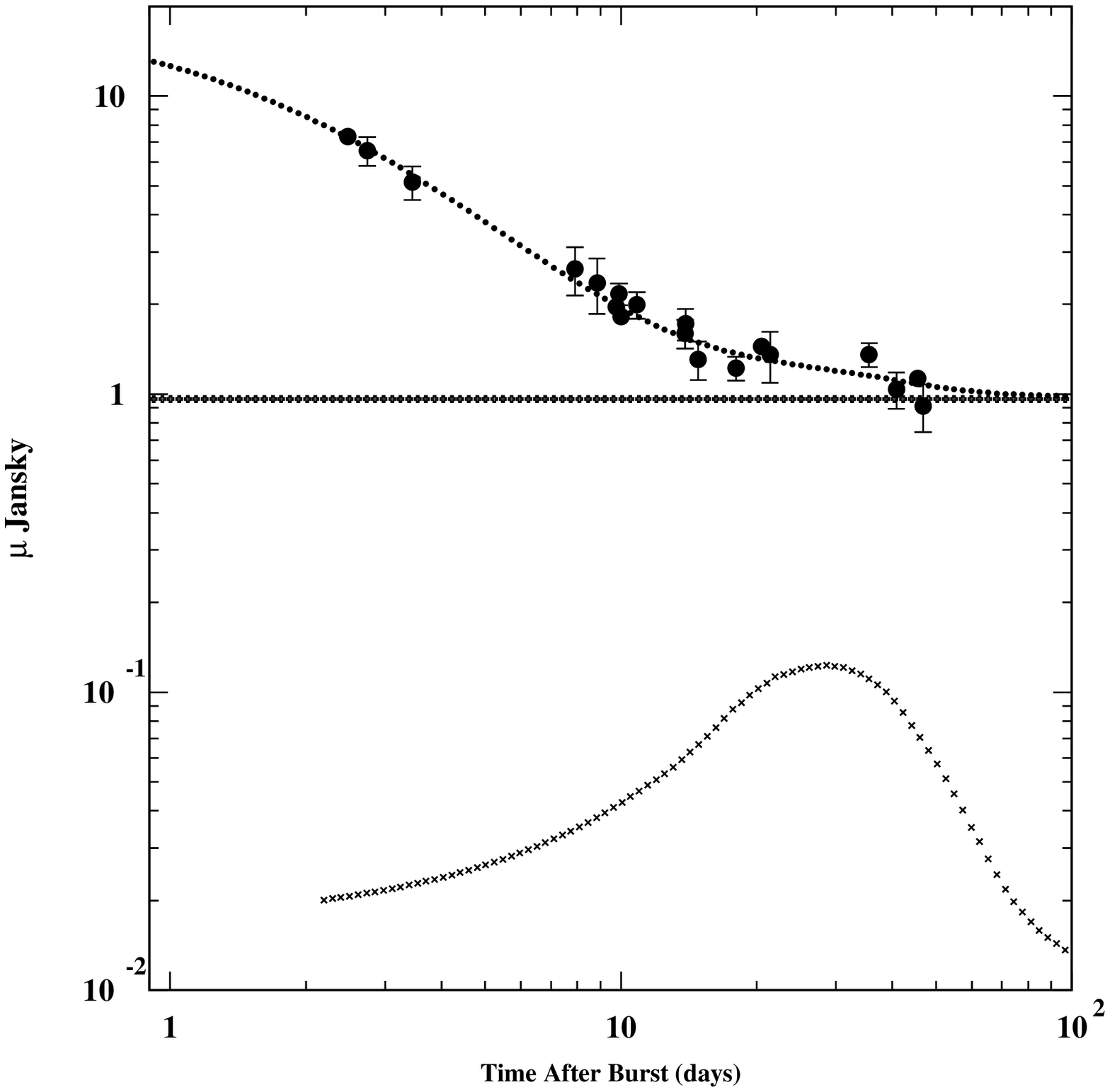, width=8cm} \\ 
\hspace*{.5cm} 
\epsfig{file=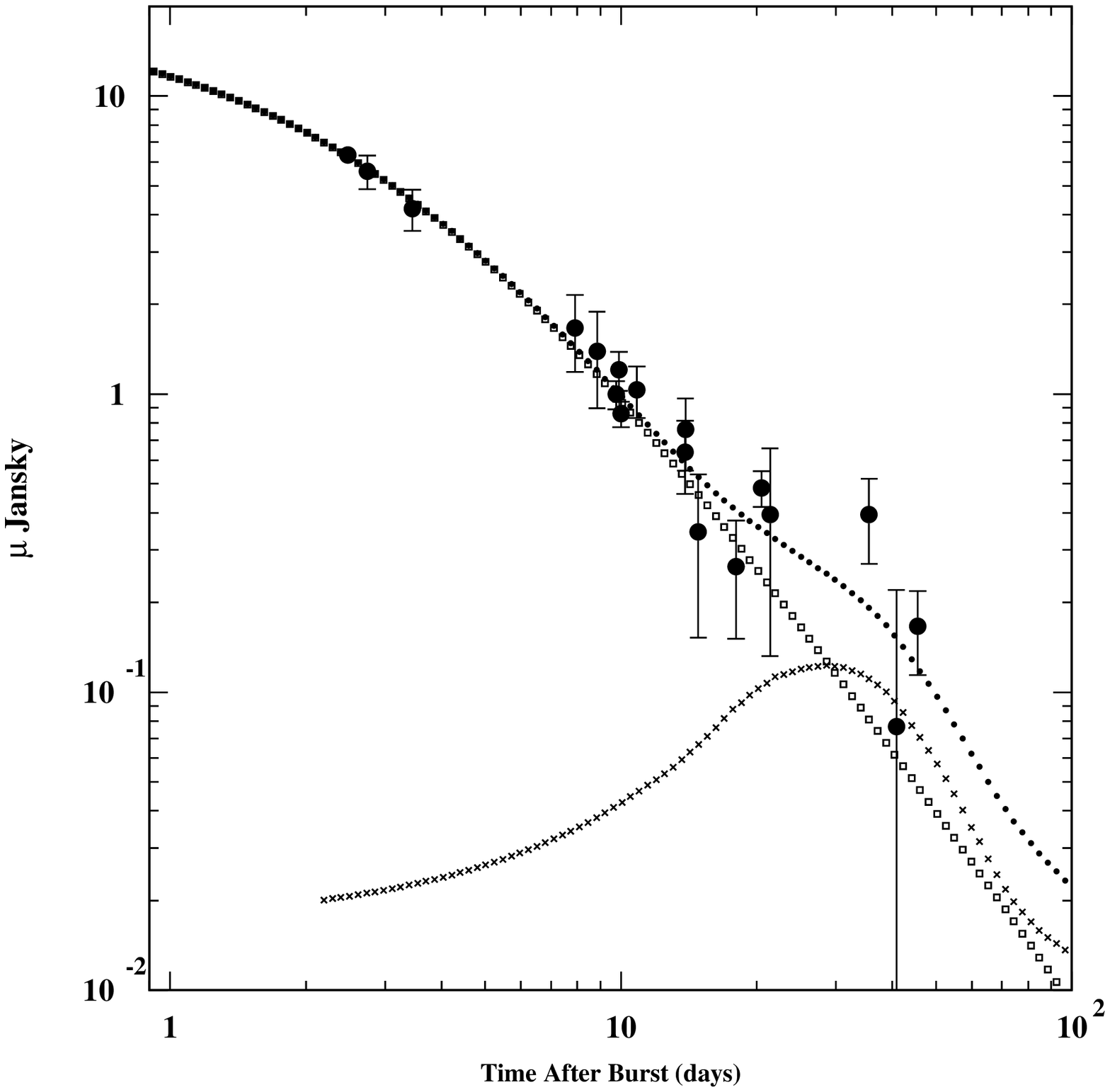, width=8cm} 
\end{tabular} 
\caption{Comparisons between our fitted R-band afterglow 
(upper curves) and the observations listed in Table II, 
not corrected for  extinction, 
for GRB 000418, at $\rm z=1.119$. 
Upper panel: Without subtraction of the host 
galaxy's contribution (the straight line). 
Lower panel: With the host galaxy subtracted, and the 
CB's AG (the line of squares) given by Eqs.~(\ref{fluxdensity}) and 
(\ref{cubic}). The contribution 
from a 1998bw-like supernova placed at the GRB's 
redshift, Eq.~(\ref{bw}), corrected for  extinction, 
is indicated in both panels by a line of crosses. 
There is some indication of a SN1998bw-like 
contribution.} 
\label{fig418} 
\end{figure} 
\clearpage 
\begin{figure}[t] 
\begin{tabular}{cc} 
\hskip 2truecm 
\vspace*{2cm} 
\hspace*{-1.7cm} 
\epsfig{file=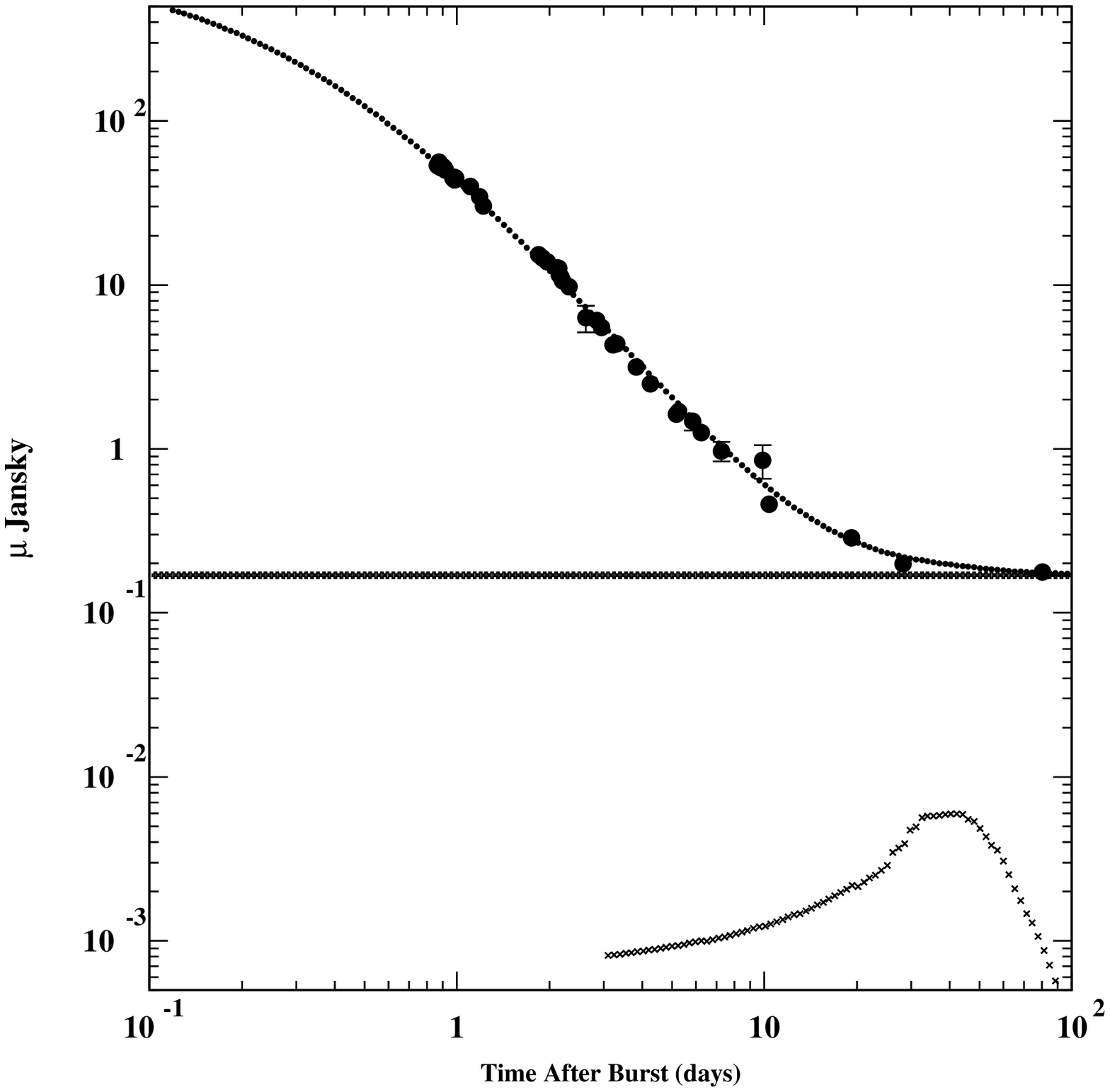, width=8cm} \\ 
\hspace*{.5cm} 
\epsfig{file=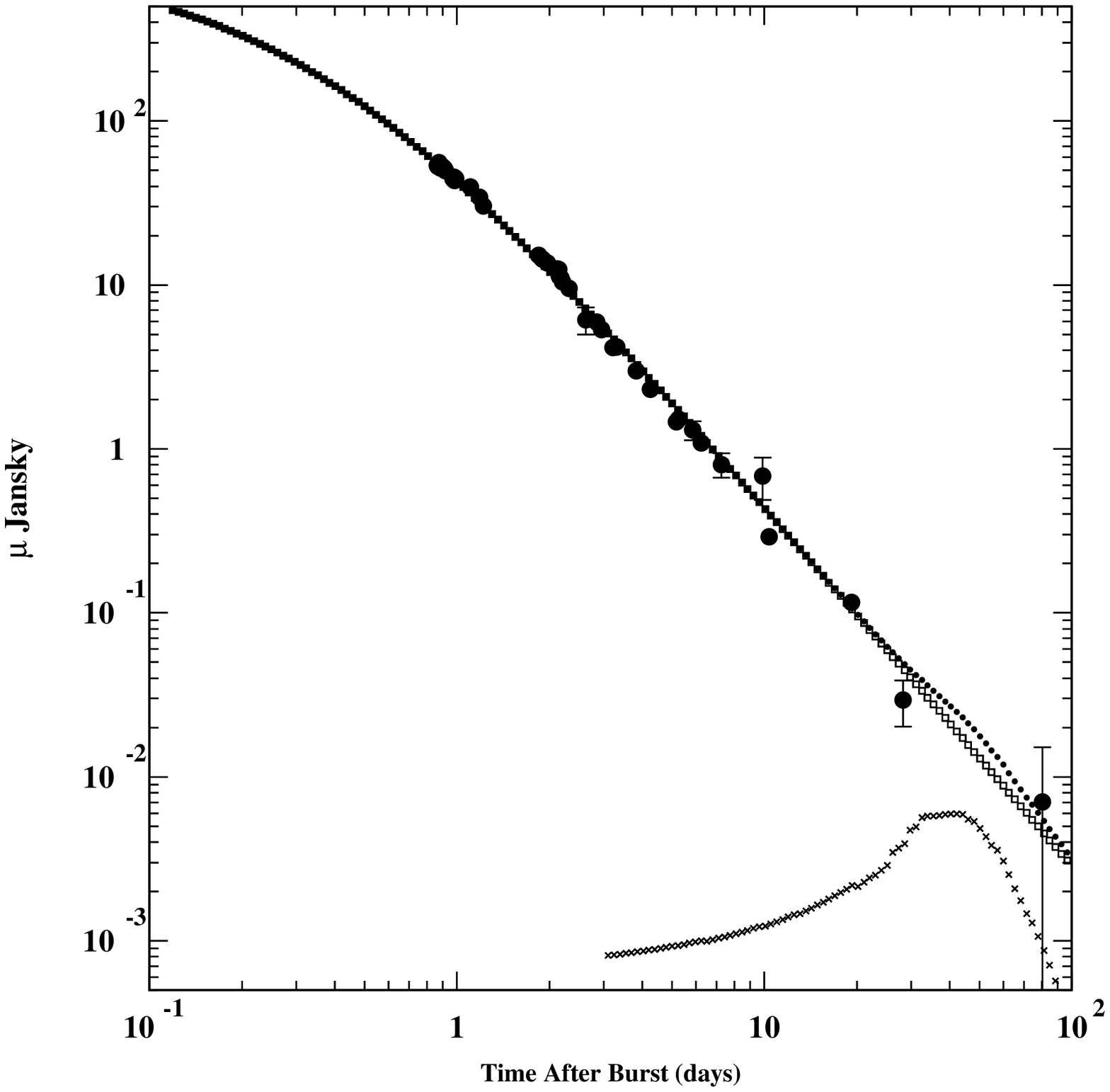, width=8cm} 
\end{tabular} 
\caption{Comparisons between our fitted R-band afterglow 
(upper curves) and the observations listed in Table II, 
not corrected for  extinction, 
for GRB 000926, at $\rm z=2.066$. 
Upper panel: Without subtraction of the host 
galaxy's contribution (the straight line). 
Lower panel: With the host galaxy subtracted, and the 
CB's AG (the line of squares) given by Eqs.~(\ref{fluxdensity}) and 
(\ref{cubic}). The contribution 
from a 1998bw-like supernova placed at the GRB's 
redshift, Eq.~(\ref{bw}), corrected for  extinction, 
is indicated in both panels by a line of crosses. 
A SN1998bw-like contribution could not have been seen.} 
\label{fig926} 
\end{figure} 

\begin{figure}[t] 
\begin{tabular}{cc} 
\hskip 2truecm 
\vspace*{2cm} 
\hspace*{-1.7cm} 
\epsfig{file=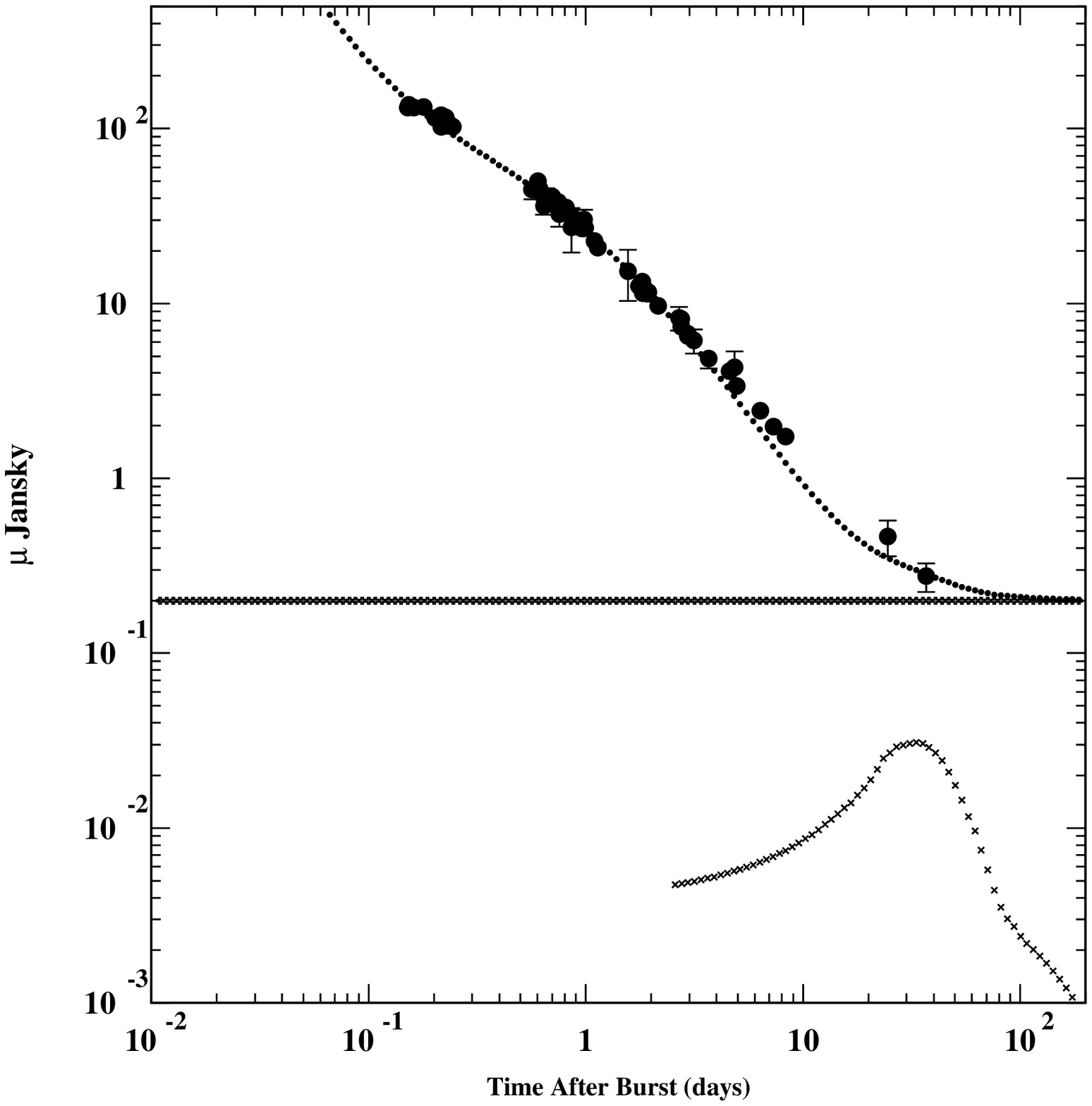, width=8cm} \\ 
\hspace*{.5cm} 
\epsfig{file=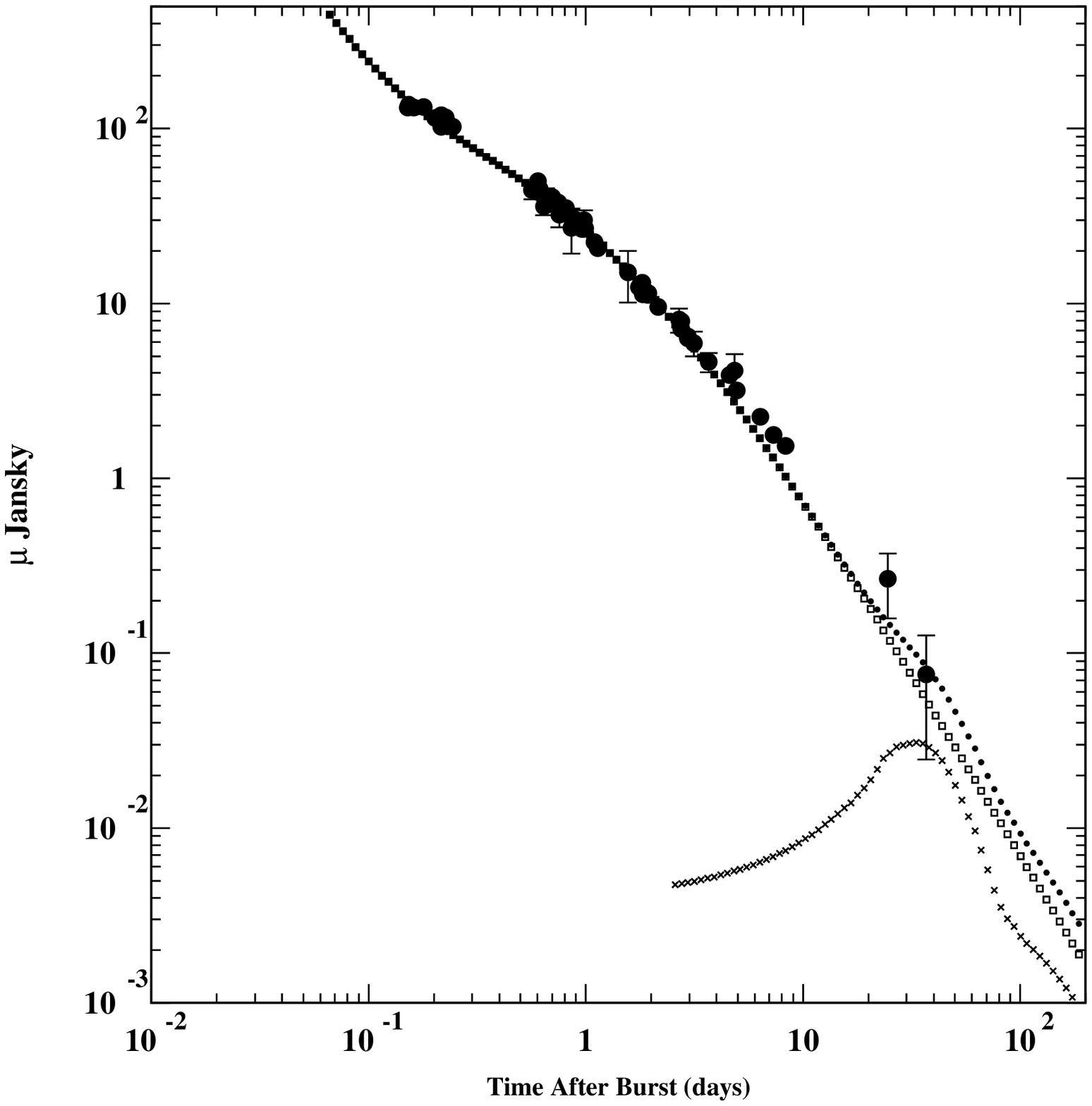, width=8cm} 
\end{tabular} 
\caption{Comparisons between our fitted R-band afterglow 
(upper curves) and the observations listed in Table II, 
not corrected for  extinction, 
for GRB 010222, at $\rm z>1.474$. 
Upper panel: Without subtraction of the host 
galaxy's contribution (the straight line). 
Lower panel: With the host galaxy subtracted, and the 
CB's AG (the line of squares) given by Eqs.~(\ref{fluxdensity}) and 
(\ref{cubic}). The contribution 
from a 1998bw-like supernova placed at the GRB's 
redshift, Eq.~(\ref{bw}), corrected for  extinction, 
is indicated in both panels by a line of crosses. 
There are no observations at the time a 1998bw-like 
supernova would have significantly contributed.} 
\label{fig222} 
\end{figure} 
 
\clearpage 
 
\begin{figure}[t] 
\begin{tabular}{cc} 
\hskip 2truecm 
\vspace*{2cm} 
\hspace*{-1.7cm} 
\epsfig{file=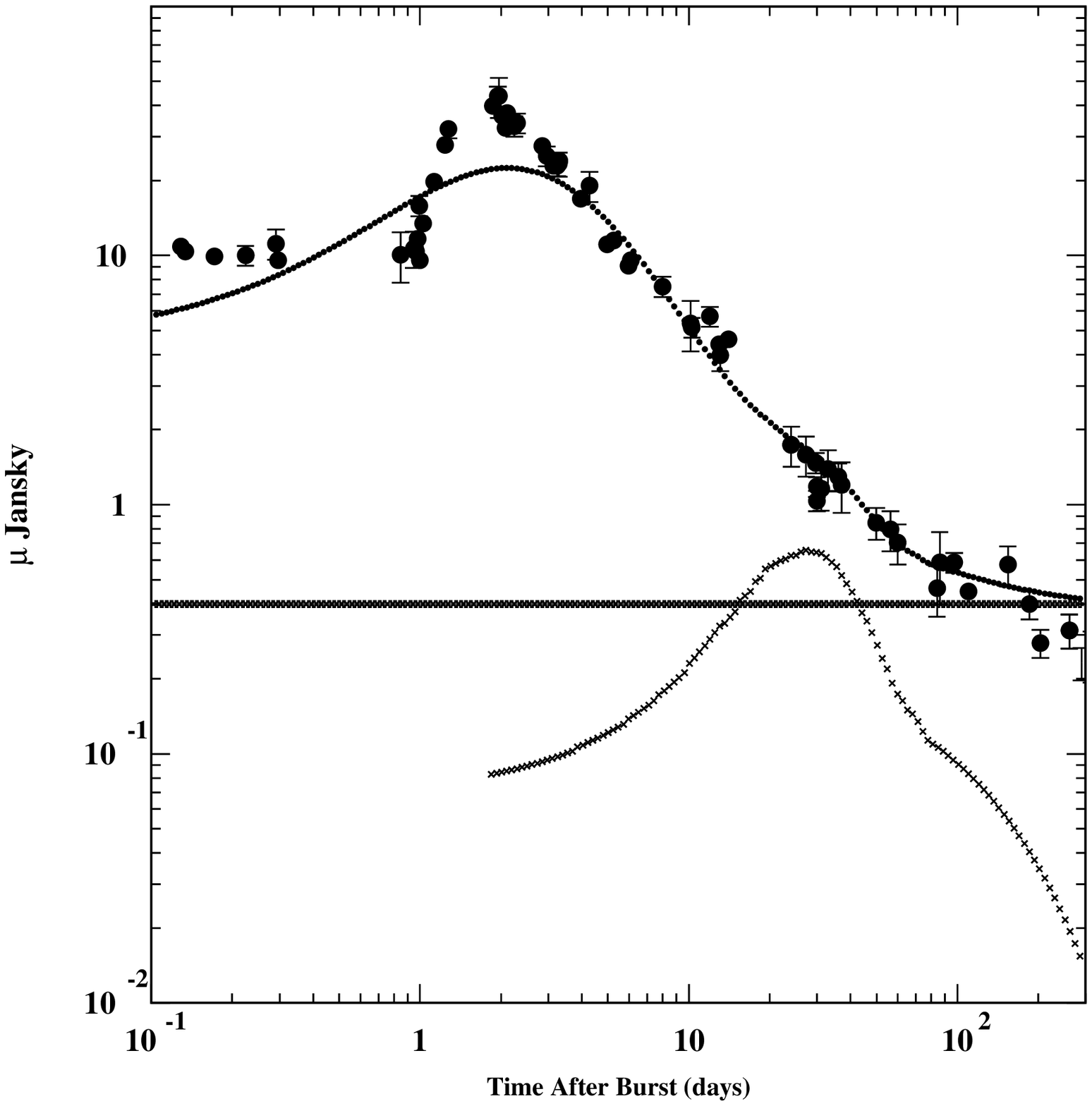, width=8cm} \\ 
\hspace*{.5cm} 
\epsfig{file=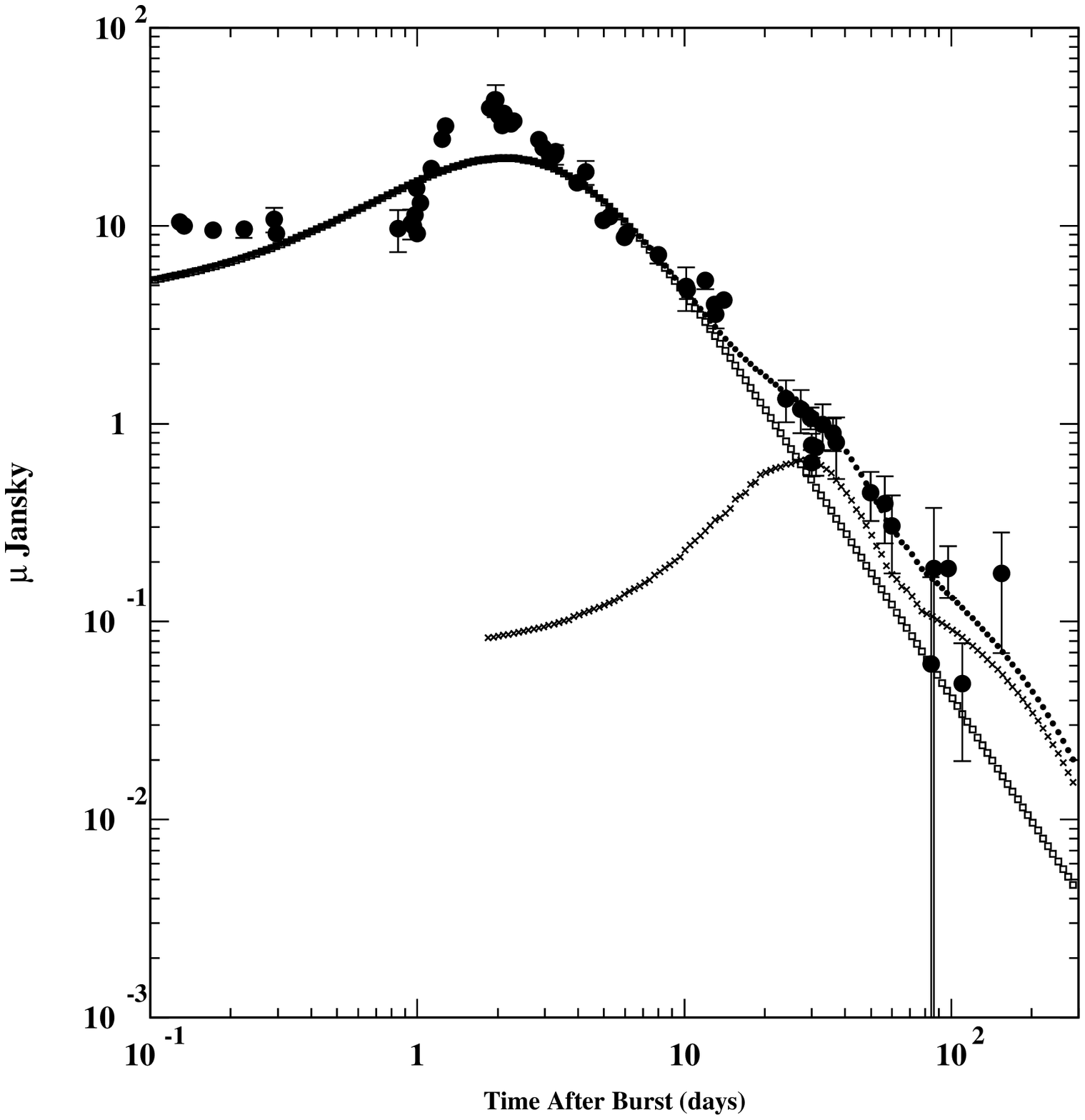, width=8cm} 
\end{tabular} 
\caption{Comparisons between our fitted R-band afterglow 
(upper curves) and the observations listed in Table II, 
not corrected for  extinction, 
for GRB 970508, at $\rm z=0.835$. 
Upper panel: Without subtraction of the host 
galaxy's contribution (the straight line). 
Lower panel: With the host galaxy subtracted, and the 
CB's AG (the line of squares) given by Eqs.~(\ref{fluxdensity}) and 
(\ref{cubic}). The contribution 
from a 1998bw-like supernova placed at the GRB's 
redshift, Eq.~(\ref{bw}), corrected for  extinction, 
is indicated in both panels by a line of crosses. 
This is the best fit not including the possibility of lensing; 
and it is terrible. } 
\label{1CB508} 
\end{figure} 
 
\begin{figure}[t] 
\begin{tabular}{cc} 
\hskip 2truecm 
\vspace*{2cm} 
\hspace*{-1.7cm} 
\epsfig{file=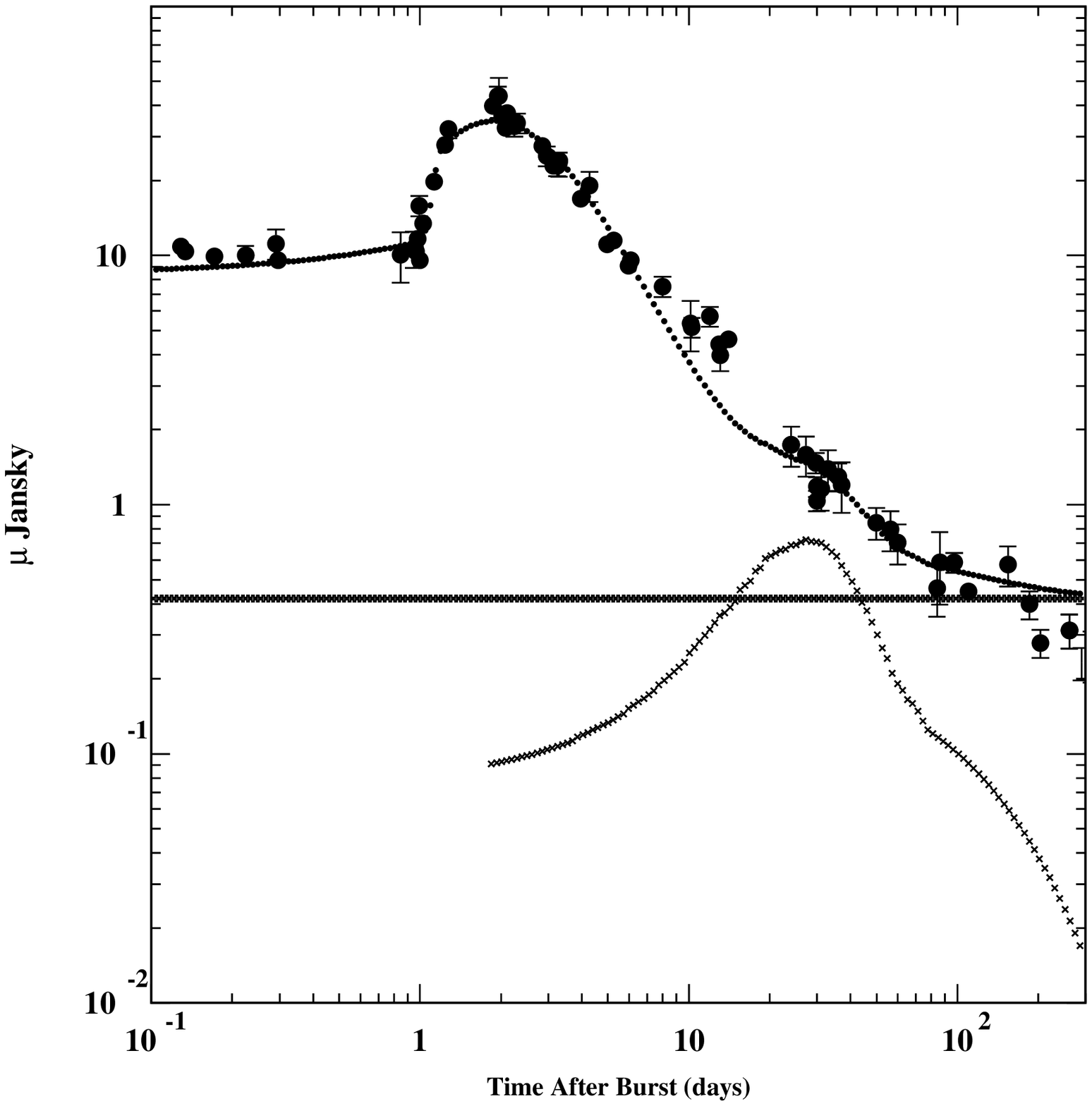, width=8cm} \\ 
\hspace*{.5cm} 
\epsfig{file=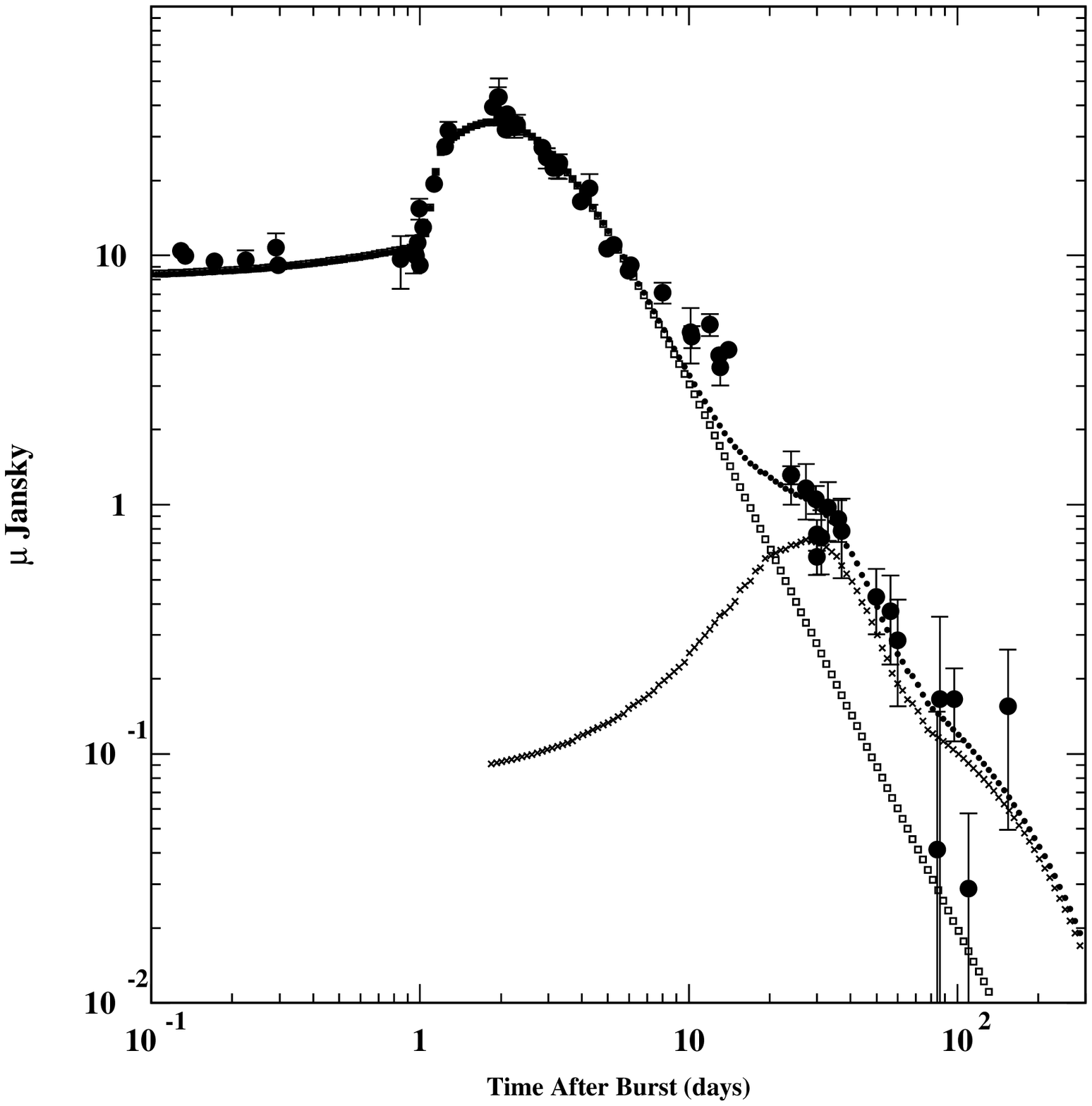, width=8cm} 
\end{tabular} 
\caption{The same as in Fig.~(\ref{1CB508}), but with a density profile
that jumps from one constant value to another, at a distance $\sim 0.24$ kpc
from the progenitor.
Upper panel: Without subtraction of the host 
galaxy's contribution (the straight line). 
Lower panel: With the host galaxy subtracted, and the 
CB's AG (the line of squares) given by Eqs.~(\ref{fluxdensity}) and 
(\ref{cubic}).
This time the fit, unlike that of Fig.~(\ref{1CB508}), is quite satisfactory. 
A SN1998bw-like contribution is necessary.} 
\label{jump508} 
\end{figure} 
\clearpage 

 
\begin{figure}[t] 
\begin{tabular}{cc} 
\hskip 2truecm 
\vspace*{2cm} 
\hspace*{-1.7cm} 
\epsfig{file=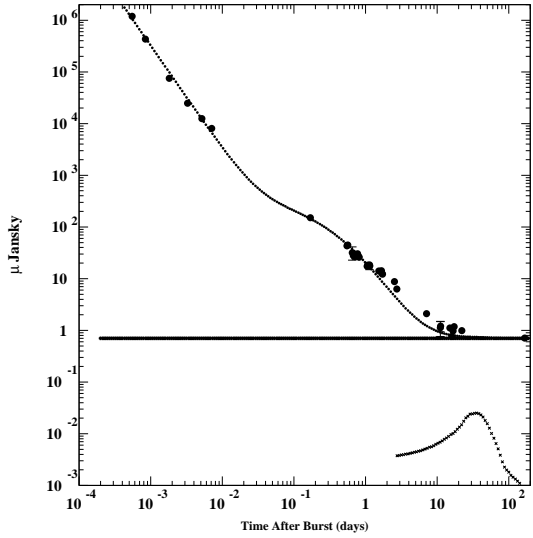, width=8cm} \\ 
\hspace*{.5cm} 
\epsfig{file=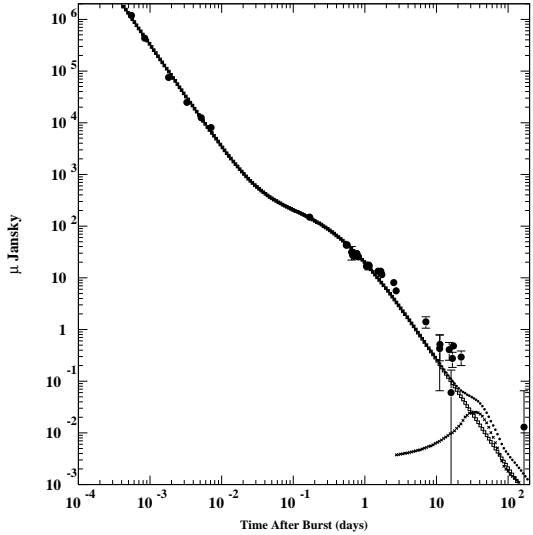, width=8cm} 
\end{tabular} 
\caption{Comparisons between our fitted R-band afterglow 
(upper curves) and the observations, 
not corrected for extinction, for GRB 990123. 
Upper panel: Without subtraction of the host 
galaxy's contribution (the straight line). 
Lower panel: With the host galaxy subtracted. 
Unlike in the previous ensemble of figures, these AG data 
start from a short time after the GRB. The starting $\rm t^{-2}$ 
behaviour is that expected if the CBs are moving trough a 
density profile, $\rm n\propto r^{-2}$, 
induced by the parent-star's pre-SN wind and 
ejections.} 
\label{early123} 
\end{figure} 
 
 
\begin{figure} 
\begin{center} 
\vspace*{.003cm} 
\hspace*{-0cm} 
\epsfig{file=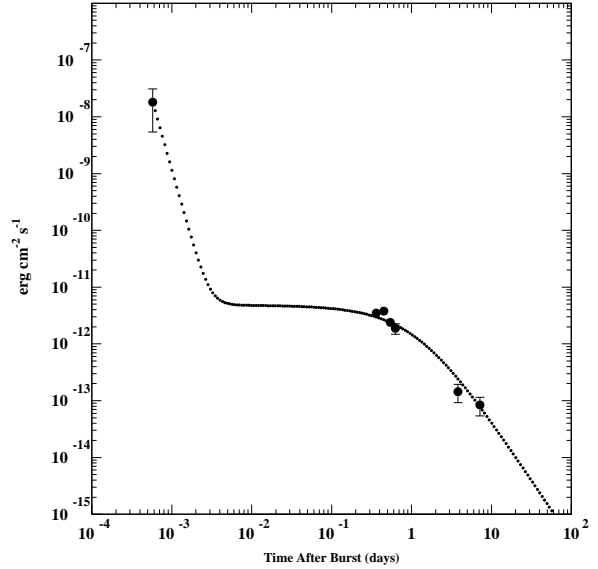,width=8.5cm} 
\caption{The early-time and late-time X-ray AG of GRB 970228 in the 2--10 
keV range 
as measured by Costa et al. (1997), fitted with 
Eq.~(\ref{Xdensity}) for a constant density along the 
CB trajectory. This assumption 
should be vastly inappropriate between 
$ \rm \sim 2\times 10^{-3}$ and $\sim 0.2$ days, 
during which $\rm n_e$ and the X-ray AG are 
expected to diminish by two or three orders of magnitude. 
Coincidentally, there are no data there.} 
\vspace*{-0.5cm} 
\label{X228} 
\end{center} 
\end{figure} 
 
 
\begin{figure} 
\begin{center} 
\vspace*{.003cm} 
\hspace*{-0cm} 
\epsfig{file=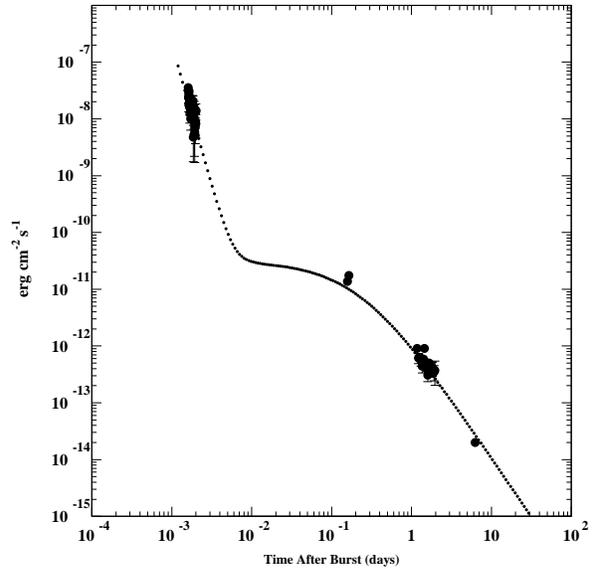,width=8.5cm} 
\caption{The early-time and late-time X-ray AG of GRB 970828 in the 2--10 
keV range as measured by Smith et al. (2001) and Yoshida et al. (2001), 
fitted with Eq.~(\ref{Xdensity}) for a constant density along the 
CB trajectory. The same comments as in Fig~(\ref{X228}) apply here.} 
\vspace*{-0.5cm} 
\label{X828} 
\end{center} 
\end{figure} 
\clearpage 
 
\begin{figure} 
\begin{center} 
\vspace*{.003cm} 
\hspace*{-0cm} 
\epsfig{file=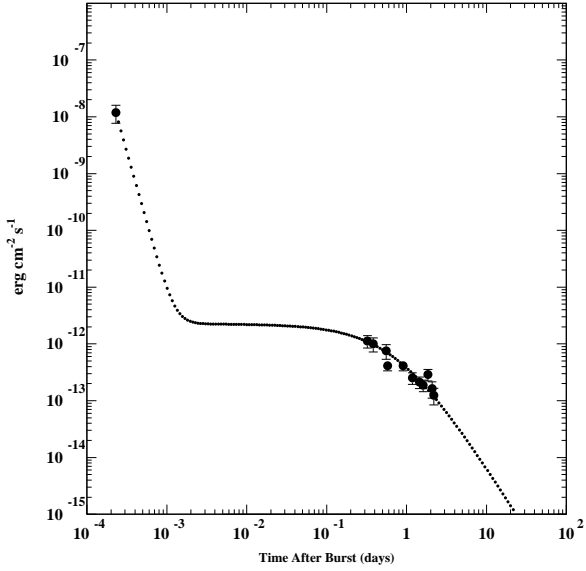,width=8.5cm} 
\caption{The early-time and late-time X-ray AG of GRB 971214 in the 2--10 
keV range as measured by Dal Fiume et al. (2000)  
fitted with Eq.~(\ref{Xdensity}) for a constant density along the 
CB trajectory. The same comments as in Fig~(\ref{X228}) apply here.} 
\vspace*{-0.5cm} 
\label{X1214} 
\end{center} 
\end{figure} 
 
 
\begin{figure} 
\begin{center} 
\vspace*{.003cm} 
\hspace*{-0cm} 
\epsfig{file=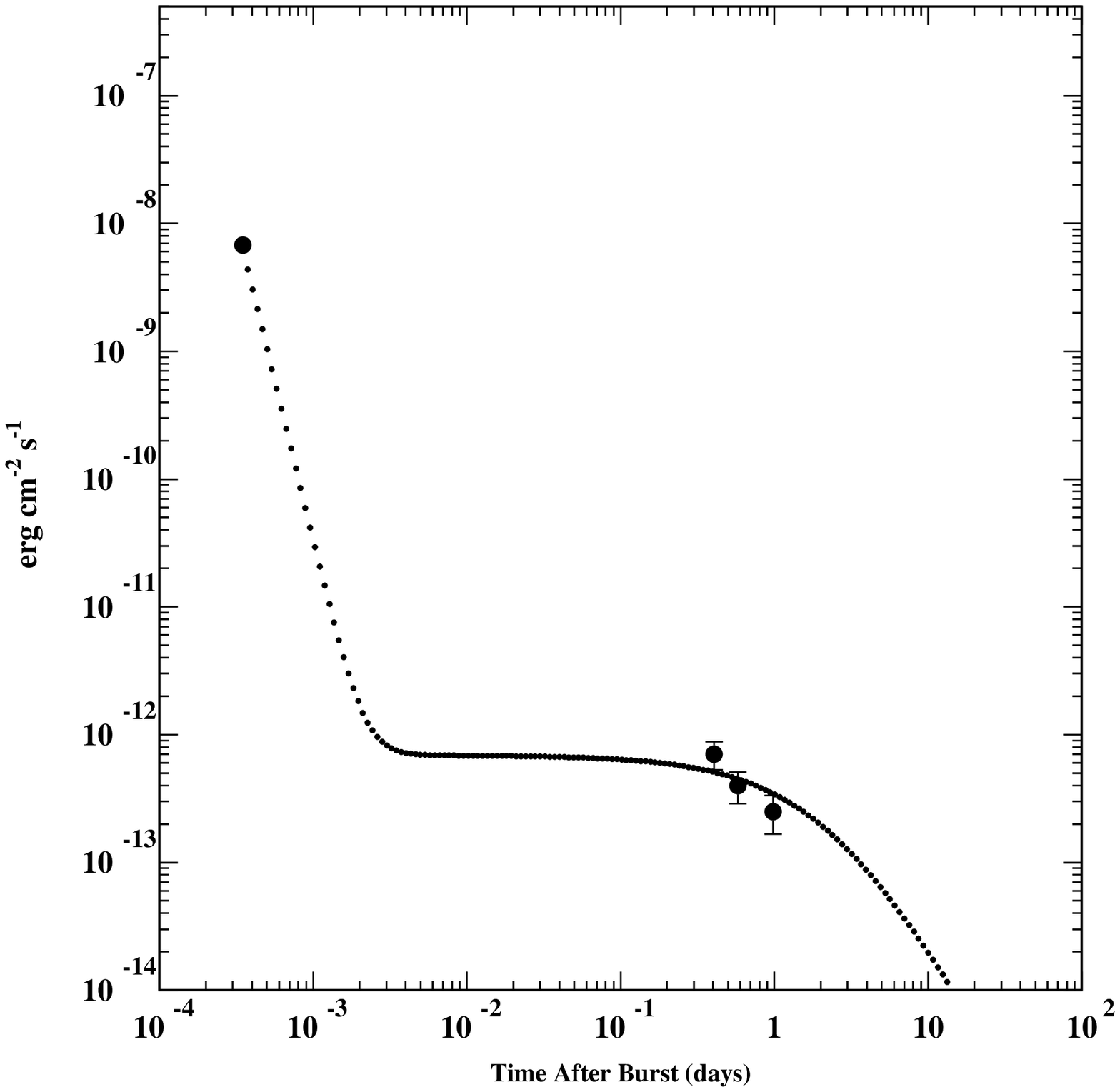,width=8.5cm} 
\caption{The early-time and late-time X-ray AG of GRB 980613 in the 2--10 
keV range (Piro 2000) 
fitted with Eq.~(\ref{Xdensity}) for a constant density along the 
    CB trajectory. The same comments as in Fig~(\ref{X228}) apply here.} 
\vspace*{-0.5cm} 
\label{X613} 
\end{center} 
\end{figure}

 
\begin{figure} 
\begin{center} 
\vspace*{.003cm} 
\hspace*{-0cm} 
\epsfig{file=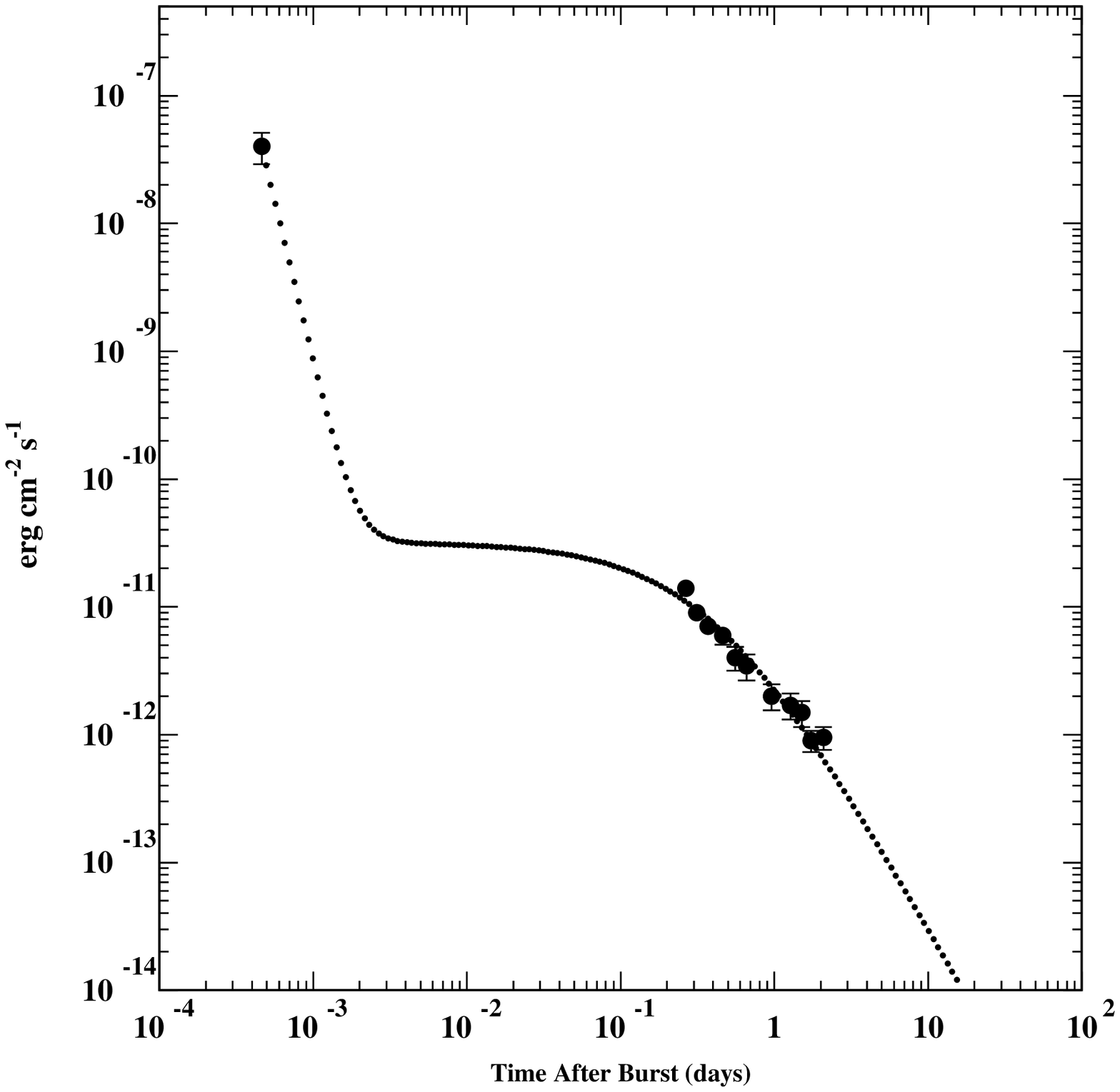,width=8.5cm} 
\caption{The early-time and late-time X-ray AG of GRB 990123 in the 2--10 
keV range (Piro 2000)  
fitted with Eq.~(\ref{Xdensity}) for a constant density along the 
    CB trajectory. The same comments as in Fig~(\ref{X228}) apply here.} 
\vspace*{-0.5cm} 
\label{X123} 
\end{center} 
\end{figure} 
 
\begin{figure} 
\begin{center} 
\vspace*{.003cm} 
\hspace*{-0cm} 
\epsfig{file=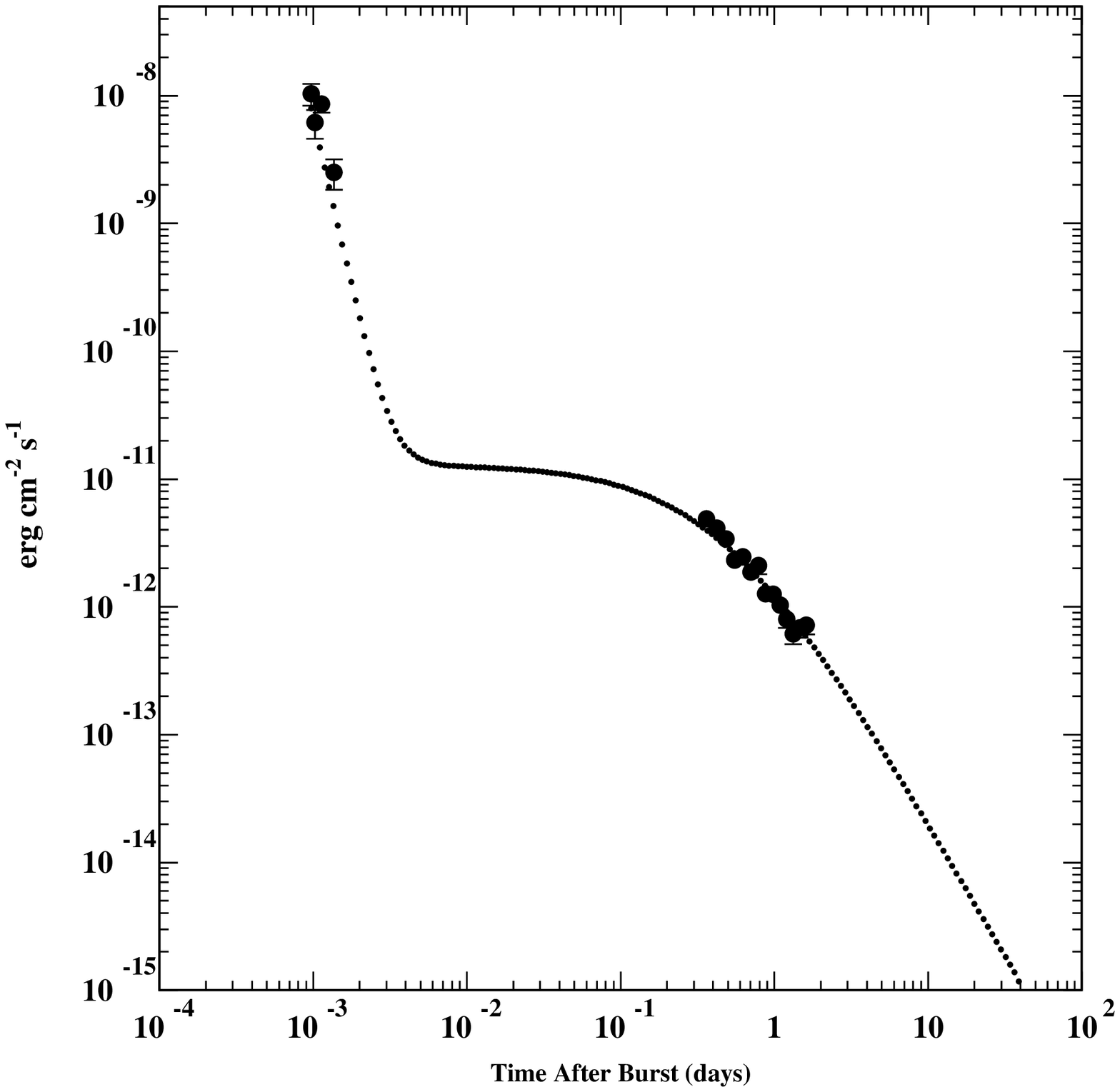,width=8.5cm} 
\caption{The early-time and late-time X-ray AG of GRB 990510 in the 2--10 
keV range as measured by Pian et al. (2001) 
fitted with Eq.~(\ref{Xdensity}) for a constant density along the 
CB trajectory. The same comments as in Fig~(\ref{X228}) apply here.} 
\vspace*{-0.5cm} 
\label{X510} 
\end{center} 
\end{figure} 
 
 
\begin{figure} 
\begin{center} 
\vspace*{.003cm} 
\hspace*{-0cm} 
\epsfig{file=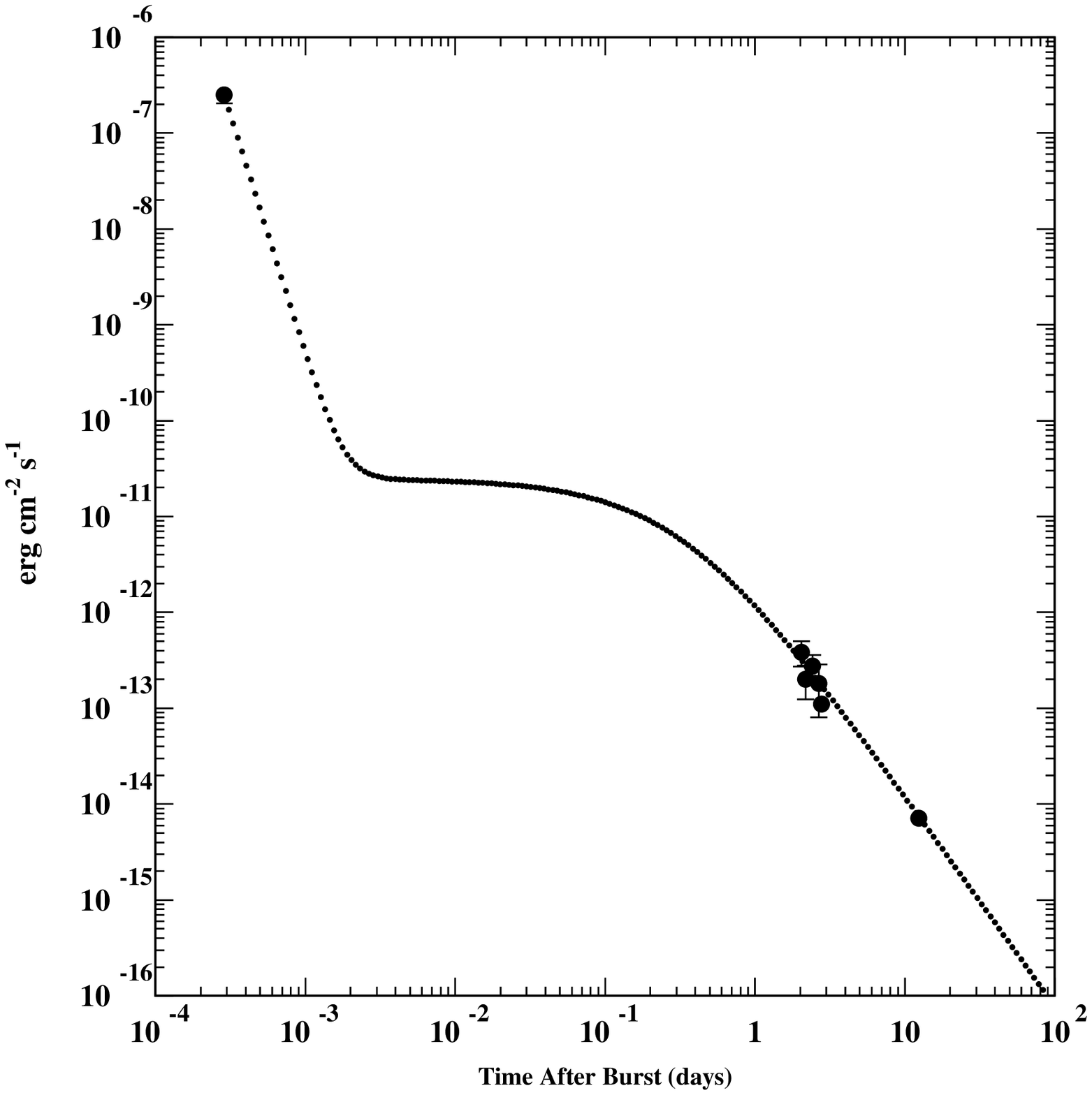,width=8.5cm} 
\caption{The early-time and late-time X-ray AG of GRB 000926 in the 2--10 
keV range as measured by Piro et al. (2001), fitted with 
Eq.~(\ref{Xdensity}) for a constant density along the 
CB trajectory. The same comments as in Fig~(\ref{X228}) apply here.} 
\vspace*{-0.5cm} 
\label{X926} 
\end{center} 
\end{figure}


\begin{figure} 
\begin{center} 
\vspace*{.003cm} 
\hspace*{-0cm} 
\epsfig{file=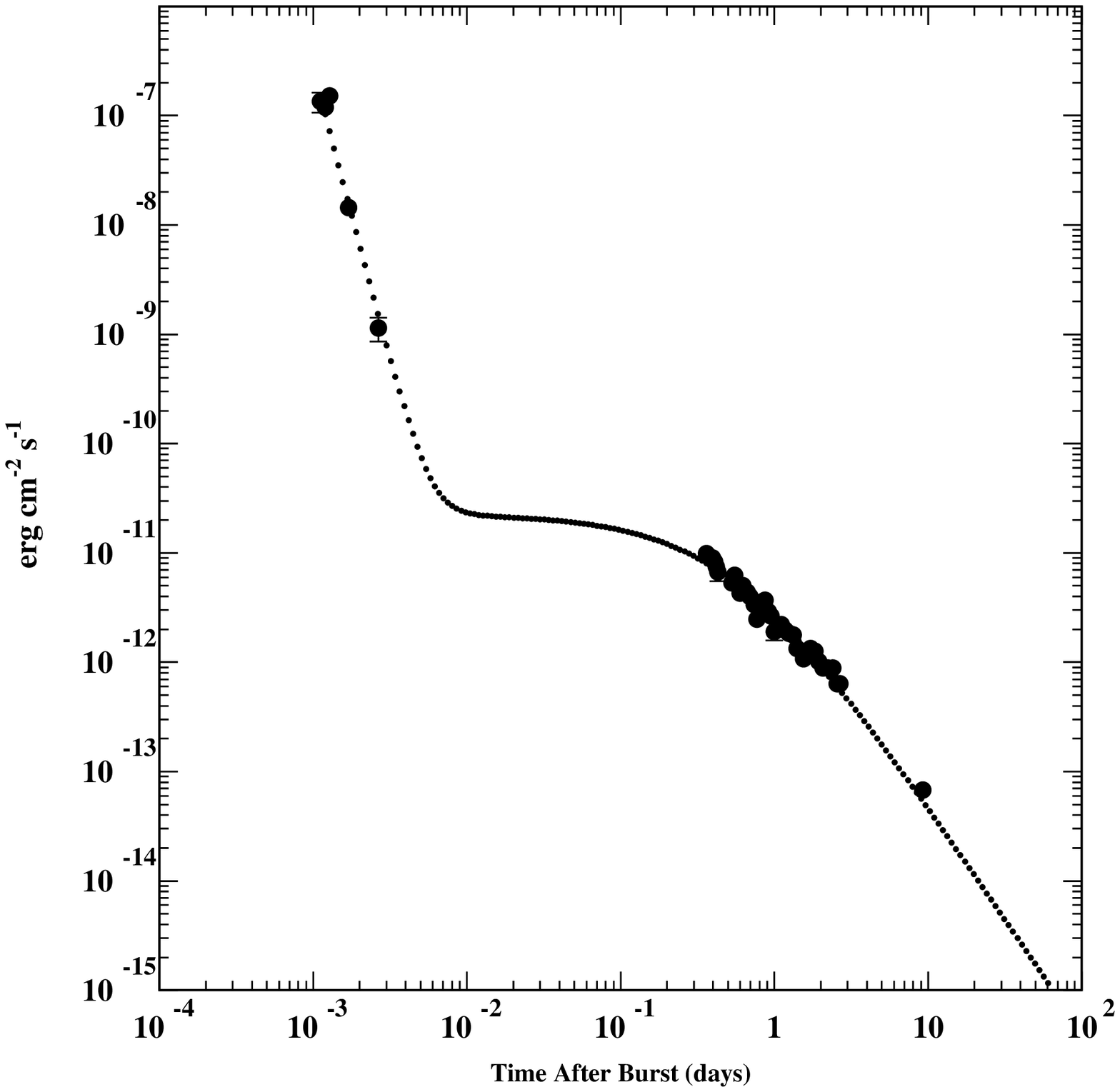,width=8.5cm} 
\caption{The early-time and late-time X-ray AG of GRB 010222 in the 2--10 
keV range as measured by In `t Zand et al. (2001), fitted with 
Eq.~(\ref{Xdensity}) for a constant density along the 
CB trajectory. The same comments as in Fig~(\ref{X228}) apply here.} 
\vspace*{-0.5cm} 
\label{X222} 
\end{center} 
\end{figure} 
 
 
\begin{figure} 
\begin{center} 
\vspace*{.2cm} 
\hspace*{-0cm} 
\epsfig{file=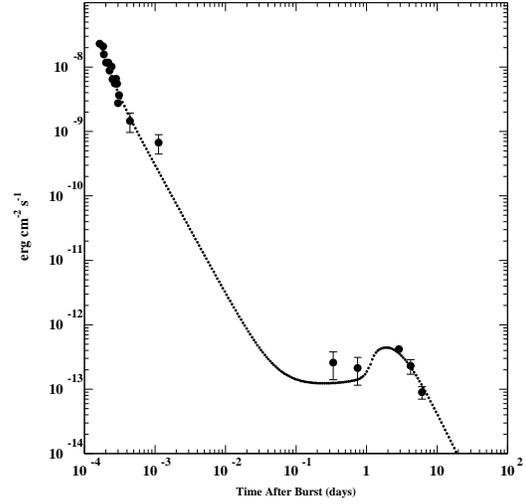,width=7.5cm} 
\caption{The early-time and late-time X-ray AG of GRB 970508 in the 2--10 
keV range as measured by Piro et al. (1998), fitted with 
Eq.~(\ref{Xdensity}) for a $\rm \sim 1/r^2$ plus constant density along the 
CB trajectory. The contribution of the X-ray line observed before 0.8 d was 
subtracted from the data. This reduces the two observed points between 
0.2 and 1 day by a factor $\sim 0.39$. 
 The overall result is compatible with an effect that, at late times, 
is achromatic, since the late optical and X-ray AGs are both
proportional to $\rm n_e$; see Fig.~(\ref{jump508}) 
for the optical counterpart.} 
\vspace*{-0.5cm} 
\label{X508} 
\end{center} 
\end{figure} 
 
\begin{figure} 
\begin{center} 
\vspace*{.003cm} 
\hspace*{-0cm} 
\epsfig{file=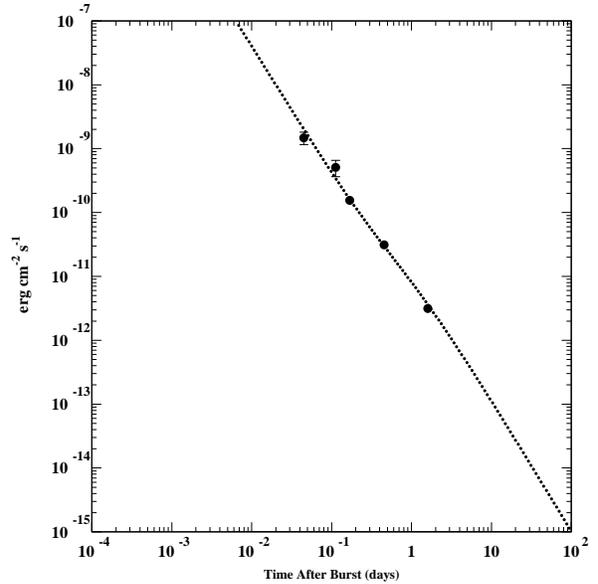,width=8.5cm} 
\caption{The early-time and late-time X-ray AG of GRB 991216 in the 2--10 
keV range as measured by Corbet and Smith (1999) and  Takeshima et al. 
(1999) fitted with Eq.~(\ref{Xdensity}) for a density declining like 
$\rm r^{-2}$ plus a constant  density along the CB trajectory.} 
\vspace*{-0.5cm} 
\label{X1216} 
\end{center} 
\end{figure}

 
\begin{figure}[t] 
\begin{tabular}{cc} 
\hskip 2truecm 
\vspace*{2cm} 
\hspace*{-2.1cm} 
\epsfig{file=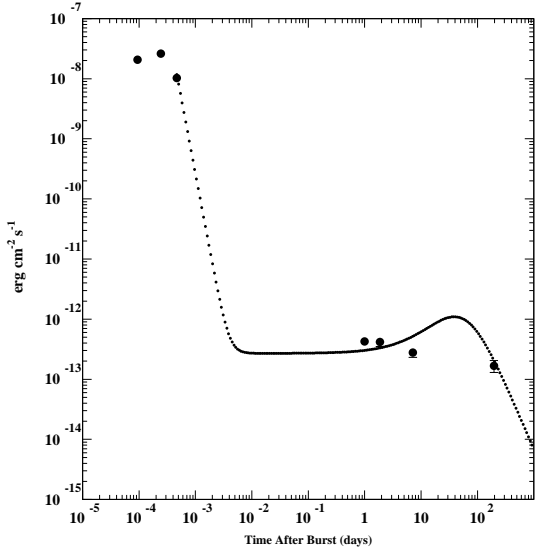, width=8cm} \\ 
\hspace*{-0.4cm} 
\epsfig{file=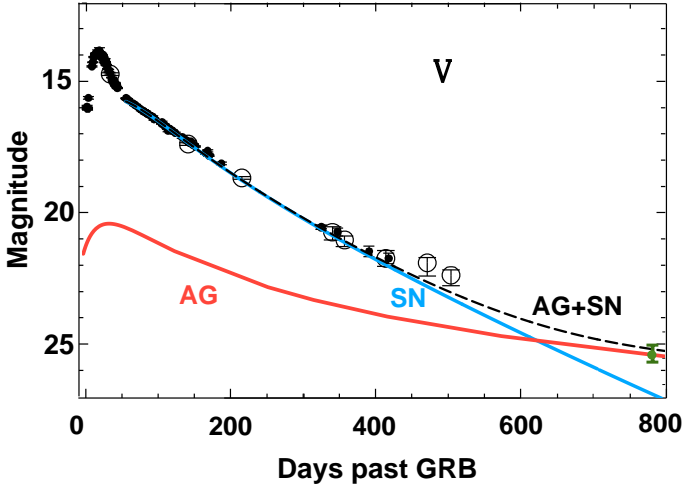,width=9cm} 
\end{tabular} 
\caption{Upper panel: A fit to the X-ray afterglow of the SN1998bw/GRB 980425 
pair. We call ``plateau'' the slowly-declining late measurements. 
Lower panel: The V-band light curve of 
the same pair, with   
the blue ``SN'' curve a fit to the SN by Sollerman et al.~(2000), 
dominated after day $\sim\!40$ by $^{56}$Co decay. 
 The red ``AG'' curve is our prediction for the CB-induced 
AG component, as given by Eq.~(\ref{fluxdensity}), 
with the parameters determined from the X-ray AG fit in the 
upper panel. The SN contribution 
dominates up to day $\sim\! 600$. The last point is an 
HST measurement at day 778, that precisely agrees with the (dashed) 
SN plus CB prediction for the total AG. For an earlier version of these 
results, see DD2000a.} 
\vspace*{-0.5cm} 
\label{425} 
\end{figure} 
 
 
\begin{figure} 
\begin{center} 
\vspace*{.003cm} 
\hspace*{-0cm} 
\epsfig{file=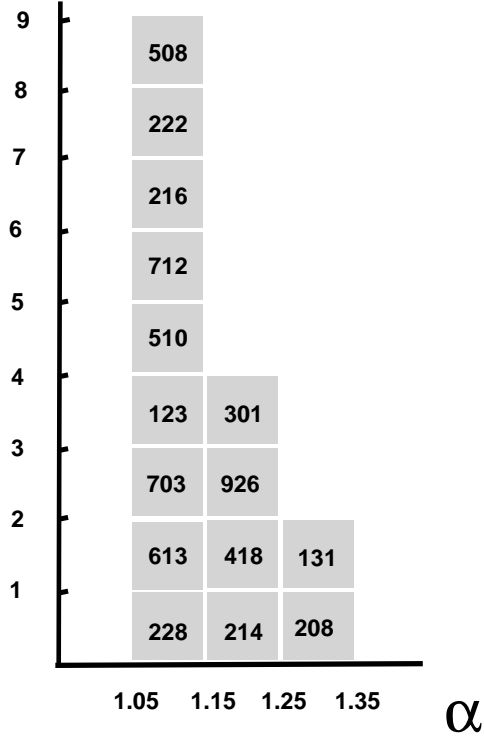,width=8.7cm} 
\caption{Distribution of $\alpha$ values, from the fits to the optical AGs. 
The prediction is $\alpha \approx 1.1$. 
Binning from $\alpha=1.06$ to 1.26 would have made this 
distribution look even more impressively narrow. 
The GRBs are labelled by the last three digits of their date.} 
\vspace*{-0.5cm} 
\label{alphadist} 
\end{center}  
\end{figure} 
 
 
\begin{figure} 
\begin{center} 
\vspace*{.003cm} 
\hspace*{-0cm} 
\epsfig{file=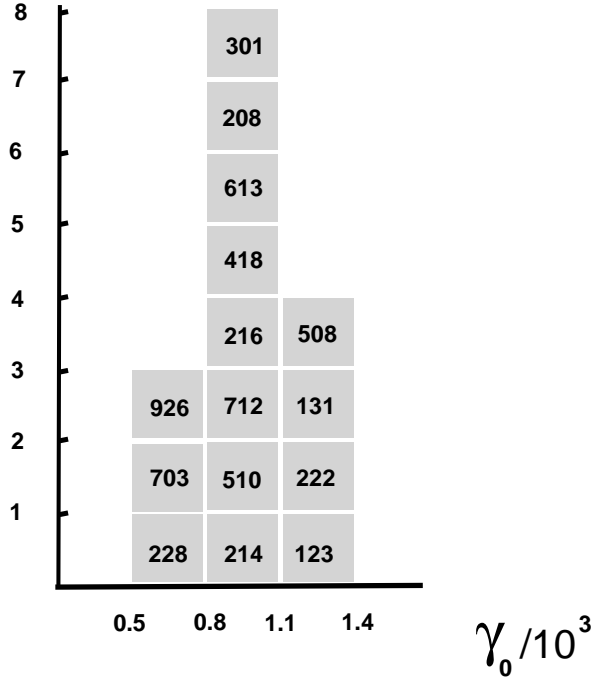,width=8.7cm} 
\caption{Distribution of $\gamma_0$ values from the fits to the optical AGs. 
The expectation is $\gamma_0\sim 10^3$. 
The GRBs are labelled by their last three digits.} 
\vspace*{-0.5cm} 
\label{gammadist} 
\end{center}  
\end{figure}

\begin{figure} 
\begin{center} 
\vspace*{.003cm} 
\hspace*{-0cm} 
\epsfig{file=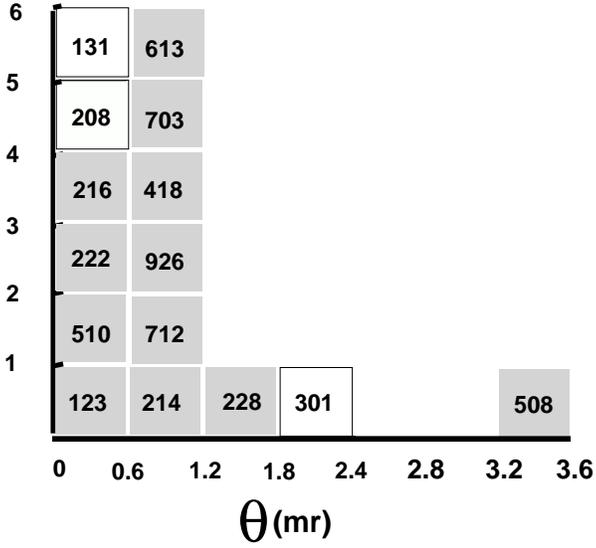,width=8.7cm} 
\caption{Distribution of $\theta$ values, in milliradians, from the fits to the 
optical AGs. 
The GRBs are labelled by their last three digits.} 
\vspace*{-0.5cm} 
\label{thetadist} 
\end{center}  
\end{figure} 
 
\begin{figure} 
\begin{center} 
\vspace*{.003cm} 
\hspace*{-0cm} 
\epsfig{file=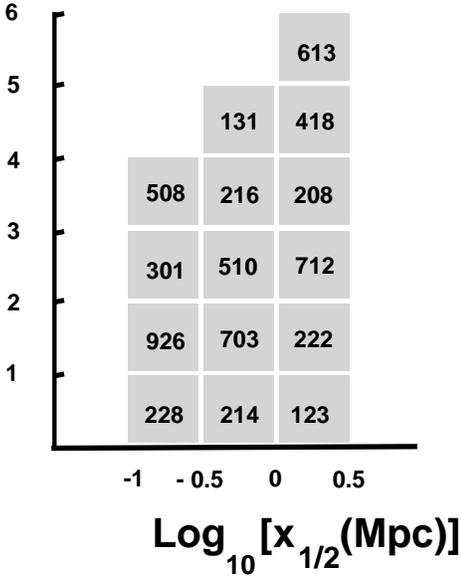,width=8.7cm} 
\caption{Distribution of $\rm 
Log_{10}[x_\infty(Mpc)]$ values, from the fits to the 
optical AGs. The expectation 
from Eq.~(\ref{gamoft}) is 0.114. 
The GRBs are labelled by their last three digits.} 
\vspace*{-0.5cm} 
\label{xdist} 
\end{center}  
\end{figure} 
 
\begin{figure} 
\begin{center} 
\vspace*{.003cm} 
\hspace*{-0cm} 
\epsfig{file=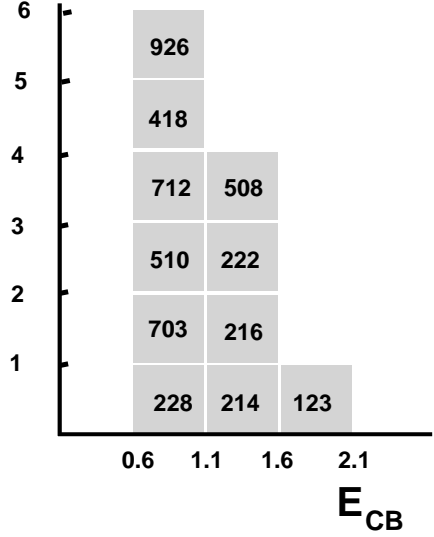,width=8.7cm} 
\caption{Distribution on $\rm E_\gamma^{CB}$ values, 
as given by Eq.~(\ref{ECBrest}), in units of $10^{44}$ erg. 
The GRBs are labelled by their last three digits.} 
\vspace*{-0.5cm} 
\label{Edist} 
\end{center}  
\end{figure} 
 
\end{document}